\documentclass[%
 reprint,
superscriptaddress,
nofootinbib,
 amsmath,amssymb,
 aps,
prb,
floatfix,
]{revtex4-2}
\usepackage{physics}
\usepackage{mdframed}

\usepackage{graphicx}
\usepackage{dcolumn}
\usepackage{bm}
\usepackage{bbold}
\usepackage{soul}
\usepackage{dsfont}

\usepackage[hidelinks,colorlinks=true,linkcolor=blue,citecolor=blue]{hyperref}

\usepackage{amsthm}
\usepackage{mathdots}
\theoremstyle{definition}
\newtheorem*{definition}{Definition}
\newtheorem{theorem}{Theorem}
\newtheorem{proposition}{Proposition}

\usepackage{tikz}
\usetikzlibrary{hobby}
\def\TrRthree{\tikz[baseline=0.5ex]{
  \foreach \c in {(0, 0), (0.5, 0), (1, 0), (1.5, 0), (2, 0), (2.5,0)} \fill \c circle (0.5mm);
  \filldraw[draw=black, opacity=0.2] (-0.1, -0.1) rectangle ++(0.7, 0.2);
  \filldraw[draw=black, opacity=0.2] (0.9, -0.1) rectangle ++(0.7, 0.2);
  \filldraw[draw=black, opacity=0.2] (1.9, -0.1) rectangle ++(0.7, 0.2);
  \draw[thick] (0.0, 0) to [curve through = {(0.73, 0.3) (1.5, 0) (2.0, 0) (2.1, 0.05)}] (2.15, 0.04);
  \draw[thick] (0, 0) to [curve through = {(0.2, -0.3)  (1.22, -0.6) (2.0, -0.6) (2.5, 0) (1.8, -0.3)}] (1.0, 0);
  \draw[thick] (1.0, 0) to [curve through = {(0.5, 0) (1.9, -0.45) (2.1, -0.4)}] (2.2, -0.04);
}
}

\def\TrRblock{\tikz[baseline=-0.2ex]{
  \foreach \c in {(0, 0.05), (0.5, 0.05)} \fill \c circle (0.5mm);
  \filldraw[draw=black, opacity=0.2] (-0.1, -0.05) rectangle ++(0.7, 0.22);
}
}

\def\TrRtwoblock{\tikz[baseline=-0.2ex]{
  \foreach \c in {(0, 0.05), (0.5, 0.05), (1, 0.05), (1.5, 0.05)} \fill \c circle (0.5mm);
  \filldraw[draw=black, opacity=0.2] (-0.1, -0.05) rectangle ++(0.7, 0.22);
  \filldraw[draw=black, opacity=0.2] (0.9, -0.05) rectangle ++(0.7, 0.22);
  \draw[thick] (1.0, 0.05) to [curve through = {(0.75, 0.2)}] (0.5, 0.05);
  \draw[thick] (0.0, 0.05) to [curve through = {(0.75, 0.3)}] (1.5, 0.05);
}
}

\def\TrRfour{\tikz[baseline=0.0]{
  \foreach \c in {(0, 0), (0.5, 0), (1, 0), (1.5, 0), (2, 0), (2.5,0), (3.0,0), (3.5,0)} \fill \c circle (0.5mm);
  \filldraw[draw=black, opacity=0.2] (-0.1, -0.1) rectangle ++(0.7, 0.2);
  \filldraw[draw=black, opacity=0.2] (0.9, -0.1) rectangle ++(0.7, 0.2);
  \filldraw[draw=black, opacity=0.2] (1.9, -0.1) rectangle ++(0.7, 0.2);
  \filldraw[opacity=0.2] (2.9, -0.1) rectangle ++(0.7, 0.2);
  \draw[thick] (0, 0) to [closed, curve through = {(1.5, 0) (2, 0) (3.5, 0)}] (1.7, -0.5);
  \draw[thick] (0.5, 0) to [closed, curve through = {(1.0, 0) (1.75, -0.12) (2.5, 0) (3.0, 0)}] (1.7, -0.25);
}
}

 \def\reddot{\tikz[baseline=-0.5ex]{
			\fill[red] (0,0) circle (1.5pt) coordinate (A);
			}
	}

\def\greendot{\tikz[baseline=-0.5ex]{
			\fill[teal] (0,0) circle (1.5pt) coordinate (A);
			}
	}

\def\bluedot{\tikz[baseline=-0.5ex]{
			\fill[blue] (0,0) circle (1.5pt) coordinate (A);
			}
	}

\def\longstate{\tikz[baseline=-0.5ex]{
			\fill[red] (3.0ex, 0) circle (1.5pt) coordinate (B);
			\fill[teal] (6.0ex,0) circle (1.5pt) coordinate (C);
			\fill[teal] (9.0ex, 0) circle (1.5pt) coordinate (D);
			\fill[red] (12.0ex, 0) circle (1.5pt) coordinate (E);
			\fill[blue] (15.0ex, 0) circle (1.5pt) coordinate (G);
			\fill[teal] (18.0ex, 0) circle (1.5pt) coordinate (H);
			\fill[teal] (21.0ex, 0) circle (1.5pt) coordinate (I);
			\fill[red] (24.0ex, 0) circle (1.5pt) coordinate (J);
			\draw[teal, thick] (C)--(D);
			\draw[red, thick] (B) .. controls (6.0ex, 1.5ex) and (9.0ex, 1.5ex) .. (E);
			\draw[teal, thick] (H) -- (I);
			\node at (3ex, -2.0ex) {$0$};
			\node at (6ex, -2.0ex) {$1$};
			\node at (9ex, -2.0ex) {$1$};
			\node at (12ex, -2.0ex) {$0$};
			\node at (15ex, -2.0ex) {$2$};
			\node at (18ex, -2.0ex) {$1$};
			\node at (21ex, -2.0ex) {$1$};
			\node at (24ex, -2.0ex) {$0$};
	}
}

\def\PFshortstate{\tikz[baseline=-0.5ex]{
			\fill[red] (0.0ex, 0) circle (1.5pt) coordinate (A);
			\fill[blue] (2.0ex,0) circle (1.5pt) coordinate (B);
			\fill[blue] (4.0ex, 0) circle (1.5pt) coordinate (C);
			\fill[red] (6.0ex, 0) circle (1.5pt) coordinate (D);
			\draw[blue, thick] (B)--(C);
			\draw[red, thick] (A) .. controls (2.0ex, 1.5ex) and (4.0ex, 1.5ex) .. (D);
	}
}

\def\dimer{\tikz[baseline=-0.5ex]{
			\fill (0,0) circle (1.5pt) coordinate (A);
			\fill (3.0ex, 0) circle (1.5pt) coordinate (B);
			\draw[black, thick] (A)--(B);}
}

\def\twodot{\tikz[baseline=-0.5ex]{
			\fill (0,0) circle (1.5pt) coordinate (A);
			\fill (3.0ex, 0) circle (1.5pt) coordinate (B);
		}
	}

\def\twodimer{\tikz[baseline=-0.5ex]{
		\fill (0,0) circle (1.5pt) coordinate (A);
		\fill (3.0ex, 0) circle (1.5pt) coordinate (B);
		\fill (6.0ex,0) circle (1.5pt) coordinate (C);
		\fill (9.0ex, 0) circle (1.5pt) coordinate (D);
		\draw[black, thick] (B)--(C);
		\draw[black, thick] (A) .. controls (3.0ex, 1.5ex) and (6.0ex, 1.5ex) .. (D);
	}
}

\begin{document}

\preprint{APS/123-QED}
\title{Highly entangled stationary states from strong symmetries}
\author{Yahui Li}
\email{yahui.li@tum.de}
\affiliation{
Technical University of Munich, 
TUM School of Natural Sciences, 
Physics Department,
Lichtenbergstr. 4,
85748 Garching,
Germany
}%
\affiliation{Munich Center for Quantum Science and Technology (MCQST), Schellingstr. 4, 80799 M{\"u}nchen, Germany}
\author{Frank Pollmann}
\email{frank.pollmann@tum.de}
\affiliation{
Technical University of Munich, 
TUM School of Natural Sciences, 
Physics Department,
Lichtenbergstr. 4,
85748 Garching,
Germany
}%
\affiliation{Munich Center for Quantum Science and Technology (MCQST), Schellingstr. 4, 80799 M{\"u}nchen, Germany}

\author{Nicholas Read}
\email{nicholas.read@yale.edu}
\affiliation{Department of Physics, Yale University, P.O. Box 208120, New Haven, CT 06520-8120, USA}
\affiliation{Department of Applied Physics, Yale University, P.O. Box 208284, New Haven, CT 06520-8284, USA}

\author{Pablo Sala}%
\email{psala@caltech.edu}
\affiliation{Department of Physics and Institute for Quantum Information and Matter, California Institute of Technology, Pasadena, CA 91125, USA}
\affiliation{Walter Burke Institute for Theoretical Physics, California Institute of Technology, Pasadena, CA 91125, USA}

\date{\today}

\begin{abstract}
We find that the presence of strong non-Abelian symmetries can lead to highly entangled stationary states even for unital quantum channels.
We derive exact expressions for the bipartite logarithmic negativity, Rényi negativities, and operator space entanglement for stationary states restricted to one symmetric subspace, with focus on the trivial subspace.
We prove that these apply to open quantum evolutions whose commutants, characterizing all strongly conserved quantities, correspond to either the universal enveloping algebra of a Lie algebra or to the Read-Saleur commutants.  
The latter provides an example of quantum fragmentation, whose dimension is exponentially large in system size.
We find a general upper bound for all these quantities given by the logarithm of the dimension of the commutant on the smaller bipartition of the chain.
As Abelian examples, we show that strong U($1$) symmetries and classical fragmentation lead to separable stationary states in any symmetric subspace.
In contrast, for non-Abelian SU$(N)$ symmetries, both logarithmic and Rényi negativities scale logarithmically with system size. 
Finally, we prove that while Rényi negativities with $n>2$ scale logarithmically with system size, the logarithmic negativity (as well as generalized Rényi negativities with $n<2$) exhibits a volume law scaling for the Read-Saleur commutants.
Our derivations rely on the commutant possessing a Hopf algebra structure in the limit of infinitely large systems, and hence also apply to finite groups and quantum groups.
\end{abstract}

\maketitle


\section{Introduction}

The interplay between entanglement and symmetries is key to providing a sharp characterization of the rich behavior of closed quantum many-body systems. On the one hand, it is essential for the characterization of quantum phases of matter at low energies, using the nature of quantum correlations as a guiding principle (see e.g., ~\cite{Sachdev_2011, Book_Wen,zeng2019quantum}). For example, one distinguishes short-range entangled symmetry-protected phases from long-range intrinsically topological ones, the former relying on the presence of symmetries~\cite{2010_chen_LRE, 2010_Pollmann_ES_topo, 2011_Chen_classification}. On the other hand, much effort has been devoted
to understanding quantum thermalization in closed systems out of equilibrium (see e.g., Ref.~\cite{2016_Alessio_ETH} for a review). Quench dynamics of generic many-body Hamiltonians at finite energy densities are expected to develop long-range quantum correlations and eventually thermalize. Yet, strong disorder can stop this spreading, keeping the system in a many-body localized phase where entanglement grows only slowly~\cite{2008_prosen_MBL_XXZ, 2012_Pollmann_MBL, 2014_Nandkishore_MBL, Abanin_2019_MBL}. Once again symmetries, especially non-Abelian ones, have something to say. 
%
It turns out that non-Abelian symmetries can increase the entanglement entropy of random pure states and highly excited states of random Hamiltonians~\cite{2023_Majidy_non-Abelian_Page, 2023_Patil_average_pure_state_entanglement_SU2}, and destabilize many-body localization~\cite{2017_Protopopov_SU2_MBL}.

Current research is extending the connection between symmetries and entanglement to open quantum systems. 
A unital quantum channel in the absence of symmetries will lead to a featureless infinite temperature mixed state, which is stripped of any quantum correlations. This matches our expectation that a structureless (high temperature) environment will completely decohere the system. Nonetheless, this fate is not necessarily unavoidable. One way to do so is to fine tune the environment to create exotic non-equilibrium dynamics~\cite{ 2008_zoller_Lindblad_pre_state, 2009_verstraete_quantum_state_engineer, 2019_buca_jaksch_non_statioanry_dissipation, 2023_symmetry_induced_DFS, 2023_google_engineered_dissipation}. More recently, it has been also shown that certain symmetric decoherence quantum channels (even when unital) can lead to non-trivial stationary states. For example, the phenomenon of strong-to-weak spontaneous symmetry breaking (sw-SSB) triggers long-range order in mixed states as measured by non-linear observables in the density matrix when imposing Abelian symmetries~ \cite{LeeYouXu2022, ma2024topological, lessa2024strongtoweak, 2024_sala_SSSB}. Yet, the latter examples involving $\mathbb{Z}_2$ sw-SSB only lead to stationary states with zero mixed-state entanglement.

Recently, non-Abelian symmetries have been found to generate a larger amount of entanglement than their Abelian counterparts in open quantum evolutions.
For example, SU($2$) symmetric monitored circuits can produce non-trivial entanglement even in the measurement-only limit~\cite{2023_Majidy_SU2_measurement}, unlike their U$(1)$ symmetric counterparts~\cite{2022_Agrawal_U1_monitored_circuits}. 
Concurrently, Ref.~\cite{2023_li_HSF_open} studied open quantum dynamics of systems with quantum fragmentation which possesses exponentially many non-Abelian conserved quantities. 
Starting from a product initial state $\ket{\psi}=\ket{\uparrow}^{\otimes L}$, the system can evolve to highly entangled stationary states as measured by the logarithmic negativity.
There, the commutant algebra formalism was utilized as a powerful method to characterize strongly conserved quantities, i.e., those that commute with every Hamiltonian term as well as every jump or Kraus operator~\cite{2000_Zanardi_quantuminfo, 2007_DFS_Bartlett, 2008_Baumgartner_math_Lindblad_2, 2023_li_HSF_open}.
It was then conjectured that the key ingredient for such  stationary states, is the lack of a common product state eigenbasis of elements in the maximal Abelian subalgebra of the commutant.
Therefore, this condition can also be satisfied by e.g., imposing a strong SU$(2)$ symmetry.
In fact, it was already pointed out that the maximally-mixed state within the total spin-zero SU$(2)$ symmetry sector has a logarithmic scaling of the entanglement of formation~\cite{2005_Livine_SU2_mixed_state_EoF}.
Moreover, an anomaly between a strong SO$(3)$ and weak translation symmetry has been shown to lead to bipartite non-separable symmetric states~\cite{lessa2024mixedstate}.

In this work, we study open dynamics with strong symmetries characterized by different commutants, including conventional U($1$) and SU$(N)$ symmetries, as well as classical and quantum fragmentation associated with exponentially large commutants. 
Specifically, we provide exact analytical expressions for different mixed-state entanglement proxies including logarithmic negativity, Rényi negativities, and operator space entanglement, for the stationary state restricted to the one-dimensional trivial irreducible representation (irrep) of the commutant (also referred to as the \emph{singlet} subspace).
Our results are based on the key observation that the basis states for the singlet subspace have a simple real-space bipartition form. 
We rigorously prove that for the systems considered in this work, a sufficient condition is that the commutant possesses a Hopf algebra structure in the infinite size limit.
This allows us to write exact expressions for the entanglement proxies, which are simply given in terms of the dimension of the irreps of the commutant and bond algebras.

We then characterize the entanglement of the stationary states in the singlet subspace for various commutants using these exact expressions.
For systems with Abelian U($1$) symmetry or classical fragmentation (i.e., fragmentation in product state basis), one finds that resulting stationary states are separable with zero logarithmic and Rényi negativities. Nonetheless, as also found in previous works~\cite{Wellnitz_2022, 2023_li_HSF_open}, the operator space entanglement scales with system size due to classical correlations.
In contrast, for strongly symmetric non-Abelian SU$(N)$ open quantum evolutions, or those preserving the exponentially large Read-Saleur (RS) commutants associated with quantum fragmentation, we show that highly entangled mixed stationary states --- as e.g., measured by the logarithmic negativity--- can be eventually reached (although not at finite times).
Intuitively, non-Abelian symmetric systems, when restricted to a symmetric subspace, generally require the presence of entangled basis states. Hence, the corresponding symmetric stationary state becomes a classical mixture involving entangled pure states, which leads to a finite mixed-state entanglement. For example, consider an SU$(2)$ symmetric system, whose Hilbert space structure is determined by the maximal Abelian subalgebra generated by $S^z_{\textrm{tot}}$ and $(S_{\textrm{tot}})^2$. The fact that these generators do not locally commute, rules out the possibility that all their common eigenstates are product states. This, for example, makes the singlet state a building block of the singlet subspace, contributing to the entanglement of the stationary state.
Using the exact expressions, we prove that for general SU$(N)$, both the logarithmic negativity and Rényi negativities scale logarithmically with system size.
Instead, for the RS commutants~\cite{2007_Read_Commutant}, the logarithmic negativity follows a volume law scaling, while the $n$-th Rényi negativities exhibit a logarithmic scaling for integer $n>2$.
We investigate these distinct scaling behaviors by introducing a generalized Rényi-$n$ negativity defined for arbitrary real $n>0$, which showcases a transition across $n=2$. 
While similar transitions for Rényi-$n$ entropies as a function of $n$ have been found in the literature for pure states~\cite{2015_Tarun_entanglement_sign_structure, 2011_Stephan_RenyiShannon_phase_transition}, to the best of our knowledge, this is the first such transition for mixed-state entanglement proxies. 
Note that while we only analytically prove the highly entangled stationary states restricted in the singlet subspace, we expect this behavior to hold when the stationary states has weight on a finite number of different symmetry subspaces~\cite{2023_li_HSF_open}. We also provide numerical data when the number of symmetry subspaces scales with system size.

The rest of this paper is organized as follows. 
We review the commutant algebra formulation to characterize strong symmetries in local quantum channels and the resulting stationary states in Sec.~\ref{sec:review_symmetries}. 
In Sec.~\ref{sec:exact_basis}, we provide an orthonormal basis of the singlet subspace in a real-space bipartition form, which is the key to studying bipartite entanglement of the stationary states in the singlet subspace.
Then in Sec.~\ref{sec:exact_entanglement}, we summarize the general exact expressions for the logarithmic negativity, Rényi negativities, and operator space entanglement of stationary states in the singlet subspace, based on the findings in Sec.~\ref{sec:exact_basis}. 
We then provide the asymptotic finite-size scaling of the half-chain entanglement for different commutant algebras given the exact expressions. 
Section~\ref{sec:conventional_sym} investigates the stationary state entanglement for conventional U($1$) and general SU$(N)$ symmetries. 
In Sec.~\ref{sec:fragment}, we study the entanglement of stationary states in systems with classical and quantum fragmentation, specifically for the Pair-Flip (PF) ~\cite{2018_Caha_pairflip} and Temperley-Lieb (TL) models~\cite{1990_TL, 2007_Read_Commutant}, respectively. For the latter, we investigate the transition from the volume-law logarithmic negativity to the logarithmically-scaling Rényi negativity. We conclude in Sec.~\ref{sec:conclusion}, including a discussion of open questions.
Finally, we consign more technical aspects of our work to the
Appendices.

\section{Review of symmetries in open systems}\label{sec:review_symmetries}

We consider local quantum channels, i.e., completely positive trace-preserving (CPTP) maps which can be written as the composition  
\begin{equation}\label{eq:quantum_channel}
    \rho_0 \to \rho=\mathcal{E}(\rho_0) = \prod_j \mathcal{E}_j(\rho_0), 
\end{equation}
of local channels
\begin{equation}
 \mathcal{E}_j(\rho_0)=\sum_{\alpha} K_{j,\alpha} \rho_0 K_{j,\alpha}^\dagger.
\end{equation}
These are given by Kraus operators $\{K_{j,\alpha}\}$ with finite local support, and $\sum_{\alpha} K_{j,\alpha}^\dagger K_{j,\alpha} = \mathbb{1}$.
We consider Hermitian Kraus operators $K_{j,\alpha} = K_{j,\alpha}^\dagger$, which implies that the quantum channel is also unital, i.e., $\sum_\alpha K_{j,\alpha} K_{j,\alpha}^\dagger = \mathbb{1}$. 
In this case, $\mathbb{1}/\dim(\mathcal{H})$ is a stationary state with $\mathcal{E}_j (\mathbb{1}) = \mathbb{1}$ for all $j$, i.e., appearing as a fixed point under the quantum channel. 
In the absence of symmetries, $\mathbb{1}/\dim(\mathcal{H})$ is the unique stationary state, indicating that the system fully decoheres and evolves to a trivial, separable state without any quantum correlations.
However, when a quantum channel exhibits a strong symmetry $O$, i.e., when each Kraus operator is invariant under the symmetry transformation or more in general, when $[K_{j,\alpha}, O] = 0$, the system allows for a rich structure of stationary states, which could exhibit non-trivial entanglement properties. 

A powerful framework to investigate the role of strong symmetries on the structure of the stationary state is given by the bond and commutant algebra language~\cite{2007_Read_Commutant, moudgalya_fragment_commutant_2022, 2023_sanjay_commutant_symmetries}, which has been applied in the field of quantum information~\cite{2000_Zanardi_quantuminfo, 2007_DFS_Bartlett}.
In this language, we can consider not only strong unitary symmetries~\cite{2012_Buca_Prosen, 2014_Albert_symmetries_Lindblad, 2020_Buca_Zhang}, but also any operator $O$ that commutes with every (Hermitian) Kraus operator.
The Kraus operators on a system of size $L$ generate a \emph{bond algebra},
\begin{equation}
    \mathcal{A}(L) = \langle  \{K_{j,\alpha}\}, \mathbb{1}\rangle, 
\end{equation}
which is given by all linear combinations of products of $K_{j,\alpha}$ for $j=1,\ldots, L$ (in this paper, all algebras are defined as including an identity element $\mathbb{1}$).
And the set of all operators $O$ that commute with every Kraus operator  form the \emph{commutant algebra}
\begin{equation}
    \mathcal{C}(L) = \{O: [O, K_{j,\alpha}] = 0, \forall j\,\in \{1,\dots,L\},\,\alpha\}.
\end{equation}
All elements $O\in \mathcal{C}(L)$ correspond to conserved quantities as they are fixed points of the dual evolution, $\mathcal{E}^\dagger(O) = O$.
By virtue of the quantum channel being unital, they are also fixed points, namely $\mathcal{E}(O) = O$. 
The formulation in terms of bond and commutant algebras is not restricted to Hermitian Kraus operators. In general, the bond algebra is defined as $\mathcal{A}(L) = \langle \{K_{j,\alpha}\}, \{K^\dagger_{j,\alpha}\}, \mathbb{1}\rangle$
such that it is closed under taking the adjoint. 
Nonetheless, the Hermitian Kraus operators are a natural choice to realize the strong symmetries we study in this work.
For example, a SU($2$) strongly-symmetric unital quantum channel can be generated by the local and symmetric Kraus operators, $K_{j,1} = \vec{S}_j \cdot \vec{S}_{j+1}$ and $K_{j,2} = \mathbb{1}-\vec{S}_j \cdot \vec{S}_{j+1}$, which are Hermitian.
An analogous statement holds for the RS commutants.
In a recent work~\cite{2023_li_HSF_open}, this formalism was applied to the case of Lindbladian dynamics with Hermitian jump operators,
which we naturally extend to the case of Hermitian local quantum channels here.
More generally, the collection of fixed points can be a different set of operators from those given by conserved quantities due to the non-Hermiticity of the quantum channel.
Discussion of symmetries, conserved quantities, and stationary states (fixed points) for Lindblad dynamics can be reviewed in Ref.~\cite{2012_Buca_Prosen, 2020_Buca_Zhang, 2014_Albert_symmetries_Lindblad}. 

Both the bond and commutant algebras are closed under taking adjoints, and each is the commutant of the other; hence they can be viewed as finite-dimensional von Neumann algebras~\cite{1998_lecture_von_Neumann_algebra}. This implies the semisimplicity of finite-dimensional algebras.
For our purposes, semisimplicity simply means that any finite-dimensional representation of these algebras
decomposes as a direct sum of irreps.
With the double commutant theorem~\cite{1998_lecture_von_Neumann_algebra, 2007_Read_Commutant, 2017_math_von_Neumann, moudgalya_fragment_commutant_2022}, the Hilbert space decomposes into the direct sum of tensor products of irreps of $\mathcal{C}(L)$ and $\mathcal{A}(L)$, 
\begin{equation} \label{eq:DoucCom}
    \mathcal{H}^{(L)} = \bigoplus_\lambda \left( \mathcal{H}^{\mathcal{C}(L)}_\lambda \otimes \mathcal{H}^{\mathcal{A}(L)}_\lambda \right),
\end{equation}
where $\lambda$ labels irreps. 
The set of irreps $\lambda$ depends on the algebras, which depend on system size $L$. This is the range of summation over $\lambda$ in the subsequent discussions.  For example, $\lambda = 0, 1, \ldots, L/2$ for SU($2$) symmetry on a spin-$1/2$ chain with even system size $L$. 
The $\mathcal{H}^{\mathcal{A}(L)}_{\lambda}$ are $D^{(L)}_\lambda$-dimensional subspaces that correspond to irreps of the bond algebra, where $D^{(L)}_{\lambda}$ is dependent on the system size.
They are dynamically-disconnected subspaces, which are commonly called Krylov subspaces.
The $\mathcal{H}^{\mathcal{C}(L)}_\lambda$ are $d_\lambda^{(L)}$-dimensional subspaces (irreps of the commutant), which gives the degeneracy of $\mathcal{H}^{\mathcal{A}(L)}_{\lambda}$ subspaces.
In the following, $d_{\lambda}$ are taken to be independent of system size $L$, which is the case for the commutants we studied in this work (see detailed discussion in Sec.~\ref{sec:exact_basis}). 
The total number of Krylov subspaces $\mathcal{H}_\lambda^{\mathcal{A}(L)}$ is given by $K=\sum_\lambda d_\lambda$.  
The degeneracy is $d_\lambda \equiv 1$ for Abelian commutants, while it can become larger $d_\lambda \geq 1$ for non-Abelian ones. 
In the following, we label the degenerate Krylov subspaces as $\mathcal{H}_{\lambda,m}^{\mathcal{A}(L)}$ for fixed $\lambda$ and $m$. The basis states of the $\mathcal{H}_{\lambda,m}^{\mathcal{A}(L)}$ are given by $\{|\lambda, m; a\rangle\}$, with $a=1,\ldots, D_\lambda$.

For example, for an SU($2$)-symmetric spin-$1/2$ chain with length $L$, the Hilbert space decomposes into a direct sum of spin-$\lambda$ symmetry sectors $\mathcal{H} = \bigoplus_\lambda \mathcal{S}_\lambda$, with $\lambda = 0, \ldots, L/2$.
Each symmetry sector further decomposes into the tensor product $\mathcal{S}_\lambda = \mathcal{H}^{\mathcal{C}(L)}_\lambda \otimes \mathcal{H}^{\mathcal{A}(L)}_\lambda$, with $\mathcal{H}^{\mathcal{C}(L)}_\lambda$ and $\mathcal{H}^{\mathcal{A}(L)}_\lambda$ corresponds to irreps of SU($2$) and the symmetric group $S_L$, respectively. 
This well-known decomposition for the case of su$(2)$ (Lie algebra of the SU($2$) group) is known as Frobenius-Schur-Weyl duality~\cite{fulton_representation_2004}.
The degenerate Krylov subspaces can be labeled by the quantum number $m=-\lambda, -\lambda+1,\ldots, \lambda$, which are the eigenvalues of total magnetization $S^z_{\mathrm{tot}} = \sum_j S_j^z$.
The total charges $S^{\alpha}_{\mathrm{tot}}$ are non-commuting among each other, leading to a non-Abelian commutant, and hence these do not share a common eigenbasis. 
The maximal Abelian subalgebra $\mathcal{M}$ of $\mathcal{C}$ can be generated by $\{\vec{S}^2, S_{\mathrm{tot}}^\alpha\}$, i.e., by $\vec{S}^2 = (S^x_{\mathrm{tot}})^2 + (S^y_{\mathrm{tot}})^2 + (S^z_{\mathrm{tot}})^2$ and one of the total charges. 
The maximal Abelian subalgebra $\mathcal{M}(L)$ contains a maximal set of commuting conserved quantities, which can be used to uniquely label the Krylov subspaces such as $(\lambda, m)$ corresponding to the choice of $\mathcal{M}(L) = \{\vec{S}^2, S_{\mathrm{tot}}^z\}$ above.  

The advantage of using the commutant algebra formulation is that it applies to conventional symmetries, but also to the case of Hilbert space fragmentation (HSF)~\cite{2020_sala_ergodicity-breaking, 2020_khemani_local, 2020_SLIOMs, moudgalya_fragment_commutant_2022}, where conserved quantities do not in general generate a symmetry group.
The HSF systems possess exponentially many conserved quantities due to highly constrained dynamics.
Therefore, HSF can be distinguished from conventional symmetries by the scaling of the dimension of the commutant $\dim [\mathcal{C}(L)] = \sum_{\lambda} d_\lambda^2$ with system size~\cite{moudgalya_fragment_commutant_2022}.
For conventional symmetries, there is at most a polynomial number of Krylov subspaces. Therefore $\dim[\mathcal{C}(L)]$ scales at most polynomially with system size. In contrast, for systems with HSF, which have exponentially many Krylov subspaces~\cite{2020_khemani_local, 2020_sala_ergodicity-breaking, 2020_SLIOMs}, $\dim[\mathcal{C}(L)]$ scales also exponentially~\cite{moudgalya_fragment_commutant_2022}.

For quantum channels with Hermitian Kraus operators, we now show that the stationary state has a general expression given by the bond and commutant algebras that depends on the initial state. 
The Kraus operators are elements of the bond algebra, which have a matrix representation as $h_{\mathcal{A}} = \bigoplus_\lambda (\mathbb{1}_\lambda \otimes A(h_{\mathcal{A}}))$.
Therefore, the dissipative dynamics acts trivially on the subspaces $\mathcal{H}_\lambda^{\mathcal{C}(L)}$, while $\mathcal{H}_{\lambda}^{\mathcal{A}(L)}$ become fully decohered to a trivial identity within each subspace~\cite{2008_Baumgartner_math_Lindblad_2, 2023_li_HSF_open}. 
Namely, $\mathcal{H}_\lambda^{\mathcal{C}(L)}$  and $\mathcal{H}_\lambda^{\mathcal{A}(L)}$ 
are known as decoherence-free and decoherence-full subspaces, respectively~\cite{2007_DFS_Bartlett}
\footnote{Note that in Ref.~\cite{2007_DFS_Bartlett}, although a similar mathematical formulation is given, the authors study collective quantum channels with SU($2$) group elements as bond algebras.}.
Therefore, for an arbitrary initial state $\rho(t=0)$ restricted to $\mathcal{H}_{\lambda}^{\mathcal{C}(L)}\otimes \mathcal{H}_{\lambda}^{\mathcal{A}(L)}$, the stationary state is given by $\rho_{\lambda} = \sum_{m,m^\prime}\mathrm{Tr}(\Pi^\lambda_{m^\prime,m}\rho(t=0))\frac{\Pi^\lambda_{m,m^\prime}}{D_\lambda}$.
Here, $\Pi^{\lambda}_{m,m^\prime} = \sum_{a=1}^{D_{\lambda}} \ket{\lambda, m;a}\bra{\lambda, m^\prime; a}$ is a projector onto the diagonal subspace when $m=m^\prime$, and an intertwine between different degenerate subspaces when $m\neq m^\prime$.
We denote the projectors $\Pi_{m,m}^\lambda \equiv \Pi^\lambda_{m}$ in the following.
In particular, when the initial state lies within a Krylov subspace with both fixed $\lambda$ and $m$, the system evolves to the unique stationary state~\cite{2008_Baumgartner_math_Lindblad_2, 2023_li_HSF_open, 2024_Yoshida_Lindblad}
\begin{equation}\label{eq:general_stationary_state}
    \rho_{\lambda,m} = \frac{1}{D_\lambda}\Pi^\lambda_{m} = \frac{1}{D_\lambda} \sum_{a=1}^{D_{\lambda}} \ket{\lambda, m;a}\bra{\lambda, m; a}.
\end{equation}
A proof of the uniqueness of this state is provided in App.~\ref{app:stationary_state}.
Note that although Hermitian Kraus operators are considered in this work, the uniqueness of stationary states extends to non-Hermitian Kraus operators. This relies on the assumption that the set of operators $\{K_{j\alpha}\}$ \emph{or} $\{K_{j\alpha}^\dagger\}$ generates all the symmetry operators, which we will discuss in App.~\ref{app:stationary_state}.
%
In the following sections, we will mainly focus on the stationary state $\rho_0 = \Pi^{\lambda=0}/D_0$ in the singlet subspace ($\lambda = 0$) for which $d_0=1$ ($m=0$).

More generally, as mentioned above, the same stationary states can be reached by a Lindblad evolution $\dot{\rho}(t)=-i \sum_j J_j[ h_j,\rho(t)] + \sum_j \gamma_j (L_j\rho(t) L_j^\dagger - 1/2\{L_jL_j,\rho(t)\})$ with Hermitian jump operators $L_j$, and local Hamiltonian terms $h_j$, for any choice of parameters $J_j\in \mathbb{R}, \gamma_j\geq 0$.
This holds as long as the commutant associated with the bond algebra $\mathcal{A}(L) = \langle \{h_j\}, \{L_j\}, \mathbb{1}\rangle$ is the same.
In fact, for Lindbladian evolutions with random coupling and decay rates $J_j$, $\gamma_j$, a general expression of the stationary state can be found for generic (non-necessarily symmetric) initial states~\cite{2023_li_HSF_open},
\begin{equation} \label{eq:rho_gen}
    \rho  = \sum_{\lambda,m,m^\prime}\left((M^{t=0}_\lambda)_{m,m^\prime}\frac{\Pi^\lambda_{m,m^\prime}}{D_\lambda}\right). 
\end{equation}
Here $M^{t=0}_{\lambda}$ is a $d_\lambda \times d_\lambda$ matrix, with matrix elements $(M^{t=0}_{\lambda})_{m,m^\prime} = \mathrm{Tr}(\Pi_{m^\prime,m}^\lambda\rho(t=0))$. 
Therefore, $M^{t=0}_\lambda$ encodes the weight of initial states in different diagonal and degenerate subspaces.
Equation~\eqref{eq:rho_gen} is also a stationary state of the quantum channel dynamics introduced in Eq.~\eqref{eq:quantum_channel}.
Nonetheless, in general, we cannot rule out the possibility of additional off-diagonal structure when considering generic initial states.
Yet, such structures are apparently expected to only appear under non-generic conditions as suggested in Ref.~\cite {2012_Buca_Prosen}.
We extend this discussion in App.~\ref{app:stationary_state}. 
This subtlety will only be relevant when considering initial states with weight on different irreps $\lambda$ in Sec.~\ref{subsec:SU2_Haar}.

\section{Basis state in the singlet subspace as bipartition}\label{sec:exact_basis}
Open quantum dynamics in the absence of symmetries and subject to Hermitian (hence unital) dissipation leads to featureless stationary states for general initial conditions. 
In this work, we will show that in the presence of strong non-Abelian symmetries (commutants), the resulting mixed stationary states can be highly entangled.
This happens despite evolving low-entangled initial states as measured by both von Neumann entropy and all other entanglement proxies we study in this work.
We mainly focus on stationary states restricted to a non-degenerate subspace labeled by the trivial representation $\lambda_{\mathrm{tot}} = 0$ (e.g., the singlet subspace for SU($2$)) for different commutant algebras.
For open quantum dynamics, such stationary states can be (eventually) reached by arbitrary initial states lying within the singlet subspace. 
For example, the initial state can be given by the tensor product of singlets $|\psi_0\rangle = (\frac{1}{\sqrt{2}}(\ket{\uparrow \downarrow} - \ket{\downarrow \uparrow}))^{\otimes(L/2)}$ for SU($2$) symmetric dynamics, which has area-law entanglement.

\begin{figure}[bt]
\includegraphics[width=7cm, scale=1]{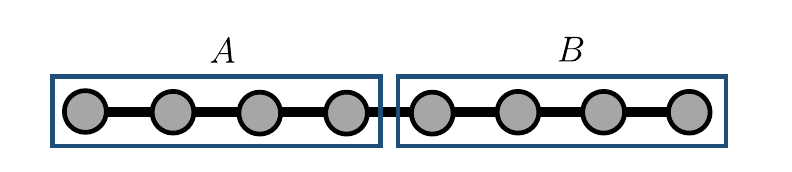}
\caption{\label{fig:half_chain} \textbf{Bipartition of a chain} to regions $A$ and $B$, with system size $L_A$ and $L_B$, respectively.}
\end{figure}

To calculate entanglement properties of stationary states, we perform a bipartition of the chain into $L=L_A+L_B$ as shown in Fig.~\ref{fig:half_chain}.
We focus on the global trivial representation $\lambda_{\mathrm{tot}} = 0$, for a compatible length of the chain $L$. 
The key observation is that quite generally ---and in particular for the commutants we study in this work--- there exists an orthonormal basis (ONB) for the $\lambda_{\mathrm{tot}} = 0$ subspace of $\mathcal{H}^{(L)}$ with dimension $D_0^{(L)}$, that has the following 
bipartitioned form:
\begin{equation} \label{eq:gen_triv0}
\begin{aligned}
    |\lambda_{\mathrm{tot}}=0;\, &\lambda; a,b\rangle \\ =& \frac{1}{\sqrt{d_\lambda}}\sum_{m=1}^{d_\lambda} \eta_{\lambda,m} \ket{\lambda, m;a}  \otimes \ket{\bar{\lambda}, \bar{m};b}.
\end{aligned}
\end{equation}
Here, $\lambda$ corresponds to an irrep of $\mathcal{C}(L_A)$ and $\bar{\lambda}$ to its dual irrep of $\mathcal{C}(L_B)$, with $\lambda$ running over irreps compatible with both lengths $L_A$, $L_B$.
As already stated below Eq.~\eqref{eq:DoucCom} and further discuss below, we assume that the dimensions $d_\lambda$ are independent of $L$ in the following.
Both irreps of the commutant ($\lambda$ and its dual $\bar{\lambda}$) have the same dimension $d_\lambda$.
Moreover, $a = 1, \ldots, D_{\lambda}^{(L_A)}$, $b = 1, \ldots, D_{\bar{\lambda}}^{(L_B)}$ run over ONBs for the irreps $\mathcal{H}_\lambda^{\mathcal{A}(L_A)} $, $\mathcal{H}_{\bar{\lambda}}^{\mathcal{A}(L_B)}$ respectively, and $D_0^{(L)}=\sum_{\lambda}D_{\lambda}^{(L_A)}D_{\bar{\lambda}}^{(L_B)}$.
The coefficient $\eta_{\lambda,m}$ is a phase with $|\eta_{\lambda,m}| = 1$, which arises depending on the choice of the dual basis. 
For example, see Eq.~\eqref{eq:SU2_basis_state_0} below for SU$(2)$ as a special case, the singlet state $\frac{1}{\sqrt{2}}(\ket{\uparrow \downarrow} - \ket{\downarrow \uparrow})$ for SU($2$) and $L=2$ can be easily written as Eq.~\eqref{eq:gen_triv0}.

Equation~\eqref{eq:gen_triv0} can be shown to hold for the universal enveloping algebra (UEA) of su$(N)$ (as well as for other Lie algebras), using the ladder operators analogous as for su($2$) (see Sec.~\ref{subsec:sun}). 
In fact, it turns out to hold in more generality (as it happens e.g., for the RS commutant) as long as the following two conditions are satisfied: (1) In the limit $L\to\infty$, the commutant possesses a Hopf algebra structure; and (2) both $\mathcal{A}(L)$, $\mathcal{C}(L)$ are semisimple.

Let's start with condition (1). In the following, we take $L^\prime$ with $L^\prime> L$. We assume there exists a system of compatible~\footnote{Here compatible means that for $L_1>L_2>L_3$, then $\phi_{L_1,L_3}=\phi_{L_2,L_3}\circ \phi_{L_1,L_2}$.} surjective algebra homomorphisms $\phi_{L^\prime,L}:\mathcal{C}(L^\prime) \to \mathcal{C}(L)$ for all $L,L^\prime$ $(L^\prime > L)$, and hence the ``inverse'' limit $ \mathcal{C}=\underleftarrow{\lim}_{L\to \infty}\mathcal{C}(L)$ exists~\cite{kassel2012quantum,2007_Read_Commutant}.
In order for this to hold, in many cases, we need to restrict the lengths $L, L^\prime$ to belong to a subsequence of integers, for example, that $L$, $L^\prime=0$ mod $N$ for general SU$(N)$ symmetric chains, while $L, L^\prime$ need to be even for the RS commutants~\cite{2007_Read_Commutant} (the use of the subsequence will be left implicit from here on; in fact, in the examples, the discussion can be extended to handle chains of all lengths, but this simplification is sufficient for our purposes).
From here it follows that any representation of $\mathcal{C}(L)$ is also a representation of $\mathcal{C}(L^\prime)$.
In particular, this also holds for irreps, i.e., an irrep of $\mathcal{C}(L)$ is also an irrep of $\mathcal{C}(L^\prime)$.
The same statements hold with $\mathcal{C}$ in place of $\mathcal{C}(L^\prime)$.
Hence all representations of $\mathcal{C}(L)$ for any $L$ can be viewed as
representations of $\mathcal{C}$. In particular, we label the isomorphism classes
of irreps of $\mathcal{C}$ (and hence also of $\mathcal{{C}}(L)$) by $\lambda$,
and note that $d_\lambda<\infty$ is independent of $L$.
Finally, we assume that $\mathcal{C}$ is not only an algebra, but it possesses a Hopf algebra structure (for a definition see App.~\ref{app:Hopf}).
That additional structure will be used to define the notion of a ``trivial" irrep $\lambda_{\mathrm{tot}}=0$ of $\mathcal{C}(L)$, to view a tensor product of irreps of $\mathcal{C}$ as a representation of $\mathcal{C}$, and to define a ``dual" representation for any given one; we will see that all of these play a role
in Eq.~\eqref{eq:gen_triv0}.

Condition (1) sounds very technical, but it actually holds for many physical systems of interest~\cite{kassel2012quantum,fulton_representation_2004}. Examples include SU$(N)$ symmetric systems, as well as those with RS commutants~\cite{2007_Read_Commutant}. Other examples of Hopf algebras that can arise in physical systems include the group algebra of any finite group, as well as quantum groups~\cite{kassel2012quantum}.

\begin{proposition} \label{th:Prop_1} Under condition (1), there is a well-defined notion of a (unique, up to isomorphism) trivial irrep $\lambda_{\mathrm{tot}}=0$
of $\mathcal{C}$, and then a normalized state within the trivial irrep can be constructed as in Eq.~\eqref{eq:gen_triv0} for any $\lambda$ with $a$ and $b$ dropped from
both sides.   
\end{proposition}
\noindent \textit{Proof.} See proof in App.~\ref{app:Prop_1}.\\

Let's now consider condition (2), which requires $\mathcal{A}(L)$ and $\mathcal{C}(L)$ to be semisimple.
In fact, the semisimplicity of $\mathcal{A}(L)$ and $\mathcal{C}(L)$ always holds, both algebras being closed under taking the adjoint of the elements. 
Semisimplicity (in this sense) also survives in the inverse limit $\mathcal{C}$ of $\mathcal{C}(L)$.
See App.~\ref{app:Prop_2} for additional details.
This leads to the second important Proposition:

\begin{proposition} \label{th:Prop_2}
Under conditions (1) and (2), the $\lambda_{\mathrm{tot}} = 0$ sector of $\mathcal{H}^{(L)}$ with dimension $D_0^{(L)}$ decomposes as
\begin{equation}
    \mathcal{H}^{\mathcal{A}(L)}_{{\lambda}_{\mathrm{tot} =0}} = \bigoplus_{\lambda}\left( \mathcal{H}_{\lambda}^{\mathcal{A}(L_A)} \otimes \mathcal{H}_{\bar{\lambda}}^{\mathcal{A}(L_B)}\right),
\end{equation}
with $D_0^{(L)}=\sum_{\lambda}D_{\lambda}^{(L_A)}D_{\bar{\lambda}}^{(L_B)}$ (the sum can be taken over all $\lambda$ because, for any $L^{\prime\prime}$, $D_\lambda^{(L^{\prime\prime})}=0$
for all except finitely many $\lambda$).
\end{proposition}
\noindent \textit{Proof.} See proof in App.~\ref{app:Prop_2}.\\

Combining Propositions \ref{th:Prop_1} and \ref{th:Prop_2}, we then conclude that
\begin{theorem} \label{th:theorem_1}
Under conditions (1) and (2) above, $\{|\lambda_{\mathrm{tot}}=0; \lambda; a,b\rangle\}$, as given by Eq.~\eqref{eq:gen_triv0}, forms an ONB of the $\lambda_{\mathrm{tot}} = 0$ sector of $\mathcal{H}^{(L)}$.
\end{theorem}

In the following section, we show that whenever the conditions of Theorem \ref{th:theorem_1} hold, one can obtain general closed-form expressions of various entanglement quantities of the stationary state within the $\lambda_{\mathrm{tot}=0}$ sector relying only on this ONB.
As we have seen, this includes systems whose commutants are the UEA of any Lie algebra, the RS commutants, the group algebra of any finite group, as well as quantum groups.

\section{Entanglement of stationary states in the singlet subspace}\label{sec:exact_entanglement}

We now use the decomposition in Eq.~\eqref{eq:gen_triv0} to exactly compute different entanglement quantities of the stationary state
\begin{equation}\label{eq:rho_singlet}
    \rho = \frac{\Pi^{\lambda_{\mathrm{tot}}=0}}{D_0},
\end{equation} 
for a bipartition of the system into two regions $A$ and $B$ as shown in Fig~\ref{fig:half_chain}. 
We show that the entanglement quantities can be fully expressed in terms of the dimensions $d_\lambda$ and $D_\lambda$ of irreps of the commutant $\mathcal{C}(L)$ and bond algebras $\mathcal{A}(L)$ respectively, which provide a general upper bound given by the dimension of the commutant. 
We denote by $L_{\min} = \min(L_A, L_B)$ the shorter length, which limits the allowed irreps; and define $\mathcal{C}_{\mathrm{min}}=\mathcal{C}(L_{\min})$.
Moreover, we assume open boundary condition (OBC) in the following.
At the end of this section, we will provide a summary of asymptotic scalings of entanglement for specific commutants in Table~\ref{tab:entanglement_asym_scaling}, and leave a detailed discussion of each case to later sections.\\

\paragraph{\textbf{Logarithmic negativity.}}

The logarithmic negativity~\cite{2002_Vidal_negativity,2005_log_neg_Plenio} is defined as
\begin{equation} \label{eq:negat}
    E_{\mathcal{N}} = \log \|\rho^{T_B}\|_1.
\end{equation}
Here $\|A\|_1 = \mathrm{Tr}\sqrt{A^\dagger A}$ is the trace norm, and $\rho^{T_B}$ is the partial transpose with respect to subsystem $B$, which is given by $\langle \psi_A, \psi_B| \rho |\psi_A^\prime, \psi_B^\prime \rangle = \langle \psi_A,\psi_B^\prime|\rho^{T_B}|\psi^\prime_A, \psi_B\rangle$ for an arbitrary orthonormal basis $\{|\psi\rangle\}$ such that $|\psi\rangle = |\psi_A\rangle\otimes |\psi_B\rangle$.
The logarithmic negativity relies on the positive partial transpose criterion~\cite{1996_PPT_Asher}, which states that the partial transpose of separable states is positive semi-definite, hence zero logarithmic negativity~\footnote{Note however that vanishing negativity does not imply that the state is separable.}. 
Moreover, it is an upper bound for the distillable entanglement contained in $\rho$~\cite{2002_Vidal_negativity}.
While its computation, both numerical and analytically, is generically challenging for large system sizes, we find an exact expression for the stationary state Eq.~\eqref{eq:rho_singlet} for all commutants considered in this work given by
 \begin{equation} \label{eq:general_En}
    E_{\mathcal{N}} = \log \left(\frac{1}{D_0^{(L)}} \sum_{\lambda} d_\lambda D_{\lambda}^{(L_A)} D_{\bar{\lambda}}^{(L_B)}\right).
\end{equation}
Here the summation of $\lambda$ ranges from all possible representations limited on system size $L_{\min} = \min (L_A, L_B)$,
and $\bar{\lambda}$ corresponds to the dual representation of $\lambda$, for which $d_{\bar{\lambda}}=d_\lambda$. 
This expression only depends on the dimension of the irreps of the bond algebra $D_{\lambda}^{(L_A)} (D_{\bar{\lambda}}^{(L_B)})$ for the left (right) partition, and the dimension of irrep of the commutant algebra $d_{\lambda}$. 

With the exact expression Eq.~\eqref{eq:general_En}, the logarithmic negativity is upper bounded by 
\begin{equation}
    E_{\mathcal{N}} \leq \log [\mathrm{dim} (\mathcal{C}_{\mathrm{min}})],
\end{equation}
where we used $\sum_{\lambda}D_\lambda^{(L_A)} D_{\bar{\lambda}}^{(L_B)} = D_0^{(L)}$ and thus  $D_\lambda^{(L_A)} D_{\bar{\lambda}}^{(L_B)} \leq D_0^{(L)}$.\\

\paragraph{\textbf{Rényi negativity.}}
A widely used and more efficiently computable entanglement proxy~\cite{2012_Calabrese_negativity, 2013_Calabrese_negativity_renyi, 2020_Grover_MC_renyi_neg}, e.g., using tensor network simulations~\cite{2020_Wybo_MBL}, is the $n$-Rényi negativity defined as 
\begin{equation}\label{eq:Rn_definition}
    R_n = -\log \left(\frac{\mathrm{Tr}[(\rho^{T_B})^n]}{\mathrm{Tr}(\rho^n)}\right),
\end{equation}
for integer $n$.
For pure states $|\psi\rangle$, $R_n \propto S_n$ for odd $n$, and $R_n \propto S_{n/2}$ for even $n$, where $S_n$ is the $n$-th Rényi entropy~\cite{2012_Calabrese_negativity, 2013_Calabrese_negativity_renyi}. 
Moreover, for even $n$, the analytic continuation of $R_n$ leads to $-E_{\mathcal{N}}$ as $n\rightarrow 1$~\cite{2012_Calabrese_negativity, 2013_Calabrese_negativity_renyi}.
To study the interpolation between $R_n$ and $E_{\mathcal{N}}$, we introduce a natural generalization of the Rényi negativity which is now defined for arbitrary integer and non-integer $n \neq 2$ as
\begin{equation}\label{eq:tilde_Rn}
    \tilde{R}_n = \frac{1}{2-n} \log \left(\frac{\mathrm{Tr}[|\rho^{T_B}|^n]}{\mathrm{Tr}(\rho^n)}\right).
\end{equation}
We have replaced $(\rho^{T_B})^n$ by $|\rho^{T_B}|^n$ to avoid negative eigenvalues of the partial transpose $\rho^{T_B}$. The factor $1/(2-n)$ is introduced to keep $\tilde{R}_n$ positive. 
Note that for pure states $|\psi\rangle$, $\tilde{R}_n(\ket{\psi}) = S_{n/2}(\ket{\psi})$ for any $n$, with $S_{n/2}$ the $n/2$-th Rényi entropy (App.~\ref{app:derive_general_expression}).
This quantity provides an analytic continuation of $R_n$ with even $n$ to arbitrary $n\in \mathbb{R}$, and it relates to both the logarithmic negativity $\tilde{R}_1 = E_{\mathcal{N}}$, as well as to the Rényi negativity $\tilde{R}_n = R_n/(n-2)$ for $n$ even and $n>2$.

For Rényi negativities with an odd $n$ index, we find
\begin{equation}\label{eq:general_Rn}
    R_n = -\log \left(\frac{1}{D_0^{(L)}} \sum_{\lambda} \frac{D_{\lambda}^{(L_A)}D_{\bar{\lambda}}^{(L_B)}}{d_{\lambda}^{n-1}} \right),
\end{equation}
obtaining those with even index $n$ via the relation $R_{n} = R_{n-1}$. 
For \emph{generalized} Rényi negativity $\tilde{R}_n$,
\begin{equation}\label{eq:general_tilde_Rn}
    \tilde{R}_n = \frac{1}{2-n}\log \left(\frac{1}{D_0^{(L)}} \sum_{\lambda} \frac{D_{\lambda}^{(L_A)}D_{\bar{\lambda}}^{(L_B)}}{d_{\lambda}^{n-2}} \right),
\end{equation}
which corresponds to the exact expression for logarithmic negativity for $n=1$ and to the Rényi negativity for $n$ even (up to a factor).  

Similarly to the negativity, we can also find the general upper bounds
\begin{equation}
    \tilde{R}_{n}	\leq 
    \begin{cases}
    \frac{1}{2-n}\log[\dim (\mathcal{C}_{\mathrm{min}})] &  n<2,\\
    \log[\max(d_{\lambda})]\leq \frac{1}{2}\log[\dim(\mathcal{C}_{\mathrm{min}})] &  n>2;
    \end{cases}
\end{equation}
limiting the amount of entanglement of the stationary state. The corresponding derivations can be found in App.~\ref{app:derive_general_expression}.\\

\paragraph{\textbf{Operator space entanglement.}}
Different from the previous quantities, the operator space entanglement (OSE)~\cite{2001_Paolo_OSE, 2007_OSE_Prosen, Dubail_2017_OSE} measures both classical and quantum correlations. This quantity corresponds to the von Neumann entropy of the vectorized mixed state $|\rho\rangle\rangle$, defined via the vectorization $|\mu\rangle\langle \nu| \rightarrow |\mu\rangle |\nu\rangle$, and relates to the efficiency of a tensor network representation of $\rho$. The OSE is given by 
\begin{equation}
    S_{\mathrm{OP}} = -\mathrm{Tr}\left(\tilde{\rho}_A \log \tilde{\rho}_A \right) = -\sum_l s_l^2 \log s_l^2,
\end{equation}
with $\tilde{\rho}_A = \mathrm{Tr}_B(|\rho\rangle\rangle \langle\langle \rho|)$. Here, $s_l$ are the Schmidt eigenvalues of vectorized state $|\rho\rangle\rangle$ for a bipartition of $L_A$ and $L_B$, normalized such that $\langle\langle \rho|\rho\rangle\rangle=1$. Note that this quantity is different from the von Neumann entanglement entropy of the mixed state $\rho$. Similar to the previous quantities, the OSE can be expressed in terms of dimensions of the irreps $d_\lambda$ and $D_\lambda$ via
\begin{equation} \label{eq:general_SOP}
    S_{\mathrm{OP}} = - \sum_{\lambda} \frac{D_{\lambda}^{(L_A)} D_{\bar{\lambda}}^{(L_B)}}{D_0^{(L)}} \log \left(\frac{D_{\lambda}^{(L_A)} D_{\bar{\lambda}}^{(L_B)}}{D_0^{(L)}d_\lambda^2}\right).
\end{equation}
Moreover, using the concavity of the logarithm one finds the general upper bound 
\begin{equation}
    S_{\mathrm{OP}}\leq \log[\mathrm{dim}(\mathcal{C}_\mathrm{min})].
\end{equation}

Therefore, we find that in general, and without knowledge of the specific commutant algebra, the exact expressions lead to two general conclusions (whenever the bipartition of basis states Eq.~\eqref{eq:gen_triv0} is valid). 
First, if the commutant algebra is Abelian, then all irreps are one-dimensional ($d_{\lambda} \equiv 1$), hence the logarithmic and the Rényi negativities vanish (recall that $\sum_{\lambda}D_\lambda^{(L_A)} D_{\bar{\lambda}}^{(L_B)} = D_0^{(L)}$). 
Tracing back to the bipartition of basis states Eq.~\eqref{eq:gen_triv0}, $d_{\lambda} \equiv 1$ indicates that all basis states are local product states and thus the stationary state is separable. 
In contrast, for non-Abelian commutants with $d_{\lambda}\geq 1$, the basis states are entangled, which can lead to mixed-entangled stationary states. 
The source of non-zero logarithmic and Rényi negativities is the fact that elements of the maximal Abelian subalgebras $\mathcal{M}(L)$ of the commutant do not share a common local product basis (such that the stationary state is a direct sum of projectors onto entangled states).
Second, the logarithmic negativity, Rényi negativities, and operator space entanglement are upper bounded by $\log [\dim( \mathcal{C}_{\mathrm{min}})]$. 
In particular, for non-Abelian commutants whose dimension does not scale with subsystem size e.g., any finite discrete non-Abelian group like the dihedral $D_n$ or symmetric $S_n$ groups with $n\geq 3$, the negativity is at most an O$(1)$ number.
Therefore, in general, finite groups have at most a $\mathrm{O}(1)$ logarithmic negativity (i.e., independent of system size), conventional symmetries have at most a logarithmic scaling, while fragmented systems with exponentially large commutants can showcase a volume law.

\renewcommand{\arraystretch}{1.5}
\begin{table}[]
\centering
\begin{tabular}{c  c  c  c} 
 \hline \hline
Commutant &   $E_{\mathcal{N}}$ & $R_3$ & $S_{\text{OP}}$ \\ [0.5ex] 
\hline
$\mathcal{C}(L)$ & \multicolumn{3}{c}{ $\leq  \log [\dim (\mathcal{C}_{\mathrm{min}})]$}\\
\hline
$G$ & $\leq$ O$(1)$     & $\leq$ O$(1)$     & $\leq$ O$(1)$ \\
 $\mathcal{C}_{\mathrm{U}(1)}$     & 0     & 0     & $\frac{1}{2}\log L$ \\ 
 U(su$(2)$)    & $\frac{1}{2}\log L$     & $\log L$    & $\frac{3}{2}\log L$ \\
 U(su$(N)$) & $\leq N(N-1)\log L$ & $ c_{\mathrm{SU}(N)}^{R_3} \log L $ & $ c_{\mathrm{SU}(N)}^{S} \log L$ \\
  \hline
 $\mathcal{C}_{\mathrm{PF}(N)}$ & 0 & 0 &   $c_{\mathrm{PF}(N)}^{S}\sqrt{L}$ \\
 $\mathcal{C}_{\mathrm{TL}(N)}$ & $\geq c_{\mathrm{TL}(N)}^{E_{\mathcal{N}}} L$ & $\leq \frac{3}{2} \log L$ & $ c_{\mathrm{TL}(N)}^{S}\sqrt{L} $ \\ [1ex] 
 \hline
 $\mathcal{C}_{\mathrm{TL}(N)}$ & $\tilde{R}_{n<2} \geq \tilde{c}_{N, n}^{\mathrm{lin}} L$ &\multicolumn{2}{l}{\, \, $\tilde{R}_{n>2} \leq \tilde{c}_{N, n}^{\log} \log L$}   \\
\hline
\end{tabular}
\caption{\textbf{Asymptotic system-size scaling of bipartite entanglement for different commutant algebras up to subleading order.}
For general commutants (with a Hopf algebra structure), the logarithmic negativity $E_{\mathcal{N}}$, Rényi negativity $R_3$, and the operator space entanglement $S_{\mathrm{OP}}$ are upper bounded by the logarithm of the dimension of the commutant on the smaller bipartition $L_{\min}$.
This indicates that for finite groups $G$, any of these entanglement quantities are at most a finite system-size independent value.
For other commutants, we evaluate the asymptotic scaling of half-chain entanglement.
For Abelian U($1$) symmetry, the stationary state has zero $E_{\mathcal{N}}$ and $R_3$, while the operator space entanglement  $S_{\mathrm{OP}}$ scales logarithmically with system size $L$ due to non-vanishing classical correlations.
For non-Abelian SU($2$) symmetry, all threee quantities $E_{\mathcal{N}}$, $R_3$ and $S_{\mathrm{OP}}$ scales logarithmically. 
In the case of higher spin SU$(N)$, $R_3$ scales logarithmically with coefficient $c_{\mathrm{SU}(N)}^{R_3} \in [\frac{N(N-1)}{2}, \frac{N^2-1}{2}]$.  We can also prove an upper bound for $E_{\mathcal{N}}$ scaling as $N(N-1)\log L$, as well as $S_{\mathrm{OP}} \sim c^S_{\mathrm{SU}(N)} \log L$, with $c^S_{\mathrm{SU}(N)}\in [\frac{N^2-1}{2}, N^2-1]$. 
For systems with Abelian commutant algebra corresponding to that of the classically fragmented model PF($3$), the stationary states are separable leading to vanishing $E_{\mathcal{N}}$ and $R_3$, similar to the behavior of U($1$) symmetric systems. 
For quantum fragmented systems with non-Abelian commutant corresponding to that of the TL($N$) model, $E_{\mathcal{N}}$ is lower bounded by a linear scaling with prefactor $c^{E_{\mathcal{N}}}_{\mathrm{TL}(N)}$ (see details in main text), while $R_3$ is upper bounded by a logarithmic scaling.
The generalized Rényi negativity $\tilde{R}_n$ shows a transition from linear scaling to logarithmic scaling across $n=2$, consistent with the distinct scaling of $E_{\mathcal{N}} = \tilde{R}_{n=1}$ and $R_{n\,\mathrm{even}} = (n-2)\tilde{R}_{n}$.
Additionally, fragmentation also leads to a large scaling $~\sqrt{L}$ of the OSE for both PF($N$) and TL($N$) systems. 
The analytical expressions of the coefficients for the TL($N$) case can be found in Sec.~\ref{sec:fragment}.
}
\label{tab:entanglement_asym_scaling}
\end{table}

In the following sections, we analytically derive the asymptotic finite-size scaling of half-chain entanglement proxies for specific commutants, i.e., for fixed values of $D_\lambda$ and $d_\lambda$. The results are summarized in Table \ref{tab:entanglement_asym_scaling}.
Note that for Rényi negativity, we focused on the scaling of $R_3$, which is the smallest non-trivial Rényi with integer index (since $R_1 = R_2 = 0$).

First, we study conventional symmetries such as Abelian U($1$) symmetry and non-Abelian SU$(N)$ symmetry in Sec.~\ref{sec:conventional_sym}. We show that with Abelian U($1$) symmetry, the stationary state is a separable state with vanishing $E_{\mathcal{N}}$ and $R_3$, which is expected for generic systems coupled to Hermitian dissipative baths. 
On the other hand, SU($2$) and higher SU$(N)$ showcase non-vanishing quantum entanglement even under dissipation. 
In addition, we investigate logarithmic negativity for SU($2$) symmetry with $\lambda_{\mathrm{tot}}\neq 0$ in Sec.~\ref{subsec:SU2_Haar}.
In Sec.~\ref{sec:fragment}, we study Hilbert space fragmentation with commutant of dimension $\dim [\mathcal{C}(L)]\sim e^{L}$. 
As examples, we study the classical fragmentation PF($N$) (with commutant $\mathcal{C}_{\mathrm{PF}(N)}$ and fragmentation in local product basis) and quantum fragmentation TL($N$) (with the RS commutant). 
The stationary states with $\mathcal{C}_{\mathrm{PF}(N)}$ are separable, while the stationary states with $\mathcal{C}_{\mathrm{TL}(N)}$ are highly entangled, which resembles the U($1$) and SU$(N)$ case.
Moreover, quantum fragmentation of TL($N$) largely enhances the quantum entanglement of the stationary states, which shows a volume law scaling of $E_{\mathcal{N}}$ and a logarithmic scaling of $R_3$. 
To understand the scaling difference between $R_n$ and $E_{\mathcal{N}}$ for the commutants $\mathcal{C}_{\mathrm{TL}(N)}$, we study the scaling of the quantity $\tilde{R}_n$ as a function of $n$, and show that the scaling exhibits a sharp change across $n=2$. 

\section{Conventional symmetries}\label{sec:conventional_sym}
In this section, we quantify the stationary state entanglement for conventional symmetries, including Abelian U($1$) and non-Abelian SU$(N)$ symmetries. We use U($1$) and SU($2$) as introductory examples to show the distinction of stationary state entanglement when restricted to a single symmetry sector. Then we discuss the effect of considering stationary states supported on a combination of different total spin sectors of SU($2$) and the generalization to SU$(N)$ symmetries with larger spin-$S$ (with $N=2S+1$). 
 
\subsection{Abelian U($1$) symmetry}
We start with an Abelian U($1$) symmetry generated by the total magnetization $S^z_{\mathrm{tot}}$ on a spin-$1/2$ chain. 
There are $L+1$ number of U($1$) symmetry sectors labeled by the eigenvalues $M = -L/2, -L/2+1, \ldots, L/2$ of the total magnetization $S^z_{\mathrm{tot}}$.
The commutant algebra is generated by the $S^z_{\mathrm{tot}}$, i.e., $\mathcal{C} = \langle\{S^z_{\mathrm{tot}}\}\rangle = \mathrm{span}\{\mathbb{1}, S^z_{\mathrm{tot}}, (S^z_{\mathrm{tot}})^2, \ldots, (S^z_{\mathrm{tot}})^{L}\}$,  with $\dim [\mathcal{C}(L)] = \sum_M d_M^2 = L+1$, which equals the number of Krylov subspaces as $d_{M}\equiv 1$ for Abelian commutants.
The dimension of the symmetry sectors (irrep of $\mathcal{A}(L)$) are given by the binomial coefficients $D_M = (\begin{smallmatrix}L\\
L/2-M
\end{smallmatrix})$.
A possible choice of local operators to generate the corresponding bond algebra (centralizer of this commutant) is $\mathcal{A}(L) = \langle \{S_j^x S_{j+1}^x+  S_j^y S_{j+1}^y\}, \{S_j^z\}, \mathbb{1} \rangle$~\cite{moudgalya_fragment_commutant_2022}, which can be used to construct the set of Kraus operators/Lindblad jump operators to reach the desired stationary state in ~Eq.~\eqref{eq:general_stationary_state}. 

We consider the stationary state in the $M_{\mathrm{tot}}=0$ sector and a chain of even length $L$ for simplicity. A set of basis states in the $M_{\mathrm{tot}}=0$ sector can be written as a tensor product of left and right partitions
\begin{equation}\label{eq:U1_basis}
    |M_{\mathrm{tot}}=0; M; a, b\rangle = |M; a\rangle  |-M; b\rangle,
\end{equation}
with $M=-L_{\min}/2, -L_{\min}/2+1, \ldots, L_{\min}/2$ as the magnetization on a chain of length $L_{\min}=\min(L_A, L_B)$, and $a=1,\ldots, D_{M}^{(L_A)}$, $b=1,\ldots, D_{M}^{(L_B)}$ labeling different states in the $\pm M$ sector on system size $L_A,L_B$ respectively. 
Notice that the basis states Eq.~\eqref{eq:U1_basis} are of the form shown in Eq.~\eqref{eq:gen_triv0} with $d_{\lambda}= 1$ (and then these are product states), with $\eta_{\lambda,m} = 1$.

For an initial state in the $M_{\mathrm{tot}}=0$ sector, the stationary state corresponds to the identity matrix within the sector
\begin{equation}
    \rho = \frac{1}{D_0^{(L)}} \sum_{M} \sum_{a,b} |M_{\mathrm{tot}}=0; M; a, b\rangle \langle M_{\mathrm{tot}}=0; M; a, b|,
\end{equation}
where $D_0^{(L)}$ is the total number of states in the $M_{\mathrm{tot}}=0$ sector for a chain with $L$ sites. Since all the basis states are local product states, the partial transpose is trivially
\begin{equation}\label{eq:U1_partial_transpose_rho}
    \rho^{T_B} = \frac{1}{D_0^{(L)}} \sum_{M, a, b} |M;a\rangle|-M;b\rangle\langle M;a|\langle-M;b| = \rho.
\end{equation}
Therefore, the stationary state is a separable state (an incoherent convex sum of product states) with positive partial transpose, and thus zero logarithmic $E_{\mathcal{N}}$ and Rényi $R_n$ negativities. 
These results agree with the exact expressions of $E_{\mathcal{N}}$ (Eq.~\eqref{eq:general_En}) and $R_n$ (Eq.~\eqref{eq:general_Rn}), which equal to zero with $d_{M} \equiv 1$ for Abelian symmetries and $\sum_{M} D_{M}^{(L_A)} D_M^{(L_B)} = D_0^{(L)}$.
Moreover, this conclusion can be generalized to stationary states in Krylov subspaces spanned by local product state basis. 
This easily extends to the other symmetry subspaces of U$(1)$ with $M_{\mathrm{tot}}\neq 0$. 
Moreover, it applies to $\mathbb{Z}_2$ symmetric systems~\cite{LeeYouXu2022, ma2024topological, lessa2024strongtoweak, 2024_sala_SSSB}, as well as other systems with Abelian finite groups or with classical fragmentation (e.g., see Sec.~\ref{subsec:classical_frag}).

\begin{figure*}[bt]
\includegraphics[width=18cm, scale=1]{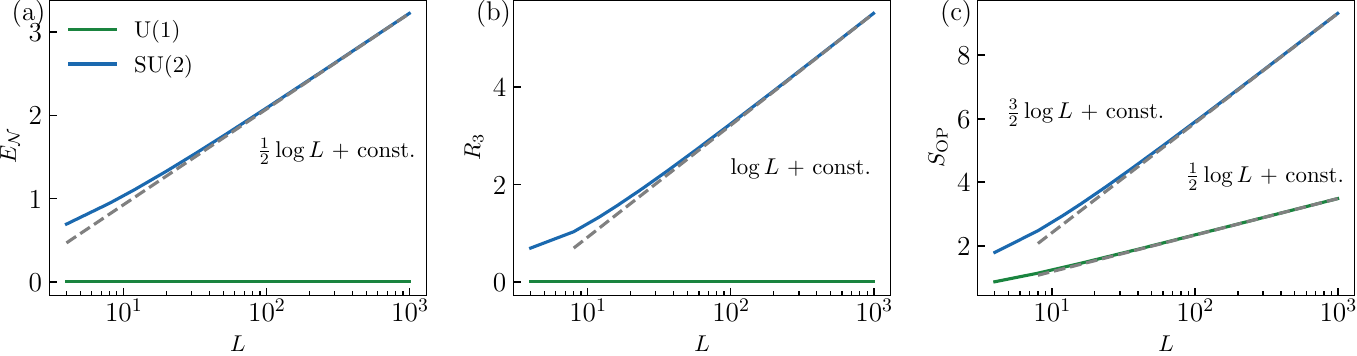}
\caption{\label{fig:Entanglement_U1_SU2} \textbf{Entanglement of stationary states with conventional U($1$) and SU($2$) symmetries.} (a) Both logarithmic $E_{\mathcal{N}}$ and (b) Rényi negativity $R_3$ vanish for U($1$) symmetry, but scales logarithmically with system size $L$ as $L\rightarrow\infty$ for SU($2$). (c) The OSE for both U($1$) and SU($2$) scales logarithmically as $L\rightarrow\infty$. Results are obtained from exact expressions in Sec.~\ref{sec:exact_entanglement}. Here dashed lines show the analytic asymptotic scaling.}
\end{figure*}

On the other hand, the OSE is non-zero due to classical correlations, which arise from the fluctuations of the conserved total magnetization $M_{\mathrm{tot}}=0$ across the bipartition.
The OSE is the entanglement entropy of the vectorized mixed state
$|\rho\rangle\rangle = \frac{1}{\sqrt{D_0}}\sum_{M,a,b} |M_{\mathrm{tot}}=0; M; a, b\rangle |M_{\mathrm{tot}}=0; M; a,b\rangle$,
properly normalized, which maps to the pure state with an equal superposition of all $M_{\mathrm{tot}}=0$ states. For each fixed $M$, there is one Schmidt value which squares to $  D_{M}^{(L_A)}D_{M}^{(L_B)}/D_0^{(L)}$. Therefore, the OSE is given by
\begin{equation}\label{eq:U1_SOP}
    S_{\text{OP}} = - \sum_{M} \frac{D_{M}^{(L_A)}D_{M}^{(L_B)}}{D_0^{(L)}} \log \left(\frac{D_{M}^{(L_A)}D_{M}^{(L_B)}}{D_0^{(L)}}\right),
\end{equation}
agreeing with Eq.~\eqref{eq:general_SOP}. 
With Eq.~\eqref{eq:U1_SOP} and the dimensions $D_{M}$, we evaluate the asymptotic scaling of OSE in the $M_{\mathrm{tot}}=0$ sector for half-chain bipartition, which is $S_{\mathrm{OP}} \sim \frac{1}{2}\log L$ as shown in Fig.~\ref{fig:Entanglement_U1_SU2}c. Details can be found in App.~\ref{app:U1}. The logarithmic scaling of OSE with the system size is also compatible with the logarithmic growth in time in Ref.~\cite{Wellnitz_2022}.

\subsection{Non-Abelian SU(2) symmetry}\label{subsec:SU2}
We now consider the case of a global non-Abelian SU($2$) symmetry on a spin-$1/2$ chain, with $\mathcal{C} = \langle \{S_{\mathrm{tot}}^x, S_{\mathrm{tot}}^y, S^z_{\mathrm{tot}} \}\rangle$, i.e., the UEA of su$(2)$. The corresponding bond algebra can be generated by $\mathcal{A}(L) = \langle \{ \vec{S}_j \cdot \vec{S}_{j+1} \}, \mathbb{1}\rangle$~\cite{moudgalya_fragment_commutant_2022, 2023_sanjay_commutant_symmetries}, which corresponds to the permutation group of $L$ elements $S_L$. With SU($2$) symmetry, the symmetry sectors are labeled by $\lambda$, which corresponds to the total spin $J$. The total magnetization $m=-\lambda, -\lambda + 1, \ldots, \lambda$ labels the  $d_\lambda$ degenerate subspaces with fixed $\lambda$. The degeneracy and dimension of the Krylov subspaces are given by
\begin{equation}\label{eq:SU2_dimension}
\begin{aligned}
    &d_\lambda = 2\lambda + 1,\,  \\&D_\lambda^{(L)} = \begin{pmatrix}L\\
    L/2+\lambda
    \end{pmatrix} - \begin{pmatrix}L\\
    L/2+\lambda+1
    \end{pmatrix}.
\end{aligned}
\end{equation}
With SU($2$) symmetry, the basis states that span a Krylov subspace $\mathcal{H}_{\lambda, m}^{\mathcal{A}(L)}$ can be entangled states, which is distinct from the case of U($1$) symmetry. 
We study the stationary state in the total spin $\lambda_{\mathrm{tot}} = 0$ sector (thus $m_{\mathrm{tot}}=0$) for a chain with length~\footnote{We choose $L=4n$, $n\in\mathbb{N}$, such that the total spin allows for $\lambda_{\mathrm{tot}} =0 $ and the bipartition with even $L_A$ and $L_B$ for simplicity. The results generalize to bipartition with odd $L_A$ and $L_B$.} $L=4n$, $n\in \mathbb{N}$. 
The basis states can be written as 
\begin{equation}\label{eq:SU2_basis_state_0}
    |\lambda_{\mathrm{tot}} = 0; \lambda; a, b \rangle = \sum_{m=-\lambda}^{\lambda} c_{m}(\lambda) |\lambda, m; a\rangle |\lambda, -m; b\rangle.
\end{equation}
For the $\lambda_{\mathrm{tot}} = m_{\mathrm{tot}} = 0$ subspace, the left and right partitions correspond to the same irrep $\lambda$, since irreps of SU$(2)$ are self-dual.
Moreover, the left and right bipartition has magnetization $m$ and $-m$ respectively, with $\lambda = 0, 1,\ldots, L_{\min}/2$ (for even length $L$), and $m = -\lambda, \ldots, \lambda$.
For SU($2$) with $\lambda_{\mathrm{tot}}=0$, the Clebsch–Gordan (CG) coefficients $c_{m}(\lambda) = \langle \lambda, m; \lambda, -m | \lambda_{\mathrm{tot}}=0, m_{\mathrm{tot}}=0\rangle$ have the exact expression
\begin{equation}\label{eq:CG_SU2}
    c_{m}(\lambda) = \frac{(-1)^{\lambda-m}}{\sqrt{2\lambda+1}}= \frac{(-1)^{\lambda-m}}{\sqrt{d_\lambda}}.
\end{equation}
It shows that for a fixed $\lambda$, the CG coefficients take the same value up to a minus sign for different $m$, compatible with the general expression Eq.~\eqref{eq:gen_triv0}.
The stationary state is the maximally mixed state in the $\lambda_{\mathrm{tot}}=0$ sector, i.e., an equal sum of projectors onto basis states.
\begin{equation}
\begin{aligned}
    &\rho  =\frac{1}{D_0^{(L)}}\sum_{\lambda=0}^{L_{\min}/2}\\
    & \times \sum_{a=1}^{D_{\lambda}^{(L_A)}} \sum_{b=1}^{D_{\lambda}^{(L_B)}} |\lambda_{\mathrm{tot}} = 0;\lambda;a,b\rangle \langle \lambda_{\mathrm{tot}} = 0;\lambda;a,b|, 
\end{aligned}
\end{equation}
with $D_0^{(L)}$ the dimension of the singlet subspace for a chain of $L$ sites, while $D_{\lambda}^{(L_A)}$ is the dimension of the $\lambda$ subspaces on $L_A$ sites and similar for $D_{\lambda}^{(L_B)}$.
Due to the entangled basis states in Eq.~\eqref{eq:SU2_basis_state_0}, the partial transpose of $\rho$ is non-trivial.

We first analyze the eigenvalues of $\rho^{T_B}$ to obtain the logarithmic negativity. 
The operator $\rho^{T_B}$  can be block-diagonalized into the form $\rho^{T_B} = \oplus_{\lambda, a, b} \rho^{T_B}_{\lambda,a,b}$, with
\begin{equation}\label{eq:SU2_rhoTB_subblock}
\begin{aligned}
    \rho^{T_B}_{\lambda,a, b}& = \frac{1}{D_0^{(L)}}\sum_{m,m^\prime} c_{m}(\lambda) c_{m^\prime}^{*} (\lambda) \\
     &\times|\lambda, m; a\rangle |\lambda, -m^\prime; b\rangle \langle \lambda, m^\prime; a|\langle \lambda, -m; b|,
\end{aligned}
\end{equation}
where $\lambda=0,\ldots, \min(L_A, L_B)$, $a=1, \ldots, D_{\lambda}^{(L_A)}$ and $b = 1, \ldots, D_{\lambda}^{(L_B)}$.
Each $\rho^{T_B}_{\lambda,a, b}$ squares to $\mathbb{1}_{d_{\lambda}^2}/(D_0^{(L)} d_\lambda)^2$, which implies there are in total $d_{\lambda}^2$ number of eigenvalues $\pm \frac{1}{D^{(L)}_0 d_{\lambda}}$.
With $E_{\mathcal{N}} = \log \|\rho^{T_B}\|_1 = \log \sum_i |\lambda_i|$, where $\lambda_i$ are the eigenvalues of $\rho^{T_B}$, one finds
\begin{equation} \label{eq:EN_SU2}
    E_{\mathcal{N}} = \log \frac{1}{D_0^{(L)}} \sum_\lambda d_{\lambda} D_{\lambda}^{(L_A)}  D_{\lambda}^{(L_B)}.
\end{equation}
For a half-chain bipartition $L_A = L_B=L/2$, we obtain
\begin{equation} \label{eq:EN_SU2_L}
    E_{\mathcal{N}} = \log\left((\frac{L}{2}+1) \frac{\left(\begin{smallmatrix}L/2\\
    L/4\end{smallmatrix}\right)^2}{\left(\begin{smallmatrix}L\\
    L/2\end{smallmatrix}\right)}\right).
\end{equation}

For the Rényi negativities $R_n$, we can calculate the value of $\mathrm{Tr}[(\rho^{T_B})^n]$ in Eq.~\eqref{eq:Rn_definition} diagrammatically. For example, for $n=3$ and $n=4$,
\begin{equation}\label{eq:Rn_diagram}
\begin{aligned}
    &\mathrm{Tr}[(\rho^{T_B}_{\lambda, a, b})^3] : \TrRthree \propto \sum_{m} |c_{m}|^6, \\
    &\mathrm{Tr}[(\rho^{T_B}_{\lambda, a, b})^4] : \TrRfour \propto (\sum_{m} |c_{m}|^4)^2.
\end{aligned}
\end{equation}
The rules of the diagrammatic expression can be understood as follows: As mentioned for logarithmic negativity, $\rho^{T_B}$ decomposes into $\rho^{T_B}_{\lambda, a, b}$, thus  $\mathrm{Tr}[(\rho^{T_B})^n] = \sum_{\lambda,a,b}\mathrm{Tr}[(\rho^{T_B}_{\lambda, a, b})^n]$. 
Each $\rho^{T_B}_{\lambda, a, b}$ contains terms as $|m\rangle |-m^\prime\rangle\langle m^\prime|\langle -m|$ (omitting the labels $\lambda, a, b$). 
In the diagram, every grey block $\TrRblock$ denotes one copy of $\rho^{T_B}_{\lambda, a, b}$, with the two dots denoting $m$ and $m^\prime$, respectively.
Taking product of copies of $\rho^{T_B}_{\lambda, a, b}$ or taking the trace are represented as connecting two dots such as $m$ and $m^{\prime\prime}$, which give the relation $\langle m| m^{\prime\prime}\rangle = \delta_{m, m^{\prime\prime}}$.
Therefore, every closed loop gives a factor $\sum_{m} |c_{m}(\lambda)|^{l} = d_\lambda (1/\sqrt{d_\lambda})^{l}$, with $l$ the number of dots $\bullet$ passed by the loop. 
For $n$ odd, there is one loop passing through $2n$ dots, which gives a factor of $d_{\lambda}^{1-n}$; while for $n$ even, there are two loops passing through $n$ dots, which gives $(d_\lambda^{1-\frac{n}{2}})^2 = d_\lambda^{2-n}$.
All together and including other prefactors, $R_n$ is given by
\begin{equation} \label{eq:Rn_SU2}
    R_n 
    = -\log \frac{1}{D_0^{(L)}}  \sum_{{\lambda}} \frac{D_{\lambda}^{(L_A)}D_{\lambda}^{(L_B)}}{d_{\lambda}^{n-1}}
\end{equation}
for odd $n$, and $R_{n} = R_{n-1}$ for even $n$. 
Evaluating $R_3$ for a half-chain bipartition, we obtain the simpler expression
\begin{equation}
    R_3 = \log \left(\frac{(L+2)^2}{4(L+1)}\right).
\end{equation}

Finally, we calculate the OSE of the stationary state $\rho$, which is the entanglement entropy of the corresponding vectorized state $|\rho\rangle\rangle$ 
\footnote{We notice that $|\rho\rangle\rangle$ is the ground state of the SU($4$)-symmetric spin-$3/2$ Hamiltonian $H_{\mathrm{SU}(4)} = \sum_j (1-P_{j,j+1})$, with $P_{j,j+1}$ permutations on neighboring sites~\cite{2023_moudgalya_superoperator}.}.
The vectorized state can be written as $|\rho\rangle\rangle = \sum_{\lambda} \sum_{A,B}\Psi^\lambda_{A,B} |\psi_A\rangle |\psi_B\rangle$, where $\{|\psi_{A}\}\rangle$ is the ONB $\{|\lambda, m; a\rangle |\lambda, m^\prime; a\rangle\}$ on partition $A$ and $\{|\psi_{B}\}\rangle$ is the ONB $\{|\lambda, -m; b\rangle |\lambda, -m^\prime; b\rangle\}$  on partition $B$.
The matrix $\Psi^\lambda_{A,B}$ has $d_\lambda^2$ number of Schmidt values that square to $D_{\lambda}^{(L_A)}D_{\lambda}^{(L_B)}/(D_0^{(L)}d_\lambda^2)$ for each $\lambda$. Therefore, the OSE is given by  
\begin{equation} \label{eq:SOP_SU2}
    S_{\mathrm{OP}} = - \sum_{\lambda} \frac{D_{\lambda}^{(L_A)}D_{\lambda}^{(L_B)}}{D_0^{(L)}} \log \frac{D_{\lambda}^{(L_A)}D_{\lambda}^{(L_B)}}{D_0^{(L)} d^2_{\lambda}}.
\end{equation}

The previously derived exact expressions of $E_\mathcal{N}$, $R_n$, and OSE for SU($2$) symmetry coincide with those in Sec.~\ref{sec:exact_entanglement}, given that $\lambda = \bar{\lambda}$.
And the derivation generalizes to other cases, with more details given in App.~\ref{app:derive_general_expression}.

We obtain the asymptotic scaling using the explicit expressions for the dimensions of irreps of SU($2$) in Eq.~\eqref{eq:SU2_dimension}. 
We study the logarithmic negativity $E_{\mathcal{N}}$, the third Rényi negativity, and the operator space entanglement $S_{\mathrm{OP}}$.
We obtain that $E_{\mathcal{N}} \sim \frac{1}{2}\log L + O(1)$, $R_3 \sim \log L + O(1)$, and $S_{\mathrm{OP}}\sim\frac{3}{2}\log L + O(1)$ as $L\rightarrow \infty$  (see App.~\ref{app:SU2} for details).
In Fig.~\ref{fig:Entanglement_U1_SU2}, we compare the numerical values obtained by evaluating the exact expressions (shown as a blue solid line) for $E_\mathcal{N}$ (panel a)), $R_3$ (panel b) and $S_{\mathrm{OP}}$ (panel c)) and their asymptotic logarithmic scaling (corresponding to the dashed grey line), finding a good quantitative agreement. For comparison, we also include the corresponding values in the presence of a U$(1)$ symmetry.

Notice that based on our results, one could infer that Abelian commutants necessarily give rise to separable stationary states. 
However, this might not hold in general, when e.g., considering a SU($2$) dynamical symmetry~\cite{2019_buca_jaksch_non_statioanry_dissipation, 2020_Dieter_dynamical_sym}. 
The corresponding commutant is Abelian at the cost of adding a non-local term $S^z_{\mathrm{tot}}$ to the bond algebra $\mathcal{A}_{\mathrm{SU}(2)}(L) = \langle \{\vec{S}_j \cdot\vec{S}_{j+1}\},\mathbb{1}\rangle$~\cite{2023_sanjay_commutant_symmetries}. 
Therefore, the maximal Abelian subalgebra of dynamical SU($2$) and SU($2$) are equal, leading to the same common basis states that spanned the Krylov subspaces and thus similar entangled stationary states. 
For example, they have the same highly entangled stationary state in the $\lambda_{\mathrm{tot}}=0$ subspace. 
However, whether its commutant has a Hopf algebra structure is left as an open question.

\subsection{SU($2$) with $\lambda_{\mathrm{tot}} \geq 0$}\label{subsec:SU2_Haar}
While we mainly focus on stationary states restricted to the singlet subspace, in this section we consider two different scenarios. %
First, we study stationary states restricted to one Krylov subspace $\mathcal{H}^{\mathcal{A}(L)}_{\lambda_{\mathrm{tot}}, m_{\mathrm{tot}}=0}$ with $\lambda_{\mathrm{tot}}>0$. The stationary states are given by Eq.~\eqref{eq:general_stationary_state} for $m_{\mathrm{tot}}=0$.
Second, we consider initial states with non-vanishing overlap on multiple Krylov subspaces, $\bigoplus_{\lambda_{\mathrm{tot}}=0}^{\lambda_{\mathrm{max}}} \mathcal{H}_{\lambda_{\mathrm{tot}}, m_{\mathrm{tot}}=0}^{\mathcal{A}(L)}$. 
Here we consider the stationary state of Eq.~\eqref{eq:rho_gen} appearing under Lindblad evolutions with random decay rates. As explained in Sec.~\ref{sec:review_symmetries}, this rules out possible additional structure between different irreps.
When restricted to $m_{\mathrm{tot}}=0$ subspace, the stationary state becomes
\begin{equation}\label{eq:SU2_rho_ss_multi_lambda}
    \rho = \sum_{\lambda_{\mathrm{tot}}}p_{\lambda_{\mathrm{tot}}} \frac{\Pi^{\lambda_{\mathrm{{tot}}}}_{m_{\mathrm{tot}=0}}}{D_{\lambda_{\mathrm{tot}}}},
\end{equation}
with $p_{\lambda_{\mathrm{tot}}} = \mathrm{Tr}[\Pi^{\lambda_{\mathrm{tot}}}_{m_{\mathrm{tot}}=0}\rho(t=0)]$ the weight of initial state in the $\lambda_{\mathrm{tot}}$ and $m_{\mathrm{tot}}=0$ subspace.

To study the entanglement, we give the bipartite form of basis states in $\lambda_{\mathrm{tot}}>0$ and $m_{\mathrm{tot}}=0$ subspace, 
\begin{equation}\label{eq:SU2_general_lambda_basis_state}
\begin{aligned}
    &|\lambda_{\mathrm{tot}}, m_{\mathrm{tot}}=0;\lambda_A, \lambda_B; a, b \rangle \\
    &= \sum_{m=-\min(\lambda_A,\lambda_B)}^{\min(\lambda_A,\lambda_B)}c_{m}(\lambda_{\mathrm{tot}};\lambda_A, \lambda_B) |\lambda_A, m; a\rangle |\lambda_B, -m; b\rangle.
\end{aligned}
\end{equation}
The CG coefficients are $c_{m_A,m_B}(\lambda_{\mathrm{tot}},\lambda_A,\lambda_B) = \langle \lambda_A, m_A, \lambda_B, m_B | \lambda_{\mathrm{tot}}, m_{\mathrm{tot}}=0\rangle$, with the triangular condition $|\lambda_A-\lambda_B|\leq \lambda_{\mathrm{tot}} \leq |\lambda_A+ \lambda_B|$, and $m_B=-m_A$ such that $m_{\mathrm{tot}} = 0$. 
The CG coefficients of SU($2$) vanish when the triangular condition or the selection rules are not satisfied.
Then the logarithmic negativity is given by 
\begin{equation}\label{eq:SU2_En_lambdas}
\begin{aligned}
    &E_{\mathcal{N}} =\log \left(\sum_{\lambda_{\mathrm{tot}}}p_{{\lambda_{\mathrm{tot}}}} \sum_{\lambda_A = 0}^{L_A/2} \sum_{\lambda_B = |\lambda_{A}-\lambda_{\mathrm{tot}}|}^{\lambda_A + \lambda_{\mathrm{tot}}} \frac{D_{\lambda_A}^{(L_A)}D_{\lambda_B}^{(L_B)}}{D^{(L)}_{\lambda_{\mathrm{tot}}}} \right. \\ 
    &\left. \times \sum_{m,m^\prime} 
|c_m(\lambda_{\mathrm{tot}};\lambda_A, \lambda_B) c^{*}_{m^\prime}(\lambda_{\mathrm{tot}};\lambda_A, \lambda_B)|\right).
\end{aligned}
\end{equation}
The sum of $\lambda_A, \lambda_B$ satisfies the triangular condition, and $m, m^\prime = - \min (\lambda_A, \lambda_B), \ldots, \min (\lambda_A, \lambda_B)$. 
We obtain this expression with eigenvalues of $\rho^{T_B}$ using the same approach as in the singlet subspace of SU($2$) (see details in App.~\ref{app:SU2}).

\begin{figure}[pt]
\centering
\includegraphics[width=1\columnwidth]{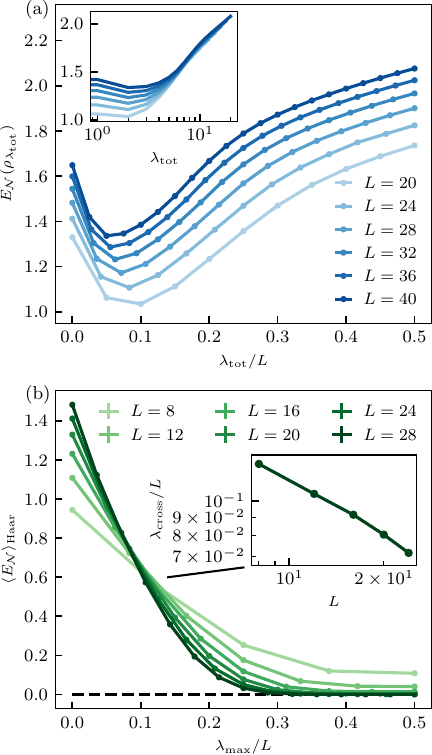}
\caption{\label{fig:SU2_neg_random_Haar} \textbf{Logarithmic negativity for stationary states in SU($2$) symmetry subspaces with total spin $\lambda_{\mathrm{tot}}\geq 0$ and $m_{\mathrm{tot}}=0$.}
(a) Logarithmic negativity $E_{\mathcal{N}}$ of the stationary state $\rho_{\lambda_{\mathrm{tot}}} = \Pi^{\lambda_{\mathrm{tot}}}_ {m_{\mathrm{tot}}=0}/D_{\lambda_{\mathrm{tot}}}$. 
For the stationary state restricted to one symmetry subspace with 
$\lambda_{\mathrm{tot}}$ and $m_{\mathrm{tot}}=0$, 
the $E_{\mathcal{N}}$ is non-zero and increases with system size for fixed density $\lambda_{\mathrm{tot}}/L$.
For $\lambda_{\mathrm{tot}} = 0, L/2$, we analytically show that $E_{\mathcal{N}}(\rho_{\lambda_{\mathrm{tot}}}) \sim \frac{1}{2}\log L$.
\textbf{Inset.} 
The logarithmic negativity $E_{\mathcal{N}}(\rho_{\lambda_{\mathrm{tot}}})$ scales logarithmically with $\lambda_{\mathrm{tot}}$ for large $\lambda_{\mathrm{tot}}$.
Moreover, for small $\lambda_{\mathrm{tot}}$, $E_{\mathcal{N}}$ increases with system size. While for large $\lambda_{\mathrm{tot}}$, $E_{\mathcal{N}}$ approximately stays constant with $L$.
(b) The average $E_{\mathcal{N}}$ of stationary states corresponding to Haar random initial states sampled from the Hilbert space $\bigoplus_{\lambda_{\mathrm{tot}}=0}^{\lambda_{\mathrm{max}}}\mathcal{H}^{\mathcal{A}(L)}_{\lambda_{\mathrm{tot}},m_{\mathrm{tot}}=0}$ constrained within the zero total magnetization $S^z_{\mathrm{tot}}=0$ (i.e., $m_{\mathrm{tot}}=0$) subspace. 
For $\lambda_{\mathrm{max}}$ not scaling with system size $L$, the average $E_{\mathcal{N}}$ grows with $L$.
However, our data suggests that at any finite ratio $\lambda_{\mathrm{max}}/L$, this value vanishes with increasing system size.  
Each data point is averaged over more than $100$ random states.2
\textbf{Inset.} 
The crossing point $\lambda_{\mathrm{cross}}/L$ of the curves with $L$ and $L+4$ in the main plot decays polynomially with system size. This indicates that in the large system size limit, the Haar random initial states supported in a finite fraction of $\lambda$ subspaces ($\lambda_{\mathrm{max}}/L = O(1)$) gives an average zero $E_{\mathcal{N}}$.
}
\end{figure}

First, we study the $E_{\mathcal{N}}$ for stationary states $\rho_{\lambda_{\mathrm{tot}}}$ restricted to one single Krylov subspace $\lambda_{\mathrm{tot}}$ with $ m_{\mathrm{tot}}=0$. This means that the weight $ p_{\lambda_{\mathrm{tot}}}=1$ for one particular value of $\lambda_{\mathrm{tot}}$ in Eq.~\eqref{eq:SU2_rho_ss_multi_lambda}.
For $\lambda_{\mathrm{tot}} = L/2$ and $m_{\mathrm{tot}}=0$, the subspace is one-dimensional, and thus the stationary state is a pure state, which is given by $\ket{\psi} = (S_{\mathrm{tot}}^{-})^{L/2}\ket{\uparrow}^{\otimes L} \propto \sum_{\phi}\ket{\phi_{m=0}}$ as an equal superposition of all $m=0$ states, with $\ket{\phi_{m=0}}$ given in Eq.~\eqref{eq:U1_basis}.
The negativity is $E_{\mathcal{N}}(\rho_{\lambda_{\mathrm{tot}=L/2}}) = S_{1/2} (\ket{\psi}) \sim \frac{1}{2}\log L$ (App.~\ref{subsec:SU2_Haar}).
Hence, we analytically showed that $E_{\mathcal{N}}(\rho_{\lambda_{\mathrm{tot}}}) \sim \frac{1}{2}\log L$  for both the smallest $\lambda_{\mathrm{tot}} = 0$ (singlet) as well as the largest $\lambda_{\mathrm{tot}} = L/2$ irreps.

We now numerically evaluate Eq.~\eqref{eq:SU2_En_lambdas} for any other value of ${\lambda_{\mathrm{tot}}}$. The results are shown in 
Figure~\ref{fig:SU2_neg_random_Haar}a. 
For arbitrary fixed density $\lambda_{\mathrm{tot}}/L$, the $E_{\mathcal{N}}(\rho_{\lambda_{\mathrm{tot}}})$ increases with increasing system size.  
This indicates that the stationary states restricted to one subspace with fixed $\lambda_{\mathrm{tot}}/L$ are also highly entangled as captured by the faster-than-area-law scaling of the logarithmic negativity.
The inset shows that $E_{\mathcal{N}}(\rho_{\lambda_{\mathrm{tot}}})$ approximately scales logarithmically with $\lambda_{\mathrm{tot}}$ for large $\lambda_{\mathrm{tot}}$.

Second, we consider stationary states corresponding to Haar random initial states sampled within a direct sum of subspaces, $\bigoplus_{\lambda_{\mathrm{tot}}=0}^{\lambda_{\mathrm{max}}} \mathcal{H}_{\lambda_{\mathrm{tot}},m_{\mathrm{tot}=0}}^{\mathcal{A}(L)}$ (see Fig.~\ref{fig:SU2_neg_random_Haar}b).
The stationary state takes the form $\rho = \sum_{\lambda_{\mathrm{tot}} = 0}^{\lambda_{\mathrm{max}}} p_{\lambda_{\mathrm{tot}}} \Pi^{\lambda_{\mathrm{tot}}}_{m_{\mathrm{tot}}=0}/D_{\lambda_{\mathrm{tot}}}$. 
We expect that with increasing $\lambda_{\mathrm{max}}$, the stationary state has weight on a larger fraction of the full Hilbert space, and thus increasingly resembles the trivial infinite-temperature state which has trivial quantum entanglement.
Evaluating Eq.~\eqref{eq:SU2_En_lambdas}, we study the average $E_{\mathcal{N}}$ for such stationary states, $\langle E_{\mathcal{N}} \rangle_{\mathrm{Haar}}$, for a half-chain bipartition of spin-$1/2$ chain with $L=4n$.
Figure~\ref{fig:SU2_neg_random_Haar}b shows that, with increasing $\lambda_{\mathrm{max}}/L$, the $\langle E_{\mathcal{N}} \rangle_{\mathrm{Haar}}$ indeed decreases, and that it also decreases faster for larger system sizes.
The inset in Fig.~\ref{fig:SU2_neg_random_Haar}b shows the crossing points $\lambda_{\mathrm{cross}}/L$ of two consecutive system sizes (i.e., of the curves for $L$ and $L+4$ in the main plot), which decreases as the system size increases. 
This suggests that a finite fraction of total spin ($\lambda_{\mathrm{max}}/L = O(1)$) leads to zero $\langle E_{\mathcal{N}} \rangle_{\mathrm{Haar}}$ as $L\rightarrow\infty$.
For $\lambda_{\mathrm{max}}/L = 1/2$, the initial states are sampled from the full $m_{\mathrm{tot}}=0$ sector, with $p_{\lambda_{\mathrm{tot}}}$ approximately given by $ {D_{\lambda_{\mathrm{tot}}}}/{D_{m_{\mathrm{tot}=0}}}$ up to fluctuations that decrease exponentially with the system size.
Therefore, the stationary state $\rho \approx \Pi^{m_{\mathrm{tot}}=0}/D_{m_{\mathrm{tot}=0}}$ as $L\rightarrow \infty$, which only manifests a U($1$) symmetry and hence leads to vanishing $\langle E_{\mathcal{N}}\rangle_{\mathrm{Haar}}$ (App.~\ref{app:SU2}).

\subsection{SU$(N)$ symmetry and higher spin}\label{subsec:sun}
Our results for SU($2$) symmetry can be generalized to any SU$(N)$ symmetry on a spin-$(N-1)/2$ chain (local Hilbert space dimension $N$). 
When imposing a SU$(N)$ symmetry on such Hilbert space, the Schur-Weyl duality~\cite{fulton_representation_2004} leads to a decomposition of the Hilbert space into the irreps of SU$(N)$ and the symmetric group $S_L$
\begin{equation}
    \mathcal{H}^{(L)} = \bigoplus_\lambda \left(\mathcal{H}^{\mathrm{SU}(N)}_{\lambda} \otimes \mathcal{H}^{S_{L}}_\lambda \right),
\end{equation}
which can be understood as a special case of the commutant and bond algebra language, with
\begin{equation}
    \mathcal{C} = U(su(\mathrm{N})),\,\,\,  \mathcal{A}=\mathbb{C}[S_L],
\end{equation}
with $U$(su$(N)$) as the UEA of su$(N)$.
The bond algebra can be generated by permutation operators, $\mathcal{A}(L)= \langle \{P_{j,j+1}\}, \mathbb{1}\rangle$, where $P_{j,j+1} = \sum_{\sigma \sigma^\prime} (|\sigma \sigma^\prime\rangle\langle \sigma^\prime \sigma|)_{j,j+1}$ is the permutation operator of two neighboring spins (see e.g., Ref.~\cite{202_classenhowes_SY_RS_commutant}). 
For SU$(N)$ symmetry, irreps are labeled by a set of non-negative integers $\lambda = (\lambda_1, \ldots, \lambda_N)$, with $\lambda_1 \geq \lambda_2 \geq \ldots \geq \lambda_N$ and $\lambda_1 + \lambda_2 + \ldots \lambda_N = L$. These $\lambda_j$'s correspond to the number of $\ell$-cycles in a permutation with $\ell\geq j$. For example, for the permutation $(123)(45)(67)(8)$, $(\lambda_1,\lambda_2,\lambda_3)=(4, 3, 1)$. 
On the other hand, $m$ labels the degenerate $\lambda$ irreps, which can be given by the so-called Gelfand–Tsetlin (GT) patterns~\cite{1950_Gelfand_Tsetlin, 1968_Biedenharn_GT, 2011_alex_numerical_SUN}. 
The dimensions of the irreps of SU$(N)$ and $S_L$ are given by~\cite{fulton_representation_2004}
\begin{equation}\label{eq:SUN_irrep_dim}
\begin{aligned}
    d_\lambda &= \frac{1}{(N-1)!(N-2)!\ldots 1!}\prod_{1\leq i < j\leq N} (\tilde{\lambda}_i - \tilde{\lambda}_j),\\
    D_\lambda^{(L)} &= \frac{L!}{\tilde{\lambda}_1!\tilde{\lambda}_2! \ldots \tilde{\lambda}_N!} \prod_{1\leq i < j\leq N} (\tilde{\lambda}_i - \tilde{\lambda}_j),
\end{aligned}
\end{equation}
where $\tilde{\lambda}_i = \lambda_i + N-i$.
For $N=2$, the total spin is simply given by $J = (\lambda_1-\lambda_2)/2$, $m$ reduces to the total magnetization, and the corresponding $d_\lambda$ and $D_{\lambda}^{(L)}$ are given by Eq.~\eqref{eq:SU2_dimension}.

Consider the singlet subspace of the chain with length $L=2nN$ and $n\in \mathbb{N}$. 
The singlet subspace is labeled by $\lambda = (L/N, \ldots, L/N)$, which is equivalent to $\lambda=(0,\ldots,0)$ and denoted as $\lambda = 0$ in the following~\footnote{The SU$(N)$ irrep $\lambda = (\lambda_1, \ldots, \lambda_N)$ are equivalent up to a global constant, $\lambda+c = (\lambda_1+c, \ldots, \lambda_N+c)$ with $c \in \mathbb{Z}$. Thus SU($2$) irrep can be uniquely labeled by the $J=(\lambda_1-\lambda_2)/2$.}.
While the UEA of su$(N)$ is an example for which the general expressions of basis states Eq.~\eqref{eq:gen_triv0} in Sec.~\ref{sec:exact_basis} hold due to its Hopf algebra structure, we provide here a sketch of a more specific proof for su$(N)$. Additional details are provided in App.~\ref{app:SUN}. 
To do so, one considers the set of operators
$S^{\pm}_{(l)}$, $S^z_{(l)}$ for $1\leq l \leq N-1$, with commutation relation~\cite{1950_Gelfand_Tsetlin, 2011_alex_numerical_SUN}
\begin{equation}\label{eq:SUN_commutation_main}
    [S^{+}_{(l)}, S^{-}_{(l)}] = 2 S^z_{(l)}, \,\,[ S^z_{(l)}, S^{\pm}_{(l)}] = \pm  S^{\pm}_{(l)},
\end{equation}
which recovers the commutation relation of $S^\pm_{\mathrm{tot}}$, $S^z_{\mathrm{tot}}$ when $N=2$.
The basis states of the singlet subspace satisfy
\begin{equation}\label{eq:SUN_loweringraising_op_main}
    S^{\alpha}_{(l)} |\lambda_{\mathrm{tot}}=0\rangle = 0,  \forall 1\leq l \leq N-1, \alpha \in \{\pm,z\}.
\end{equation}
Moreover, the expressions of operators $S^{\alpha}_{(l)}$ acting on other $\lambda_{\mathrm{tot}}\neq0$ basis states given by the GT patterns are also known~\cite{1986_raczka_GT_theory, 2011_alex_numerical_SUN}.
Therefore, using similar strategy for the derivation of SU$(2)$ CG coefficients, we can prove that:
(i) the singlet states are composed of an irrep $\lambda$ and its dual $\bar{\lambda}$, where $\bar{\lambda} = (L/N-\lambda_N, \ldots, L/N-\lambda_1)$ for $\lambda = (\lambda_1, \ldots, \lambda_N)$. (ii) The singlet states are equal superpositions (up to minus signs) of pairs of basis states labeled by $\lambda, m$ and their dual basis $\bar{\lambda}, \bar{m}$. The dual $\bar{m}$ is also uniquely determined by $m$ (see details in App.~\ref{app:SUN}). 
Thus, the CG coefficients are given by $c_m(\lambda) = \frac{1}{\sqrt{d}_\lambda}$ up to a minus signs. 
Therefore,
\begin{equation}\label{eq:SUN_basis_state}
    |\lambda_{\mathrm{tot}} = 0;\lambda; a, b \rangle 
    =  \frac{1}{\sqrt{d_\lambda}} \sum_{m} \eta_{\lambda,m} |\lambda, m; a\rangle |\bar{\lambda}, \bar{m}; b\rangle,
\end{equation} 
with $|\eta_{\lambda,m}| = 1$.
The corresponding stationary state in the singlet subspace is given by $\rho = \Pi^{\lambda_{\mathrm{tot}}=0}/D_0^{(L)}$. 
Hence, with Eq.~\eqref{eq:SUN_basis_state}, we can use the same techniques as in Sec.~\ref{subsec:SU2} to obtain the same exact expressions for the different entanglement proxies in Sec.~\ref{sec:exact_entanglement} as a function of the dimensions of the irreps of the bond and commutant algebras.

\begin{figure}[bt]
\centering
\includegraphics[width=1\columnwidth]{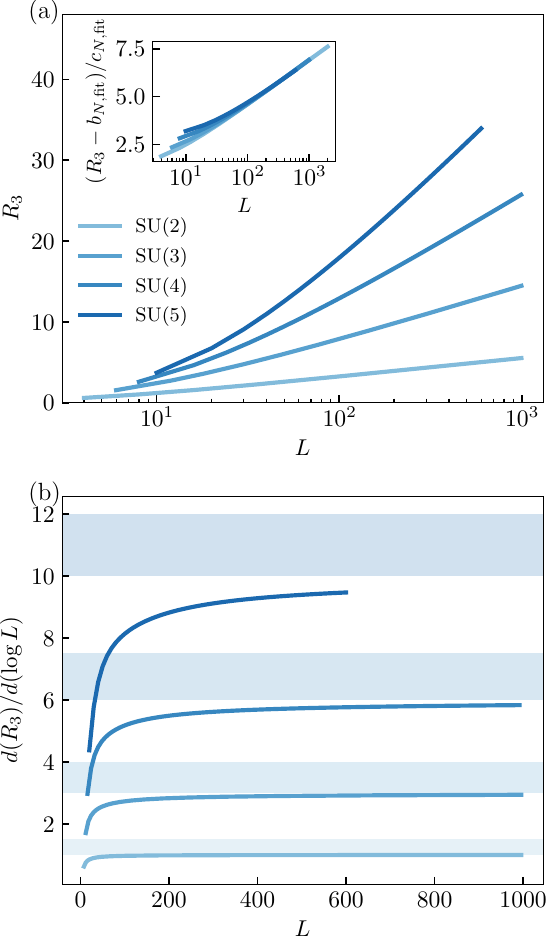}
\caption{\label{fig:R3_SUd} \textbf{Rényi negativity $R_3$ for SU$(N)$ symmetry.} (a) Results from the exact expression of $R_3$ for SU$(N)$ in Eq.~\eqref{eq:general_Rn} and Eq.~\eqref{eq:SUN_irrep_dim}. Asymptotically, $R_3$ scales logarithmically with system size for different $N$. \textbf{Inset.} Rescaling of $R_3$ to obtain data collapse with $c_{N,\mathrm{fit}} = \frac{1}{2} N^2 + c_1 N + c_0$. We obtain numerically that $c_1 \approx -0.66$, $c_0 \approx 0.29$. The fitting coefficient $b_{N,\mathrm{fit}}$ is chosen such that $|R_3-c_{N,\mathrm{fit}}\log L|$ is minimized for large $L$. (b) Scaling coefficients of $R_3$ for different $N$. We prove that $R_3 \sim c \log L$ with $c \in [\frac{N(N-1)}{2}, \frac{N^2-1}{2}]$. This interval is shown in the shade regions with $N=2, 3, 4, 5$ increasing from bottom to top. The solid lines show the derivative of $R_3$ with respect to $\log L$ from the curves in (a). The derivative is converging to the lower bound $\frac{N(N-1)}{2}$ as $L$ increases.}
\end{figure}

We study the asymptotic finite-size scaling of the different entanglement quantities in the limit $ L\gg N$ for half-chain bipartition with the dimensions of irreps $d_\lambda$, $D_\lambda^{(L)}$ given in Eq.~\eqref{eq:SUN_irrep_dim}.
For example, we obtain that $R_3$ scales logarithmically with system size
\begin{equation}\label{eq:SUn_R3_asy_scaling}
\begin{aligned}
    R_3 \sim \,& c_{\mathrm{SU}(N)}^{R_3} \log L + O(1), \\
    &\text{ with }c_{\mathrm{SU}(N)}^{R_3} \in [\frac{N(N-1)}{2}, \frac{N^2-1}{2}].
\end{aligned}
\end{equation}
Notice that $N^2-1$ corresponds to the dimension of the algebra su$(N)$, while  $N(N-1)$ corresponds to the codimension of its maximal Abelian subalgebra.  
Figure~\ref{fig:R3_SUd} shows the numerical evaluation of the exact expressions of $R_3$ (Eq.~\eqref{eq:general_Rn}) in panel a, and their derivatives $d(R_3)/d(\log L)$ in panel b. The inset in Figure~\ref{fig:R3_SUd}a shows the scaling collapse of the data from where we obtain the scaling coefficients $c_{N,\mathrm{fit}} \approx \frac{1}{2} N^2 + c_1 N + c_0$ with $c_1\approx -0.65$ and $c_0 \approx 0.29$. The hypothesis for the fitting expression of the scaling coefficient is based on the analytic upper and lower bound for $c_{\mathrm{SU}(N)}^{R_3}$. On the other hand, panel b shows that the derivatives are converging to the lower bound $\frac{N(N-1)}{2}$ as the system size increases. Here the blue bands span the range $c_{\mathrm{SU}(N)}^{R_3} \in [\frac{N(N-1)}{2},\frac{N^2-1}{2}]$ for different $N$.
We also find that the operator space entanglement scales logarithmically with system size, 
\begin{equation}
\begin{aligned}
    S_{\mathrm{OP}} \sim & c_{\mathrm{SU}(N)}^{S} \log L + O(1), \,\,\, \\
    &\text{ with } c_{\mathrm{SU}(N)}^{S} \in [\frac{N^2-1}{2}, N^2-1].
\end{aligned}
\end{equation}
Finally, we can obtain an upper bound for the logarithmic negativity 
\begin{equation}
    E_{\mathcal{N}} \leq N(N-1)\log L.
\end{equation}
We provide details of the derivation in App.~\ref{app:SUN}.
By numerically evaluating the exact expressions, we observe that the three entanglement proxies scale logarithmically and are compatible with the upper and lower bounds. 
Therefore, we conclude that for SU$(N)$ symmetry with $N\ll L$, the entanglement proxies scale logarithmically, similar to the case of SU($2$).

In App.~\ref{app:numerics_dynamics}, we include numerical simulations showing that starting from a symmetric initial state, the system eventually saturates to these analytical predictions of SU($N$).

\section{Hilbert space fragmentation}\label{sec:fragment}
As stated in the introduction, fragmented systems are those whose Hilbert space decomposes into exponentially many Krylov subspaces~\cite{2020_sala_ergodicity-breaking, 2020_khemani_local, 2020_SLIOMs, moudgalya_fragment_commutant_2022}. 
They possess a large number of conserved quantities, leading to an exponentially large commutant $\dim[\mathcal{C}(L)]\sim e^L$. 
Moreover, fragmentation can be classified as either classical or quantum~\cite{moudgalya_fragment_commutant_2022}. 
Classical fragmentation refers to the existence of a local \emph{product} basis as the common eigenbasis for all elements in a maximal Abelian subalgebra of the commutant.
For quantum fragmentation, no such basis exists. 
In this section, we study the effect of exponentially many conserved quantities, especially for quantum fragmentation, on the entanglement of stationary states.

\subsection{Classical fragmentation}\label{subsec:classical_frag}

Similar to the U($1$) case, with a local product basis spanning every symmetry subspace, the partial transpose of the stationary states in Eq.~\eqref{eq:general_stationary_state} is trivial (similar to Eq.~\eqref{eq:U1_partial_transpose_rho}). 
Therefore, the stationary states are separable, with zero logarithmic and Rényi negativities. 
However, the strong dynamical constraints in fragmented systems can lead to large classical correlations, as measured by the operator space entanglement~\cite{2023_li_HSF_open}.

\begin{figure*}[t]
\includegraphics[width=18cm, scale=1]{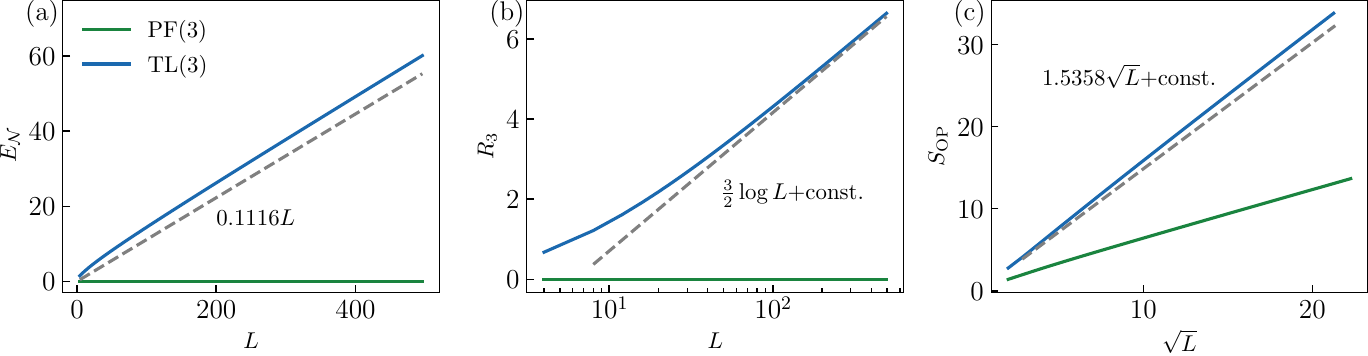}
\caption{\label{fig:Entanglement_frag} \textbf{Entanglement of stationary states for exponentially large commutants corresponding to Hilbert space fragmentation.} (a) The logarithmic negativity $E_{\mathcal{N}}$ is zero for the PF($3$) model with classical fragmentation. For the commutant $\mathcal{C}_{\mathrm{TL}(3)}$ with quantum fragmentation, $E_{\mathcal{N}}$ is lower bounded by $\sim 0.116L + O(\log L)$, showcasing a volume law scaling. (b) The $R_3$ of PF($3$) model vanished, while it scales logarithmically for TL($3$) model. (c) The OSE scales as $O(\sqrt{L})$ as proven in Ref.~(PF) for the PF($3$) model. For TL($3$), the OSE scales as $\sim 1.5358\sqrt{L} + O(\log L)$ (App.~\ref{app:fragment}).} 
\end{figure*}

We take the PF($N$) chain with spin-$(N-1)/2$ as an example, and consider $L=4n$ with $n\in\mathbb{N}$ and OBC~\cite{2018_Caha_pairflip}.
Unlike in the previous sections, we now make explicit use of the bond algebra to define the commutant.
As we already explained, each of these algebras is completely characterized once a Hilbert space and the other algebra (i.e., its centralizer) are specified. 
The bond algebra is generated by the local terms as follows~\cite{moudgalya_fragment_commutant_2022}
\begin{equation}
    \mathcal{A}_{\mathrm{PF}(N)}(L) = \langle\{  |\sigma \sigma\rangle \langle \sigma^\prime \sigma^\prime|)_{j,j+1} + \text{h.c.}\}, \{S_j^z\}, \mathbb{1} \rangle, 
\end{equation}
where $\sigma = 1,\ldots, N$ are different spin-$z$ components, and $S_j^z$ is the local spin-$z$ operator for spin-$(N-1)/2$. 
The first local term is the pair-flip term, which flips a pair of neighboring spins of the same sign to another sign.
The commutant $\mathcal{C}_{\mathrm{PF}(N)}(L)$ contains $N-1$ independent U($1$) charges, $N^{\sigma} = \sum_j (-1)^j N_j^{\sigma}$, with $N_j^{\sigma} = (|\sigma\rangle\langle \sigma|)_j$, as $\sum_{\sigma=1}^N N^{\sigma} = \mathbb{1}$.  
Nonetheless, due to the additional dynamical constraints, it has dimension $\dim [\mathcal{C}_{\mathrm{PF}(N)}(L)] \sim e^L$ for $N>2$, which are constructed in Ref.~\cite{moudgalya_fragment_commutant_2022}.
The Krylov subspaces can be constructed by identifying the `dot patterns' for each product state as follows: First, we map the spins to different colors, e.g., $\ket{0} = \ket{\,\reddot\,}$, $\ket{1} = \ket{\,\greendot\,}$, $\ket{2} = \ket{\,\bluedot\,}$ for $N=3$. Then for a product state, connect every two neighboring spins with the same color as a pair from left to right. Remove the paired spins, and repeat the previous step until all the unpaired spins have different colors from their neighboring spins. 
The unpaired spins are the dot patterns of length $M$, $\Sigma_M \equiv (\sigma_1, \sigma_2, \ldots, \sigma_M)$ with $\sigma_i \neq \sigma_{i+1}$ and even number of dots $M =0, 2, \ldots, L$ for even $L$. 
For example, the product state
\begin{equation}
|\longstate\rangle\label{dotpattern}
\end{equation}
has the dot pattern $(\bluedot\,\,\reddot)$.
Note that spin pairs do not cross each other.
Each Krylov subspace is labeled by a dot pattern $\Sigma_M$, i.e., the Krylov subspace is spanned by all the product states with the same dot pattern, which is invariant under the pair-flip dynamics.
The number of Krylov subspaces is thus given by the number of different dot patterns, $K=1+ \sum_{M} N (N-1)^{M-1} \sim O((N-1)^L)$, which scales exponentially with system size. 
The commutant being Abelian, its irreps are one-dimensional $d_{\Sigma_{M}} \equiv 1$, and the dimension of the Krylov subspaces with $M$ dots is $D_{\Sigma_M} = D_{M}^{\mathrm{PF}(N)}$, which only depends on the length of the dot patterns. The dimension $D_{M}^{\mathrm{PF}(N)}$ can be calculated via generating functions in Ref.~\cite{2018_Caha_pairflip}.
Consider the Krylov subspace labeled by the zero dot ($M_{\mathrm{tot}}=0$). The basis states of this sector are given by
\begin{equation}\label{eq:PF_basis}
    |M_{\mathrm{tot}}=0; \Sigma_{M}; a, b\rangle = |\Sigma_{M}; a\rangle |\bar{\Sigma}_{M}; b\rangle,
\end{equation}
where $\bar{\Sigma}_{M} = (\sigma_{M}, \ldots, \sigma_1)$ are the dot patterns with reverse order of $\Sigma_{M} = (\sigma_1, \ldots, \sigma_{M})$ to form a state with zero dots.
For example, a zero-dot product state $|\,\PFshortstate\rangle$ has $\Sigma_{M} = (\reddot\,\,\,\bluedot)$ and $\bar{\Sigma}_{M} = (\bluedot\,\,\,\reddot)$ for half-chain biparition.
The corresponding stationary state is an equal sum of projectors onto the set of basis states in Eq.~\eqref{eq:PF_basis}, $\rho = \frac{1}{D_{0}}\sum_{\Sigma_{M}; a, b}|M_{\mathrm{tot}}=0;\Sigma_{M}; a, b\rangle\langle M_{\mathrm{tot}}=0; \Sigma_{M}; a, b|$.
As the basis states are local product states, the logarithmic negativity and the Rényi negativity are zero, the same as with U($1$) symmetry.
Similar to the U($1$) case, the operator space entanglement is given by 
\begin{equation}\label{eq:PF_OSE}
    S_{\mathrm{OP}} = \sum_{\Sigma_{M}} \frac{D_{\Sigma_{M}}^{(L_A)}D_{\bar{\Sigma}_{M}}^{(L_B)}}{D_0^{(L)}} \log \frac{D_{\Sigma_{M}}^{(L_A)}D_{\bar{\Sigma}_{M}}^{(L_B)}}{D_0^{(L)}}.
\end{equation}
As we show in Ref.~\cite{2023_li_HSF_open} for $N=3$, the vectorized stationary state can be mapped to the ground state of the PF($3$) model, which can be generalized to arbitrary $N$. The ground state has a half-chain von Neumann entropy that scales as $O(\sqrt{L})$ (and thus OSE of the stationary state) as found in Ref.~\cite{2018_Caha_pairflip} and shown in Fig.~\ref{fig:Entanglement_frag}c (green line).
The OSE of the PF($N$) model scales parametrically faster than for U($1$) symmetry, due to the extensive number of conserved dot pattern configurations.
This contributes to large classical correlations. 
Recall that in general grounds one can show $S_{\mathrm{OP}}\leq \log[\dim(\mathcal{C}_{\mathrm{min}})]\sim O(L)$.

\subsection{Quantum fragmentation}

Now we turn to quantum fragmentation, where the Hilbert space fragments into exponentially many Krylov subspaces in an entangled basis.
An example is the TL($N$) model for spin-$S$ chain with local Hilbert space dimension $N=2S+1$ and OBC~\cite{1990_TL}. The bond algebra is given by
\begin{equation}
    \mathcal{A}_{\mathrm{TL}(N)}(L) = \langle \{e_{j,j+1}\}, \mathbb{1}\rangle,
\end{equation}
with $e_{j,j+1}\equiv\sum_{\sigma,\sigma^\prime = 1}^{N}(|\sigma\sigma\rangle\langle\sigma^\prime \sigma^\prime|)_{j,j+1}$, which is the well-known Temperley-Lieb algebra (see e.g., Ref.s~\cite{2007_Read_Commutant, moudgalya_fragment_commutant_2022}).
In this case, Ref.~\cite{2007_Read_Commutant} provided a detailed analysis of its commutant $\mathcal{C}_{\mathrm{TL}(N)}$, the Read-Saleur (RS) commutant. 
The advantage of introducing the commutant via its bond algebra, is that the latter allows for a presentation in terms of spatially local terms.
These local terms are projections onto the two-site singlet state $\frac{1}{\sqrt{N}}\sum_{\sigma=1}^N |\sigma\sigma\rangle_{j,j+1}$, which map to the spin-$1/2$ Heisenberg terms $\vec{S}_j \cdot \vec{S}_{j+1}$ for $N=2$ and a spin-$1$ purely biquadratic term $(\vec{S}_j \cdot \vec{S}_{j+1})^2$ for $N=3$ (via an on-site unitary transformation), respectively. 
The TL($3$) model can be understood as a more symmetric formulation of the PF($3$) model with e.g., an additional SU($3$) symmetry.
Similar to the PF($N$) family of models, the Krylov subspaces of the TL$(N)$ models can be labeled by an extensive number of dot patterns of length $2\lambda$, with $\lambda = 0, 1, \ldots, L/2$ for even $L$~\cite{2007_Read_Commutant}. 
The dot patterns are defined as the set of states that are annihilated by all operators $e_{j,j+1}$, i.e., $e_{j,j+1}\ket{\psi}=0$ for all $j$.
Therefore, these patterns include the product-state dot patterns of PF($N$) model, i.e., $\Sigma_{2\lambda} = (\sigma_1, \ldots, \sigma_{2\lambda})$ for $\sigma_i\neq \sigma_{i+1}$, as well as entangled dot patterns, e.g., $(|\sigma\sigma\rangle - |\sigma^\prime \sigma^\prime \rangle)_{j,k}$ with $\sigma\neq \sigma^\prime$.
The subspaces labeled by dot patterns of the same length are $d_\lambda$ degenerate, with degeneracy~\cite{2007_Read_Commutant}
\begin{equation}
    d_\lambda = [2\lambda+1]_{q},
\end{equation}
and dimension
\begin{equation}
     D_\lambda^{(L)} = \begin{pmatrix}L\\
    L/2+\lambda
    \end{pmatrix} - \begin{pmatrix}L\\
    L/2+\lambda+1
    \end{pmatrix}.
\end{equation}
Here $[n]_q = (q^{n}-q^{-n})/(q-q^{-1})$ is the $q$-deformed integer with $q\geq 1$ defined by $N=q+q^{-1}$ for $N\geq 2$. When $N=2$ ($q=1$), $d_\lambda$ recovers the SU($2$) case as in Eq.~\eqref{eq:SU2_dimension}.

As in the previous sections, we consider the stationary state restricted to the trivial representation $\lambda_{\mathrm{tot}}=0$ sector (i.e., zero dot pattern).
While the CG series of the RS commutant is the same as for SU$(2)$~\cite{2007_Read_Commutant}  (i.e., $V_{j_1}\otimes V_{j_2} \cong \bigoplus_{j=|j_1-j_2|}^{j_1+j_2}V_j$ with $V_j$ irreps of the commutant), due to its complexity, the CG coefficients can not be easily obtained in general. 
However, the RS commutant possesses a Hopf algebra structure (in the limit $L\to \infty$)~\cite{2007_Read_Commutant}. 
Hence, Theorem \ref{th:theorem_1} applies, and the basis states in the trivial representation (corresponding to $\lambda_{\mathrm{tot}} = 0$) is given by Eq.~\eqref{eq:gen_triv0}. For $L=4n$,
\begin{equation}\label{eq:TL_basis}
    |\lambda_{\mathrm{tot}}=0; \lambda; a, b\rangle = \frac{1}{\sqrt{d_\lambda}} \sum_{m} \eta_{\lambda,m} |\lambda, m; a\rangle |\lambda, \bar{m}; b\rangle,
\end{equation}
with $|\eta_{\lambda,m}|=1$.
We also include some analytic and numeric results for small system sizes in App.~\ref{app:fragment} to provide some additional intuition.

For the RS commutant, the degeneracy scales as $d_{\lambda} \sim q^{2\lambda}$ (since $q>1$ for $N\geq 3$), which leads to a large logarithmic negativity as given by the exact expression Eq.~\eqref{eq:general_En}. 
Consider now a half-chain bipartition. We obtain that the logarithmic negativity is lower bounded by a linear scaling in the limit $L\to \infty$
\begin{equation}
    E_{\mathcal{N}} > c_{\mathrm{TL}(N)}^{\mathrm{lin}} L + O(\log L),
\end{equation}
with $c_{\mathrm{TL}(N)}^{\mathrm{lin}}$ a function of $N$ that is approximately given by $ c_{\mathrm{TL}(3)}^{\mathrm{lin}}\approx 0.1116$ for $N=3$. We derive $c_{\mathrm{TL}(N)}^{\mathrm{lin}}$ analytically in App.~\ref{app:fragment}.
This indicates that $E_{\mathcal{N}}$ grows at least with the volume of the system $L$, as shown in Fig.~\ref{fig:Entanglement_frag}a (see blue solid line for exact numerical values, and dashed line for the lower bound).
On the other hand, the $n$-th Rényi negativity scales logarithmically,  
\begin{equation}
R_n < R_{\infty} \sim \frac{3}{2}\log L + O(1),
\end{equation}
where the prefactor $3/2$ holds for all $N\geq 3$.

\begin{figure}[bt]
\centering
\includegraphics[width=1\columnwidth]{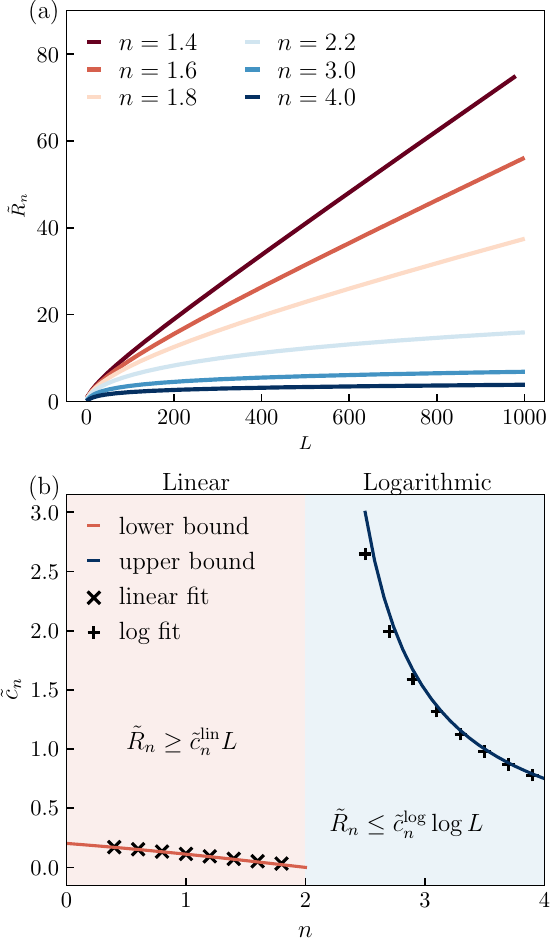}
\caption{\label{fig:TL3_Rn} \textbf{Generalized Rényi negativity $\tilde{R}_n$ of TL($3$) with $n>0$.} (a) Results of $\tilde{R}_n$ obtained from the exact expression Eq.~\eqref{eq:general_tilde_Rn}. For $0<n<2$, $\tilde{R}_n$ scales linearly with system sizes (red solid lines). For $n>2$, $\tilde{R}_n$ scales logarithmically (blue solid lines). 
(b) Scaling prefactors for $\tilde{R}_n$. 
Solid lines are analytically derived bounds on the asymptotic scaling of $\tilde{R}_n$. The crosses are the fitting coefficients $\tilde{c}_{n}^{\mathrm{fit}}$ from curves similar in (a). 
When $0<n<2$, $\tilde{R}_n$ is lower bounded by a linear scaling, $\tilde{R}_n \geq \tilde{c}^{\mathrm{lin}}_n L$. 
The fitting coefficients $\tilde{c}^{\mathrm{fit}}_n$ (crosses) lie above the  $\tilde{c}^{\mathrm{lin}}_n$ (red line) as expected. For $n>2$, $\tilde{R}_n$ is upper bounded by a logarithmic scaling, $\tilde{R}_n \leq \tilde{c}^{\mathrm{log}}_n \log L$, and we show that the fitting coefficients $\tilde{c}_n^{\mathrm{fit}}$ (pluses) lie below $\tilde{c}_n^{\mathrm{log}}$ (blue line).
}
\end{figure}

To understand that distinct volume-law scaling of $E_{\mathcal{N}}$ from the logarithmic one for $R_n$ with $n>2$,
we study the generalized Rényi negativity $\tilde{R}_n$ defined in Eq.~\eqref{eq:tilde_Rn} for arbitrary non-integer values $n\neq 2$. This definition interpolates between the logarithmic negativity and $R_n$, i.e., $\tilde{R}_1 = E_{\mathcal{N}}$, and $\tilde{R}_n = R_n/(n-2)$ for $n$ even.
In App.~\ref{app:fragment}, we prove that
\begin{equation}
\begin{aligned}
    &\tilde{R}_{n}	\geq  \tilde{c}_{N, n}^{\mathrm{lin}} L + O(\log L), & n<2,\\
    &\tilde{R}_n \leq \tilde{c}_{N,n}^{\mathrm{log}} \log L + O(1), & n>2.
\end{aligned}
\end{equation}
The prefactor $\tilde{c}_{N, n}^{\mathrm{lin}} = \tilde{c}_{\tilde{a},n}^{^{\mathrm{lin}}}$ with $\tilde{a}$ depending on $q$ (and thus $N$), which is given by
\begin{equation}
\begin{aligned}
    \tilde{c}_{\tilde{a},n}^{^{\mathrm{lin}}} =& -(\frac{1}{2}+2\tilde{a})\log(\frac{1}{4}+\tilde{a})-(\frac{1}{2}-2\tilde{a})\log(\frac{1}{4}-\tilde{a})\\
    &\,\, +2(2-n)\tilde{a} \log q -2\log2,
\end{aligned}
\end{equation}
with 
\begin{equation}
    \tilde{a} = \frac{1}{4} \frac{q^{2-n}-1}{q^{2-n}+1}.
\end{equation}
Note that value $\tilde{c}_{N,n}^{\mathrm{lin}} = c_{\mathrm{TL}(N)}^{\mathrm{lin}}$ for $n=1$. 
And for $n>2$, $\tilde{c}_{N,n}^{\mathrm{log}} = \frac{3}{2(n-2)}$.
The exact dependence of $\tilde{R}_n$ with system size when numerically evaluating Eq.~\eqref{eq:general_tilde_Rn}) is shown in Fig.~\ref{fig:TL3_Rn}a for TL($3$) and several values of $n$. 
Therefore, there is a transition from volume law to logarithmic law for the quantity $\tilde{R}_n$ as a function of $n$ across $n=2$. Figure~\ref{fig:TL3_Rn}b shows that indeed for $n<2$ ($n>2$), $\tilde{R}_n$ scales linearly (logarithmically) with $L$, with a prefactor $\tilde{c}_n^{\mathrm{lin}}$ ($\tilde{c}_n^{\mathrm{log}}$) that lies close to its lower (upper) bound. 

Finally, we find that the OSE of the TL$(N)$ model has the same scaling as the PF($N$) model,
\begin{equation}
    S_{\text{OP}} \sim  \sqrt{\frac{8}{\pi}} (\log q)\sqrt{L} + O(\log L),  
\end{equation}
which is shown in Fig.~\ref{fig:Entanglement_frag}c.
The detailed derivation of the asymptotic scalings can be in App.~\ref{app:fragment}. 
Moreover, in App.~\ref{app:numerics_dynamics} we include numerical simulations for $N=3,4$ showing that starting from a symmetric initial state, the system eventually saturates to these analytical predictions.

\section{Conclusion and Discussion}\label{sec:conclusion}

In this work, we characterized the stationary state of various strongly-symmetric evolutions in terms of their bond and commutant algebras.
We derived exact closed-form expressions for the logarithmic negativity $E_{\mathcal{N}}$, Rényi negativities $R_n$, and operator space entanglement (OSE) for stationary states $\rho = \Pi^{\lambda_{\mathrm{tot}}=0}/D_{0}^{(L)}$ restricted to one symmetric subspace --- the trivial subspace of the commutant.
Our derivations made use of the orthonormal basis constructed in Eq.~\eqref{eq:gen_triv0} within the global $\lambda_{\mathrm{tot}}=0$ subspace.
In Sec.~\ref{sec:exact_basis}, we proved that a sufficient condition for this decomposition to hold is that the commutant possesses a Hopf algebra structure in the limit $L\to \infty$.
Instances include many systems of interest, like e.g., those whose commutant corresponds to the UEA of any Lie algebra and the RS commutants considered in this work, as well as the group algebra of any finite group, and quantum groups.
Assuming that this structure is satisfied, we found that all the entanglement quantities we considered are upper bounded by the logarithm of the dimension of the commutant on the smaller bipartition, without specific knowledge of the commutants.
Moreover, whenever the commutant is Abelian (i.e., $d_{\lambda}=1$ for all $\lambda$), $E_\mathcal{N}$ and all $R_n$ exactly vanish.

The previous consequence leads to the following general conclusions: (i) for finite Abelian groups and classical fragmentation, the stationary state in the singlet subspace is separable. (ii) The negativity is \emph{at most} an O$(1)$ number for any symmetric evolution where the dimension of the commutant does not scale with system size, as it e.g., happens for non-Abelian finite groups. Moreover, (iii) the negativity can scale as fast as the logarithm of the system size for conventional continuous symmetries, e.g., SU$(N)$; while (iv) it can showcase a volume-law scaling for quantum fragmented systems. 
In particular, we provided a detailed analysis of systems with conventional U$(1)$ symmetries and SU$(N)$ symmetries, as well as classical and quantum fragmentation, the latter realized via Read-Saleur commutants. 
In the case of U$(1)$ symmetric systems, and those with classical fragmentation, we analytically found that while the stationary states are separable, the OSE asymptotically scales as $\log(L)$ and $\sqrt{L}$ respectively. The non-zero OSE is related to the classical fluctuations of the conserved charges. In contrast, for SU$(N)$ symmetric systems, we found that all quantities asymptotically scale as $\log(L)$, with the scaling coefficient depending on $N$. For the RS commutants, we proved that the $E_{\mathcal{N}}$ exhibits a volume-law scaling, while $R_n$ with integer Rényi index $n$ scales only logarithmically in system size. 
We further characterized this novel transition by introducing a generalized Rényi negativity $\tilde{R}_n$ defined for any real $n > 0$, leaving it as an open question whether a similar behavior can be observed in other systems.
Overall, our work shows that (non-Abelian) strong symmetries can extensively enhance the mixed-state entanglement of the stationary states, but also limit how entangled they can be within the singlet subspace.

Our work also gives rise to several interesting directions regarding open quantum dynamics and mixed-state phases. 
First, Theorem \ref{th:theorem_1}, which provides an ONB in the trivial subspace with an explicit bipartite structure, and the resulting exact expressions of mixed-state entanglement quantities, directly applies to many other physical systems of interest.
Second, while we focused on three particular such quantities, it would be relevant to understand whether similar closed-form expressions can be found for other mixed-state entanglement measures~\cite{1996_entanglement_distillation, 1997_relative_entanglement, 2002_entanglement_of_purification,   2004_squash_entanglement, Brand_o_2011} using Theorem \ref{th:theorem_1}. In particular, if a general (upper or lower) bound in terms of the dimension of the commutant can be found (For a comparison of several mixed-state entanglement measures including faithful ones, see Table I in Ref.~\cite{Brand_o_2011}.).
If that applies, one could exactly compute various mixed-state entanglement measures for a large family of physical systems.
Such exact results for many-body wave functions are scarce in the literature. Hence, our results could provide a deeper understanding of the distinction between different mixed-state entanglement measures, as well as the effect of different symmetries.

In addition, the bond and commutant algebras formalism can be applied to broader contexts. 
Firstly, this can be directly carried out for larger representations of SU($N$), i.e., larger local Hilbert space dimension, other than the fundamental representations we considered in this work. 
Second, as mentioned in Sec.~\ref{sec:review_symmetries}, this analysis can be extended to non-Hermitian Kraus operators~\cite{2008_Baumgartner_math_Lindblad_2, 2024_Yoshida_Lindblad}, which hence can lead to non-unital quantum channels. Can a similar analysis of mixed-state entanglement be extended to the resulting stationary states?
Another interesting direction would be to explore the possible effects of weak symmetries, i.e., conserved quantities preserved by the combination of the system and the environment, and extend its formulation in the commutant algebra language.

As a more practical application of our findings, it would be relevant to understand whether such highly entangled mixed stationary states can be used for any quantum information task, such as entanglement resources for quantum teleportation~\cite{1993_Bennet_teleportation}.
In fact, the non-Abelian commutants provide both diagonal and off-diagonal decoherence-free subspaces (i.e., irreps of the commutant)~\cite{1998_DFS_Lidar_Chuang_Whaley, 2000_Zanardi_quantuminfo, 2000_Knill_DFS_noiseless_subsystem, 2001_DFS_fault_tolerant_quantum_computation, 2003_DFS_review, 2007_DFS_Bartlett, 2023_symmetry_induced_DFS}, and hence the stationary state could potentially be utilized as a quantum memory~\cite{ 2022_Wildeboer_DFS_symmetry_protected_quantum_memory}.
For them to be exploited, it would be essential to: (i) be able to engineer dissipative environments~\cite{2008_zoller_experiment_quantum_state_pre_open_system, 2008_zoller_Lindblad_pre_state, 2009_verstraete_quantum_state_engineer, 2023_google_engineered_dissipation} that preserve the strong symmetries of interest, e.g., SU$(2)$ symmetry; (ii) understand the robustness of our results to small symmetric-breaking perturbations, and (iii) try to (parametrically) shorten the preparation time of such highly entangled stationary states by e.g., using non-local quantum channels or classical communication~\cite{Raussendorf05, Bolt16, piroli2021locc,PhysRevX.14.021040, verresen2021efficiently,Bravyi22,lu2022shortcut,friedman2023feedback,hierarchy,iqbal2023topological,chen2023nishimori,lee2022decoding,Iqbal2024nonabel,smith2023aklt}.
Also, as discussed in the main text, the stationary state entanglement remains for initial states that are not perfectly prepared within the singlet subspaces, e.g., for SU$(2)$ and quantum fragmentation~\cite{2023_li_HSF_open}.
Hence, preparing completely symmetric initial states is not essential.

Finally, we leave as an interesting open direction whether a strong-to-weak spontaneous symmetry breaking (sw-SSB) transition can occur at a finite error rate $p$ (or finite time for a Lindbladian time evolution).
Unlike previous works~\cite{LeeYouXu2022, ma2024topological, lessa2024strongtoweak, 2024_sala_SSSB}, some of the systems we considered can lead to highly entangled stationary states that cannot be reached by a local quantum channel on a finite time. These give rise to many interesting questions: Is it possible that a sw-SSB occurs at a finite time, as measured by (symmetric) Rényi-2 correlations or fidelity measures? And if so, can this transition be characterized as a thermal phase transition of a classical (and symmetric!) stat-mech model? And finally, could we find a purification of the stationary state~\cite{2024_sala_SSSB} that corresponds to a symmetry-protected topological order?

\begin{acknowledgments}
We are grateful to Berislav Buca, Yujie Liu, Sanjay Moudgalya, Sara Murciano, Subhayan Sahu, Thomas Schuster, Robijn Vanhove, Ruben Verresen, Hironobu Yoshida and Yizhi You for helpful discussions. Also to Tarun Grover for making his lecture notes for the Boulder School 2023 available online.
P.S. acknowledges support from the Caltech Institute for Quantum Information and Matter, an NSF Physics Frontiers Center (NSF Grant PHY-1733907), and the Walter Burke Institute for Theoretical Physics at Caltech.
This work was supported by the European Research Council (ERC) under the European Union’s Horizon 2020 research and innovation program under Grant Agreement No. 85116 and No. 771537, the Deutsche Forschungsgemeinschaft (DFG, German Research Foundation) under Germany’s Excellence Strategy EXC-2111- 390814868, TRR 360 (project-id 492547816), FOR 5522 (project-id 499180199), and the Munich Quantum Valley, which is supported by the Bavarian state government with funds from the Hightech Agenda Bayern Plus.

\end{acknowledgments}

\textit{Note added.} While this work was being completed, we became aware of a related work by Moharramipour, Lessa, Wang, Hsieh and Sahu~\cite{PI_paper}.\\

\textbf{Data and materials availability.}
Data analysis and simulation codes are available on Zenodo upon reasonable request~\cite{Data_Set_Zenodo}.

\appendix
\section{Stationary states of open system dynamics}\label{app:stationary_state}
In this section, we sketch a proof of the stationary state in Eq.~\eqref{eq:general_stationary_state} for dissipative quantum channels and Eq.~\eqref{eq:rho_gen} for Lindblad dynamics with random decay rates.

For quantum channels,
$\mathcal{E} = \prod_{j}\mathcal{E}_j$, $\mathcal{E}_j(\rho) = \sum_{\alpha }K_{j,\alpha} \rho K_{j,\alpha}^\dagger $ with Hermitian Kraus operators $K_{j,\alpha}^\dagger = K_{j,\alpha}$.
The dynamics and the conserved quantities are given by the bond and commutant algebras,
\begin{equation}
\begin{aligned}
    &\mathcal{A}(L) = \langle \{K_{j,\alpha} \}, \mathbb{1}\rangle, \\
    &\mathcal{C}(L) = \{O: [O, K_{j,\alpha}] = 0, \forall j, \alpha\},  
\end{aligned}
\end{equation}
respectively.  
The Hilbert space of strongly symmetric quantum channels can be decomposed as 
\begin{equation}
    \mathcal{H}^{(L)} = \bigoplus_\lambda \left( \mathcal{H}^{\mathcal{C}(L)}_\lambda \otimes \mathcal{H}^{\mathcal{A}(L)}_\lambda \right),
\end{equation}
for a system of length $L$.
Being the Kraus operators Hermitian, elements of the commutant are simultaneously the conserved quantities and fixed points of the quantum channel, i.e., $\mathcal{E}(O) = O$ and $\mathcal{E}^\dagger (O) = O$ for $O\in\mathcal{C}$. The full commutant can be spanned by the operators $\Pi_{m,m^\prime}^\lambda = \sum_{a} |\lambda, m; a\rangle \langle \lambda, m^\prime; a|$ with $a=1,\ldots,D_\lambda$. They are projectors on a Krylov subspace when $m=m^\prime$ and intertwine operators between degenerate subspaces when $m\neq m^\prime$. 
Therefore, within each subspace labeled by $\lambda, m, m^\prime$, there is a fixed point $\Pi_{m,m^\prime}^\lambda/D_\lambda$, which is a stationary state for the diagonal subspace $m=m^\prime$, and stationary coherence for off-diagonal subspaces $m\neq m^\prime$ between degenerate subspaces~\cite{2023_li_HSF_open}. 

The stationary state restricted to a Krylov subspace is unique, which can be proven as follows: All density matrices in the subspace form a convex set $\mathcal{S}$. If there exist two stationary states $\rho_1$ and $\rho_2$ in the same subspace $\mathcal{K}$, due to the linearity of the quantum channel, an arbitrary state $\rho_\mu = \mu\rho_1 + (1-\mu)\rho_2$ is also a stationary state. This line of stationary states intersects with the boundary of the convex set $\partial S$ for a particular $\mu_0$.  Therefore $\rho_{\mu_0}$ has a lower rank. This indicates that there exists a $|\psi\rangle$, with $\rho_{\mu_0}|\psi\rangle = 0$. 
Therefore, $0 = \langle \psi| \rho_{\mu_0} |\psi\rangle
=  \langle \psi| K_{j,\alpha} \rho_{\mu_0}K_{j,\alpha}^\dagger |\psi\rangle = \|\sqrt{\rho_{\mu_0}} K_{j,\alpha}^\dagger |\psi\rangle \|^2$, for all $j, \alpha$, such that $\rho_{\mu_0}K_{j,\alpha}|\psi\rangle = 0$~\cite{2024_Yoshida_Lindblad}. 
Here, we used that $\rho_{\mu_0}$ is a density matrix and thus positive semi-definite.
Note that $K_{j,\alpha}|_{\mathcal{B}(\mathcal{K})}$ generates all operators bounded linear operators $\mathcal{B}(\mathcal{K})$ by definition of bond algberas, where $\mathcal{B(K)}$ is the operator space of Krylov subspace $\mathcal{K}$. 
This means that $\rho_{\mu_0} |\phi \rangle = 0$ for all $|\phi\rangle \in \mathcal{K}$ and thus $\rho_{\mu_0} = 0$. This then implies that $\rho_1$ and $\rho_2$ have to be linearly dependent, which leads to a contradiction. 
For the off-diagonal subspace $m\neq m^\prime$, the stationary coherence is also unique, as the fixed points in off-diagonal subspace and diagonal subspace are one-to-one related by $\Pi_{m,m^\prime}$~\cite{2008_Baumgartner_math_Lindblad_2}. 
Then, with $\Pi_{m,m^\prime}^\lambda$ as the conserved quantities, we have $\mathrm{Tr}(\mathcal{E}^\dagger (\Pi_{m,m^\prime}^\lambda) \rho_0) = \mathrm{Tr}( (\Pi_{m,m^\prime}^\lambda)^\dagger \mathcal{E}(\rho_0)) = \mathrm{Tr}( (\Pi_{m,m^\prime}^\lambda)^\dagger \rho)$. This indicates that the weights of the fixed points in different subspaces are given by the weight of initial states.
Therefore, we obtain that for an initial state with fixed $\lambda$, there is a unique stationary state
\begin{equation}
    \rho_{\lambda} = \sum_{m, m^\prime} \left((M_\lambda)_{m,m^\prime} \frac{\Pi_{m,m^\prime}^\lambda}{D_\lambda}\right),
\end{equation}
where $M_\lambda$ is a $d_\lambda \times d_{\lambda}$ matrix with trace one, and it encodes information about the initial state. The matrix element is given by $(M_\lambda)_{m,m^\prime} = \mathrm{Tr}[\Pi^\lambda_{m^\prime,m} \rho(t=0)]$.

Note that the uniqueness of the stationary state extends to non-Hermitian Kraus operators, with bond algebra $\mathcal{A}=\langle \{K_{j,\alpha}, \{K_{j\alpha}^\dagger\}, \mathbb{1}\} \rangle$, and the commutant algebra $\mathcal{C}(L)=\{O:[O, K_{j,\alpha}] = [O, K_{j,\alpha}^\dagger] = 0\}$. Assuming that $\{K_{j\alpha}^\dagger|_{\mathcal{B}(\mathcal{K})}\}$ generates all the operators in $\mathcal{B}({\mathcal{K}})$~\cite{PI_paper}, the proof of uniqueness applies.

More generally, consider Lindblad dynamics with random coupling strengths (see additional details in ~\cite{2023_li_HSF_open}), 
\begin{equation}
    \mathcal{L}(\rho) = -i\sum_j J_j [h_j, \rho] + \sum_j \gamma_j (L_j\rho L_j^\dagger - \frac{1}{2}\{L_j^\dagger L_j, \rho\}),
\end{equation}
with $J_j$ and $\gamma_j\geq 0$ as random coefficients dependent on space and time, the bond and commutant algebras are given by 
\begin{equation}
\begin{aligned}
    &\mathcal{A}(L)=\langle\{h_j\}, \{L_j\}, \mathbb{1}\rangle, \\
    &\mathcal{C}(L) = \{O:[O,h_j]=[O,L_j]=0, \forall j\}.
\end{aligned}
\end{equation}
We can prove that there is a unique stationary state for general initial states~\cite{2023_li_HSF_open}
\begin{equation}\label{eq:app_stationary_state}
    \rho = \sum_{\lambda, m, m^\prime} \left((M_\lambda)_{m,m^\prime} \frac{\Pi_{m,m^\prime}^\lambda}{D_\lambda}\right) = \bigoplus_{\lambda}\left( M_\lambda \otimes \frac{\mathbb{1}_\lambda}{D_\lambda}\right)
\end{equation}
for a given initial state in the full Hilbert space.
The stationary state is an element of $\mathcal{C}(L)$, which commutes with all elements in the $\mathcal{A}(L)$ as guaranteed by random coefficients in the Lindblad dynamics.
For Lindblad dynamics without random coefficients or quantum channels, there can exist additional structures of the stationary state. More specifically, there can be fixed points in the subspaces spanned by $\ket{\lambda, m; a}\bra{\lambda^\prime, m^\prime; a^\prime}$ for $\lambda \neq \lambda^\prime$. These fixed points can add to the structure of stationary states. However, we haven't observed these additional structures in numerical evolutions, and these are expected to happen for non-generic cases~\cite{2012_Buca_Prosen}.

\section{Decomposition of the $\lambda_{\mathrm{tot}}=0$ subspace assuming Hopf algebra structure}
\label{app:Hopf}

In this Appendix, we will prove Propositions \ref{th:Prop_1} and \ref{th:Prop_2},
and hence Theorem \ref{th:theorem_1}.
The proofs use basic concepts of Hopf algebras, for which see the book by Kassel~\cite{kassel2012quantum}. It requires knowledge of the following concepts: (i) the structure of algebra $(A, \mu, \eta)$, $A$-modules and $A$-linearity; (ii) the use of tensor products $\otimes$, the dual of a vector space $V^*$, and the coevaluation map $\delta_V$ which will play a central role in the proof; and (iii), the definition of coalgebra $(C,\Delta, \varepsilon)$, bialgebra structure $(H,\mu,\eta, \Delta,\varepsilon)$, the antipode $S:H\to H$, and finally Hopf algebras $(H,\mu,\eta, \Delta,\varepsilon, S)$. The content of points (i), (ii) and (iii) can be found in Chapters 1, 2 and 3 respectively~\cite{kassel2012quantum}. 
In Sec.~\ref{app:B_key_concepts}, we start by setting up definitions for the proof in terms of algebras and modules. In Sec.~\ref{app:Prop_1}, we give the proof of Proposition~\ref{th:Prop_1}. Note that we do not use
the inner product or complex conjugation until near the end of this
part. In Sec~\ref{app:Prop_2}, we give the proof of Proposition~\ref{th:Prop_2} in two steps.

\subsection{Key concepts}\label{app:B_key_concepts}
Let us start by introducing some basic concepts to set up the proof, where clarity will be prioritized over completeness. In the following we consider the complex numbers $\mathbb{C}$ as the field of scalars throughout. The basic component is an algebra structure, to which
the remaining structures of a Hopf algebra will be added.

\begin{definition}{\textbf{Algebra.}}
An algebra $(A, \mu,\eta)$ is a ring, together with a ring map $\eta:\mathbb{C}\to A$ whose image $\eta(A)$ belongs to the center of $A$ (i.e., it commutes with every element of $A$), and a bilinear multiplication $\mu:A\times A \to A$ bilinear over the scalars. $\eta$ turns the ring into a vector space, and satifies $\eta(1)=1$.
\end{definition}

\begin{definition}{\textbf{(Left) $A$-module.}}
It is a vector space $V$ together with a bilinear map $A\times V \to V:\, (a,v)\to av$, such that $a(a^\prime v)=(aa^\prime)v$ for all $v\in V$, $a,a^\prime\in A$ and with $1v=v$.
\end{definition}

One can similarly define a \textbf{right $A$-module} using instead a bilinear map $V\times A \to V$ with the algebra acting on $V$ from the right, i.e., $(v,a)\to va$, such that $(va)a^\prime=v(aa^\prime)$. 
The dual of a left $A$-module $V$ corresponds to the set of linear maps $V^*=\mathrm{Hom}(V,\mathbb{C})$. In general, the dual of a left $A$-module is a right $A$-module namely, with the algebra acting from the right. The usual right module structure on $V^*$ is given by $(fa)(v)=f(av)$ for $f\in V^*$ and $v\in V$.  

The final basic concept is that of $A$-linearity.
\begin{definition}{\textbf{$A$-linearity.}}
A linear map $f:V\to V^\prime$ is $A$-linear if $f(av)=af(v)$ for all $v\in V$ and $a\in A$.  In other words, $f$ is a homomorphism of left $A$-modules.
\end{definition}

We are working with vector spaces and homomorphisms (linear maps) of them.

We will now introduce the central object of the proof, the so-called coevaluation map $\delta_V$. But in order to so, we first remind the reader that as e.g., shown in Corollary II.2.2 in Kassel's book, the map $\lambda_{U,V}:V\otimes U^* \to \mathrm{Hom}(U,V)$ given for all $u\in U$ and $v\in V$ by $\lambda_{U,V}(v\otimes \alpha)=\alpha(u)v$ is an \emph{isomorphism}, i.e., $V\otimes U^* \cong \mathrm{Hom}(U,V)$, provided
$U$ or $V$ is finite dimensional.
The use of this map is that it allows us to map a linear function $f:V\to V$, into a state in $V\otimes V^*$ via $f=\lambda_{V,V}(\sum_{i,j}f_j^i v_i\otimes v^j)$. Here, $\{v_i\}$ and  $\{v^i\}$  are bases of $V$ and $V^*$ respectively, satisfying $v^j(v_i)=\delta_{ij}$, such that $f(u_j)=\sum_i f^i_j v_i$. In particular, when considering the identity map on a finite-dimensional $V$, one finds
\begin{equation}
    \mathrm{id}_V = \lambda_{V,V}(\sum_i v_i\otimes v^i).
\end{equation}

\textbf{Main object.} The previous ingredients allow us to introduce the (linear) \textbf{\emph{coevaluation map}} $\delta_V:\mathbb{C}\to V\otimes V^*$, defined via
\begin{equation} \label{eq:delta_V}
    \delta_V(1)= \lambda_{V,V}^{-1}(\mathrm{id}_V)=\sum_i v_i \otimes v^i,
\end{equation}
which is independent of the choice of basis $\{v_i\}$ (and hence its dual which is uniquely defined via $v^j(v_i)=\delta_{ij}$).
The coevaluation map $\delta_V$ produces a certain element of the tensor product of
vector spaces. We want to show that, if $V$ is a left $A$-module,
then the coevaluation map is, in a natural way, a map into a tensor product of $A$-modules,
and that the image is an $A$-module that is trivial.
This is the singlet state of the commutant that we wish to study. In order to
make such statements, additional structure is required, beyond the simple
statement that the commutant is an associative algebra. 
In order to make tensor products and duals of left-modules into left modules, and to define trivial left-modules, we need it to possess
the structure of a Hopf algebra (if it helps the reader to get a more intuitive understanding, the word $A$-module can be replaced by representation in the previous discussion).

 Hence, in the following we introduce the concepts appearing in point (iii). We first need to introduce the notion of coalgebra, which basically corresponds to the definition of an algebra but where all arrows are reversed.

\begin{definition}{\textbf{Coalgebra.}}
A coalgebra $(C, \Delta,\varepsilon)$ is given by a vector space $C$, a comultiplication $\Delta:C\to C\otimes C$, and a counit $\varepsilon:C\to \mathbb{C}$, satisfying $(\varepsilon \circ \mathrm{id})\circ \Delta = (  \mathrm{id} \circ \varepsilon)\circ \Delta $.
\end{definition}

Given an element $a$ of the coalgebra, $\Delta(a)\in C\otimes C$ is given by $\Delta(a)=\sum_i a_i^\prime\otimes a_i^{\prime\prime}$ where the sum runs over some elements of $C$ depending on $a$. At this point it is useful to introduce the so-called Sweedler's sigma notation to get rid of subscripts, and agree that the previous sum will take the form $\Delta(a)=\sum_{(a)} a^\prime\otimes a^{\prime\prime}$ with the sum running over some coalgebra elements depending on $a$, and $a^\prime$, $a^{\prime\prime}$ different elements in general. Hence, the comultiplication enables to ``share" the action of a coalgebra element $a\in C$ in the tensor product $C\otimes C$. Moreover, the comultiplication is coassociative, which allows us to iterate over the tensor product of representations in a consistent way. For example, in the case of the su$(2)$ algebra, $\Delta(\bm{J})=  \bm{J}_A\otimes \mathbb{1} + \mathbb{1}\otimes \bm{J}_B $ corresponds to the addition of angular momenta, which can be extended to many particles. However, unlike for SU$(N)$ symmetries, comultiplication is not necessarily cocommutative. The main motivation is that as we said at the beginning, we would like to take tensor product of modules, and in particular we would like those to be $A$-modules themselves. 

On the other hand, the counit $\varepsilon: C \to \mathbb{C}$, equips any vector space $V$ with a \emph{trivial} $C$-module structure by $av=\varepsilon(a)v$ where $a\in C$ and $v\in V$. For example, $\varepsilon(a)=0$ for any non-trivial element of the UEA of a Lie algebra, as well as for any non-trivial element of the Read-Saleur commutant, as it happens for the trivial representation. A more detailed discussion can be found on page 47 of Kassel's~\cite{kassel2012quantum}. 

For both algebra and coalgebra structures one can define morphisms $f: (A, \mu,\eta)\to (A^\prime,\mu^\prime,\eta^\prime)$ and $g:(C,\Delta,\varepsilon) \to (C^\prime,\Delta^\prime,\varepsilon^\prime) $, that preserve the structure and are compatible to taking tensor products: $\mu^\prime\circ (f\otimes f)=f\circ \mu$ together with $f\circ \eta=\eta^\prime$; and $(f\otimes f)\circ \Delta = \Delta^\prime\circ f$ together with $\varepsilon=\varepsilon^\prime\circ f$. These two structures can be consistently combined within the same algebra acquiring a bialgebra structure.

\begin{definition}{\textbf{Bialgebra.}}
A bialgebra is a quintuple $(H,\mu,\eta,\Delta,\varepsilon)$ where $(H,\mu,\eta)$ is an algebra, and $(H,\Delta,\varepsilon)$ is a coalgebra, such that $\Delta$ and $\varepsilon$ are morphisms of algebras. Moreover, one can define bialgebra morphisms as morphisms of both the underlying algebra and coalgebra structures defined above.
\end{definition}

The last necessary piece to achieve our goal is the \textbf{\emph{antipode}}, which can turn a right $A$-module into a left one.
To introduce it we first need to define the convolution $\star$ of two algebra homomorphisms.

\begin{definition}{\textbf{Convolution.}}
Given an algebra $(A, \mu,\eta)$ and a coalgebra $(C,\Delta, \varepsilon)$, and two maps $f,g\in \mathrm{Hom}(C,A)$, the convolution $f\star g$ of $f$ and $g$ is defined in Sweedler's notation via $(f\star g)(a)=\sum_{(a)}f(a^\prime)g(a^{\prime\prime})$. 
\end{definition}

\begin{definition}{\textbf{Antipode.}}
Let $(H, \mu,\eta,\Delta, \varepsilon)$ be a bialgebra. An endomorphism $S$ of $H$ as a vector space is called an antipode for the bialgebra $H$ if (in Sweedler's notation) 
\begin{equation}
\sum_{(a)}a^\prime S(a^{\prime\prime})=\varepsilon(a)1 = \sum_{(a)}S(a^\prime)a^{\prime\prime}.
\end{equation}
\end{definition}

\begin{definition}{\textbf{Hopf algebra.}} A \emph{Hopf algebra} is a bialgebra with an antipode (if an antipode exists, it is unique).
\end{definition}

Examples of Hopf algebras include the UEA of su$(N)$, the group algebra of any finite group, quantum groups which are not equivalent to groups or Lie algebras, and the inverse limit of the commutants of the Temperly-Lieb (TL$(N)$) algebras, referred above as Read-Saleur commutants.

\subsection{Proof of Proposition 1} \label{app:Prop_1}

\subsubsection{First steps: use of antipode and coevaluation map}
\begin{proposition}
Let $(\mathcal{C}, \mu,\eta,\Delta, \varepsilon,S)$ be a Hopf algebra. Then its antipode $S$ satisfies   
\begin{equation} \label{eq:transp}
    S(ab)=S(b)S(a),\, \text{ and }\,\, S(1)=1
\end{equation}
for all $a,b\in \mathcal{C}.$
\end{proposition}
\noindent\textit{Proof.} see Kassel~\cite{kassel2012quantum}, Thm.~III.3.4.\vspace{15pt}

Recall that the comultiplication $\Delta$ enables us to equip the $\mathcal{C}\otimes \mathcal{C}$-module $U\otimes V$ (naturally defined via $(a\otimes a^\prime)(u\otimes v)=au\otimes a^\prime v$) with a left $\mathcal{C}$-module structure via
$a(u\otimes v) = \Delta(a)(u\otimes v) = \sum_{(a)}a^\prime u\otimes a^{\prime\prime} v$.  The antipode provides a natural left $\mathcal{C}$-module structure to $\mathrm{Hom}(U,V)\cong V\otimes U^*$ [where the symbol $\cong$ means isomorphic], when $U,V$ have left $\mathcal{C}$-module structures. In particular, to turn the dual of a left $\mathcal{C}$-module into a left module. More explicitly, for any $f\in \mathrm{Hom}(U,V)$ one can show that
\begin{equation} \label{eq:Ates_mod}
    (af)(v) = \sum_{(a)}a^\prime f(S(a^{\prime\prime})v).
\end{equation}
In particular for $\alpha\in V^*$, this leads to left $\mathcal{C}$-module structure on $V^*$ via $ (a\alpha)(v) = \alpha(S(a)v)$. 
The condition in Eq.~\eqref{eq:transp} becomes necessary to make the left module well-defined.  Consider any two algebra elements $a_1, a_2$ and $\alpha\in V^*$. Then 
\begin{equation}
    (a_1 a_2 \alpha)(v)= \alpha(S(a_1 a_2)v),
\end{equation}
and associativity of the algebra requires this to agree with
\begin{equation}
    (a_1 (a_2 f))(v)=(a_2 f)(S(a_1)v)= f(S(a_2)S(a_1)v),
\end{equation}
for all $v\in V$, which holds if and only if Eq.~\eqref{eq:transp}  holds.

We have now introduced enough structure to prove that the coevaluation map $\delta_V$ is indeed a trivial module of the Hopf algebra $\mathcal{C}$ for any finite-dimensional module $V$. First of all, recall from Eq.~\eqref{eq:delta_V} that $\delta_V(1)=\lambda_{V,V}^{-1}(\mathrm{id}_V)=\sum_i v_i \otimes v^i$ for $\{v_i\}$ any basis of $V$.  Because of the use of the antipode, for any $a\in \mathcal{C}$, the algebra $\mathcal{C}$ acts on the left via $a\delta_V(1)=\sum_i \sum_{(a)}a^{\prime}v_i \otimes a^{\prime\prime}v^i$.  Moreover, $\delta_V$ is the composition of the unit $\eta:\mathbb{C}\to \mathrm{End}(V)$ and of $\lambda_{V,V}^{-1}$ given by $\delta_V = \lambda_{V,V}^{-1}\circ \eta$. By Proposition III.5.2 in Kassel's book the map $\lambda_{V,V}$ is $\mathcal{C}$-linear when $V$ is finite-dimensional, and being invertible, so is $\lambda_{V,V}^{-1}$. This means that $a(\lambda_{V,V}^{-1}(f))= \lambda_{V,V}^{-1}(af)$. On the other hand Proposition III.5.3 shows that $\eta:\mathbb{C}\to \mathrm{End}(V)$ is also $\mathcal{C}$-linear. This follows by considering the case $f=\mathrm{id}_V=\eta(1)$ in Eq.~\eqref{eq:Ates_mod}, which gives 
\begin{equation}
    (a\mathrm{id}_V)(v)=\varepsilon(a)v,
\end{equation}
for all $v\in V$, and where $\varepsilon(a)$ in a scalar. By composition, the coevaluation map $\delta_V$ is also $\mathcal{C}$-linear
\begin{equation}
\begin{aligned}
    a\delta_V(1)& = a( \lambda_{V,V}^{-1}( \mathrm{id}_V)) =  \lambda_{V,V}^{-1}( a\mathrm{id}_V)  \\ &= \lambda_{V,V}^{-1}( \varepsilon(a)\mathrm{id}_V)= \varepsilon(a) \lambda_{V,V}^{-1}( \mathrm{id}_V)  = \varepsilon(a) \delta_V(1),
\end{aligned}
\end{equation}
for all $a\in \mathcal{C}$, which implies that $\delta_V(1)=\sum_i v_i\otimes v^i$ is a trivial $A$-module of the algebra. In conclusion, we have proven following Kassel that: 
\begin{proposition}
Given a finite-dimensional left $\mathcal{C}$-module $V$ of a Hopf algebra $\mathcal{C}$, then the coevaluation map $\delta_V:\mathbb{C}\to V\otimes V^*$ gives a left $\mathcal{C}$-module
\begin{equation}
    \delta_V(1)=\sum_i v_i \otimes v^i,
\end{equation} 
which is independent of the choice of basis $\{v_i\}$ in $V$.
\end{proposition}

\subsubsection{Coevaluation map: from modules to representations}
\label{app:mod_to_rep}
The action of $\mathcal{C}$ on a left $\mathcal{C}$-module $V$ leads to a representation of $\mathcal{C}$ on $V$ $\rho:\mathcal{C}\to \mathrm{End}(V)$ acting as $\rho(a)v=av$, i.e., acting from the left. As we saw in the previous section, the antipode allows us to define a left $\mathcal{C}$-module structure on $V^*$ via $(a\alpha)(v)=\alpha(S(a)v)$. In particular, Eq.~\eqref{eq:Ates_mod} shows that the presence of the antipode can be used to turn the dual vector space $V^*$ into a representation without requiring additional structure.  This is given by $S(a)^T\alpha^T$ for all $a\in \mathcal{C}$, converting row vectors to column vectors, and right action to left action.

Hence, we can understand both left and right (simple) $\mathcal{C}$-modules as (irreducible) representations of the algebra. A simple module is one that does not containa nonzero submodule~\footnote{Being the algebras semisimple, modules (representations) are fully decomposable as direct sums.} From this perspective, $\delta_V(1)$ corresponds to a representation of $\mathcal{C}$ on $\mathcal{H}_{\lambda}\otimes \mathcal{H}_{\lambda}^*$, i.e.,  $\sum_i v_i \otimes v^i$ is just a ``vector''. Making use of the Dirac notation we can then simply write $\sum_i \ket{v_i}\otimes \ket{v^i}$, such that the action of $a\in \mathcal{C}$ is given by $ \ket{v_i}\otimes \ket{v^i}\to  \sum_{(a)}\ket{a^\prime v_i}\otimes \ket{S(a^{\prime\prime})^Tv^i}$. Finally, using (for the first time in this proof) the inner product intrinsic to the Hilbert space $\mathcal{H}_{\lambda}$ one can consider an orthonormal basis $\{v_i\}$ and its dual $\{v^i\}$ and normalize the state to find
\begin{equation}
    \ket{\psi}=\frac{1}{\sqrt{d_\lambda}}\sum_{i=1}^{d_\lambda} \ket{v_i} \otimes \ket{v^i},
\end{equation}
which proves Proposition 1.

\subsection{Proof of Proposition 2} \label{app:Prop_2}

First of all, (1) let us assume both bond $\mathcal{A}(L)$ and commutant $\mathcal{C}(L)$ algebras are self-adjoint (i.e., the algebra includes the adjoint of every element, and thus semisimple) for each length $L$ (Condition (2) in Sec.~\ref{sec:exact_basis}).  
(2) Also assume that $\mathcal{A}(L)$ is the same (isomorphic) for any interval of the same length $L$, regardless of the two endpoints (i.e., translation invariant), which is implicitly involved in the proof. 
Finally, (3) assume the commutant $\mathcal{C}(L)$ is defined for each $L$, and that the inverse limit $\underleftarrow{\lim}_{L\to \infty}\mathcal{C}(L)= \mathcal{C}$ is a Hopf algebra as we have discussed (see condition (1) in Sec.~\ref{sec:exact_basis}). 
The irreps of $\mathcal{A}(L)$ are denoted as $\mathcal{H}_{\lambda}^{\mathcal{A}(L)}$ with dimensions $D_\lambda^{(L)}$. The irreps of $\mathcal{C}(L)$ are denoted as $\mathcal{H}^{\mathcal{C}(L)}_\lambda$ with dimensions $d_\lambda$ (which are assumed to be independent of $L$).

\subsubsection{Same multiplicities for decomposition under a subalgebra}
\label{app:mul_subalg}

Let us consider a bipartition of the chain into $L=L_A+L_B$. In the following, we show that semisimplicity ensures that the fusion coefficients when decomposing an irrep of $\mathcal{C}(L_A)\otimes \mathcal{C}(L_B)$ into those of $\mathcal{C}(L)$, where $\mathcal{C}(L)\subset \mathcal{C}(L_A)\otimes \mathcal{C}(L_B) $, match those appearing when decomposing an irrep of $\mathcal{A}(L)$ into irreps of $\mathcal{A}(L_A)\otimes \mathcal{A}(L_B)\subset \mathcal{A}(L)$.
(Notice that here, we do not make use of the full Hopf algebra structure. The previous structure is all that is used in this section.)
Consider a chain of length $L$ with a Hilbert space $\mathcal{H}^{(L)} = (\mathbb{C}^m)^{\otimes L}$ (a product of factors of dimension $m$) and dimension $m^L$. Given the commutant $\mathcal{C}(L)$ and bond algebra $\mathcal{A}(L)$, the Hilbert space can be decomposed as
\begin{equation}\label{eq:app_Hilbert_decompose}
    \mathcal{H}^{(L)} = \bigoplus_\lambda \left( \mathcal{H}^{\mathcal{C}(L)}_\lambda \otimes \mathcal{H}^{\mathcal{A}(L)}_\lambda \right).
\end{equation}
 Because $\mathcal{C}(L) \subset \mathcal{C}(L_A)\otimes \mathcal{C}(L_B)$, the decomposition of a tensor product of irreps is an example of decomposition under restriction to a subalgebra. (This is connected with the comultiplication
$\Delta(\mathcal{C})\subset \mathcal{C}\otimes \mathcal{C}$). For irreps labeled by $\mu$ and $\nu$,

\begin{equation}\label{eq:app_HC_decompose}
\mathcal{H}^{\mathcal{C}(L_A)}_\mu\otimes\mathcal{H}^{\mathcal{C}(L_B)}_\nu =\bigoplus_\lambda N_{\mu\nu}^\lambda \mathcal{H}^{\mathcal{C}(L)}_\lambda,
\end{equation}
where the non-negative integers $N_{\mu\nu}^\lambda$ are the multiplicities for each $\lambda$ in the sum~\cite{fulton_representation_2004}. This expresses the ``fusion rules'', and implies that
\begin{equation}
d_\mu d_\nu = \sum_\lambda N_{\mu\nu}^\lambda d_\lambda.
\end{equation}
For a bipartition of the chain, for a subspace for irreps $\lambda_A, \lambda_B$,
it is
\begin{equation}
(\mathcal{H}^{\mathcal{A}(L_A)}_{\lambda_A} \otimes \mathcal{H}^{\mathcal{C}(L_A)}_{\lambda_A})\otimes (\mathcal{H}^{\mathcal{A}(L_B)}_{\lambda_B} \otimes \mathcal{H}^{\mathcal{C}(L_B)}_{\lambda_B}),
\end{equation}
and is an irrep of both tensor product algebras. Using Eq.~\eqref{eq:app_HC_decompose}, this can be decomposed as 
\begin{equation}
\begin{aligned}
&(\mathcal{H}^{\mathcal{A}(L_A)}_{\lambda_A} \otimes \mathcal{H}^{\mathcal{C}(L_A)}_{\lambda_A})\otimes (\mathcal{H}^{\mathcal{A}(L_B)}_{\lambda_B} \otimes \mathcal{H}^{\mathcal{C}(L_B)}_{\lambda_B})\\
&=(\mathcal{H}^{\mathcal{A}(L_A)}_{\lambda_A} \otimes \mathcal{H}^{\mathcal{A}(L_B)}_{\lambda_B} )\otimes \bigoplus_\lambda N_{\lambda_A\lambda_B}^\lambda \mathcal{H}^{\mathcal{C}(L)}_\lambda.
\end{aligned}
\end{equation}

Taking the direct sum over $\lambda_A$, $\lambda_B$, and using the decomposition Eq.~\eqref{eq:app_Hilbert_decompose}, it then follows that
\begin{equation}
\mathcal{H}^{\mathcal{A}(L)}_{\lambda} = \bigoplus_{\lambda_A,\lambda_B} N_{\lambda_A\lambda_B}^\lambda   (\mathcal{H}^{\mathcal{A}(L_A)}_{\lambda_A}\otimes\mathcal{H}^{\mathcal{A}(L_B)}_{\lambda_B}).
\end{equation}
with multiplicities the same coefficients $N_{\lambda_A\lambda_B}^\lambda$.
For the dimensions this then gives
\begin{equation}
D_\lambda^{(L)} = \sum_{\lambda_A,\lambda_B} N_{\lambda_A\lambda_B}^\lambda D_{\lambda_A}^{(L_A)}D_{\lambda_B}^{(L_B)}.
\end{equation}

Then from the above we can derive identities for the total dimension of the chain, such as
\begin{equation}
\sum_{\lambda,\lambda_A,\lambda_B} N_{\lambda_A\lambda_B}^\lambda D_{\lambda_A}^{(L_A)}D_{\lambda_B}^{(L_B)}d_\lambda
= m^L.
\end{equation}

\subsubsection{Fusion to $\lambda_{\mathrm{tot}}=0$}
We again consider a Hopf algebra $\mathcal{C}$. We will show that for finite-dimensional irreps $V$, $W$ of $\mathcal{C}$ (simple modules), the subspace of $V\otimes W$ that is a trivial module is one-dimensional if $W\cong V^*$, zero-dimensional otherwise.
For any two $\mathcal{C}$-modules $V$, $W$, define $\mathrm{Hom}(V,W)$ to be the space of linear maps $V\to W$ and its subspace $\mathrm{Hom}_\mathcal{C}(V,W)$ of linear maps that commute with the $\mathcal{C}$ action ($\mathcal{C}$ linear or $\mathcal{C}$-homomorphisms). Moreover, define $\mathrm{Hom}_{\mathrm{triv}}(V,W)\subseteq \mathrm{Hom}(V,W)$ which is the subspace of $\mathcal{C}$-trivial maps, i.e. those linear maps $f$ transform trivially as 
$af=\varepsilon(a)f$ for all $a\in \mathcal{C}$. 
First, we claim that
$\mathrm{Hom}_{\mathrm{triv}}(V,W)\subseteq \mathrm{Hom}_\mathcal{C}(V,W)$. That is, if $f$ transforms trivially then
$f$ is $\mathcal{C}$-linear.
Note that, for any $V$, $V\cong \mathrm{Hom}(k,V)$ and the isomorphism $C$-linear
(here $k$ is the field of scalars, which we take to be the complex numbers,
or can be viewed as the trivial simple module $\bf 1$).
Then for $V$ finite-dimensional, the composition map 
\begin{equation}
    \circ: \mathrm{Hom}(V,W)\otimes V \to W
\end{equation}
is $\mathcal{C}$-linear. This is a special case of Ref.~\cite{kassel2012quantum}, III.5.3(c). Recall that $\mathcal{C}$ acts on the left-hand side via
the comultiplication. Then for $a\in \mathcal{C}$, $f$ a linear map $V\to W$ that transforms trivially, and $v\in V$,

\begin{align}
a(f(v))&=\circ \sum_{(a)} a^\prime f \otimes a^{\prime\prime} v \nonumber \\
           &= \sum_{(a)}\varepsilon(a^\prime)f(a^{\prime\prime}v) \nonumber\\
           &= f(av),
\end{align}
where the last equality follows from properties of counit and comultiplication. This proves the claim.

Now for finite-dimensional $V$ and $W$, the $\mathcal{C}$-trivial subspace $(W\otimes V^*)^{\mathcal{C}}$ of $W\otimes V^*$ is isomorphic
to $\mathrm{Hom}_{\mathrm{triv}}(V,W)$ by a $\mathcal{C}$-linear isomorphism (a consequence of Prop. III.5.2 in Ref.~\cite{kassel2012quantum}).
If $V$, $W$ are irreps, then the space of $\mathcal{C}$-homomorphisms
$\mathrm{Hom}_\mathcal{C}(V,W)$ has dimension $1$ if $V=W$, $0$ otherwise (these express Schur's Lemma).
The space of interest to us is $\mathrm{Hom}_{\mathrm{triv}}(V,W)$, which is a subspace of $\mathrm{Hom}_\mathcal{C}(V,W)$. Thus $\mathrm{Hom}_{\mathrm{triv}}(V,W)$ has a dimension less than or equal to $\mathrm{Hom}_\mathcal{C}(V,W)$. Finally, note that if $V=W$, then the identity map is $\mathcal{C}$-trivial, by Kassel III.5.3(c). This proves that, for finite-dimensional irreps $V$ and $W$, and relabelling,
\begin{equation}
    \dim (V\otimes W)^\mathcal{C} = 1 \, \, \, \mathrm{if}\, W\cong V^*, \, 0 \,\,\mathrm{otherwise}.
\end{equation}
Thus the coevaluation map, which is a $\mathcal{C}$-linear map from $\bf 1$ into $V\otimes V^*$, is unique up to multiplication by a scalar. 
More generally, when $\mathcal{C}$ is semisimple, this implies that
$\mathrm{Hom}_{\rm triv}(V,W) = \mathrm{Hom}_\mathcal{C}(V,W)$
for any finite-dimensional representations $V$, $W$.

It follows from this discussion that, for $\mathcal{C}$ semisimple and
finite-dimensional irreps, the fusion rules obey
\begin{equation}
N^0_{\lambda\mu}=1\,\, \text{ if }\,\, \mu = \bar{\lambda},
\end{equation}
and $0$ otherwise. This result, together with that of Subsubsec~\ref{app:mul_subalg}, proves Proposition~\ref{th:Prop_2}. Note that it is also true that 
$N_{0\lambda}^\nu=N_{\lambda0}^\nu = 1$ if $\nu=\lambda$, and $0$ otherwise, by results in Ref.~\cite{kassel2012quantum}.

If $\mathcal{C}$ is the group algebra of a finite group, or the UEA of a semisimple Lie algebra,
then we can also prove the preceding statements about $N_{\lambda\mu}^\nu$
using group characters~\cite{fulton_representation_2004}. In fact, that derivation goes through for any compact group,
and the preceding examples are special cases.

\section{Derivation of exact expression of entanglement}\label{app:derive_general_expression}
We provide a detailed derivation of the exact expressions of the logarithmic negativity, Rényi negativities, and operator space entanglement in this section.
A key property we employed, is that the basis states of the singlet subspace ($\lambda_{\mathrm{tot}} = 0$) of the commutant algebras can be written as 
\begin{equation}\label{eq:general_basis_state}
    |\lambda_{\mathrm{tot}}=0; \lambda, \bar{\lambda}; a, b\rangle
    =\sum_{m} \frac{\eta_{\lambda,m}}{\sqrt{d_{\lambda}}} |\lambda, m; a\rangle |\bar{\lambda}, \bar{m}; b\rangle,
\end{equation}
where $|\eta_{\lambda,m}| = 1$ arises with the choice of basis. 
This is proven in App.~\ref{app:Hopf}.
The $\lambda, m$ labels the irreps of $\mathcal{C}(L_A)$, while $\bar{\lambda}$, $\bar{m}$ are the corresponding dual representations and dual basis of $\mathcal{C}(L_B)$. 
We choose system sizes that allow the singlet subspace $\lambda_{\mathrm{tot}}=0$ for different commutant algebras, and restrict to certain bipartitions specified in each case. 
For Abelian commutants (with $d_\lambda \equiv 1$), Eq.~\eqref{eq:general_basis_state} recovers Eq.~\eqref{eq:U1_basis} for U($1$) and Eq.~\eqref{eq:PF_basis} for $\mathcal{C}_{\mathrm{PF}(N)}(L)$, respectively.

We focus on the stationary state $\Pi^{\lambda_{\mathrm{tot}}=0}/D_0$, which is an equal sum of basis states Eq.~\eqref{eq:general_basis_state}.
The partial transposed matrix has block-diagonal form $\rho^{T_B} = \oplus_{\lambda, a, b} \rho^{T_B}_{\lambda, a, b}$, with

\begin{equation}\label{eq:gen_partial_transposed_rho}
\begin{aligned}
    \rho^{T_B}_{\lambda, a, b} &=  
    \sum_{m, m^\prime} \frac{\eta_{\lambda,m}\eta^{*}_{\lambda,m^\prime}}{D_0^{(L)} d_{\lambda}} \\
    &\times |\lambda, m; a\rangle |\bar{\lambda}, \bar{m}^\prime; b\rangle \langle \lambda, m^\prime; a| \langle \bar{\lambda}, \bar{m}; b|,
\end{aligned}
\end{equation}
and $a=1,\ldots,D_{\lambda}^{(L_A)}$, $b=1,\ldots,D_{\bar{\lambda}}^{(L_B)}$. 

\paragraph{\textbf{Logarithmic negativity.}}

The logarithmic negativity is given by $E_{\mathcal{N}} = \log \|\rho^{T_B}\|_1 = \log(\sum_i |\lambda_i|)$, with $\lambda_i$ as the eigenvalues of the partial transpose matrix $\rho^{T_B}$. 
As given by Eq.~\eqref{eq:gen_partial_transposed_rho}, $\rho^{T_B}$ is block-diagonal into $\rho^{T_B}_{\lambda,a,b}$. Each block squares to an $d_\lambda^2$-dimensional identity matrix, $(\rho^{T_B}_{\lambda, a, b})^2 = \mathbb{1}_{d_\lambda^2}/(D_0^{(L)} d_{\lambda})^2$, where we used $|\eta_{\lambda,m}| = 1$.
Therefore, for fixed $\lambda, a, b$, there are $d_{\lambda}^2$ number of eigenvalues with absolute values $1/(D_0^{(L)} d_{\lambda})$. 
We obtain
\begin{equation}
    E_{\mathcal{N}} = \log \sum_{\lambda, a, b}  \frac{d_{\lambda}^2}{D_0^{(L)} d_{\lambda}} = \log\left( \frac{1}{D_0^{(L)}} \sum_{\lambda} d_{\lambda} D_{\lambda}^{(L_A)} D_{\bar{\lambda}}^{(L_B)}\right),
\end{equation}
with $a=1,\ldots,D_{\lambda}^{(L_A)}$, $b=1,\ldots,D_{\bar{\lambda}}^{(L_B)}$. 

\paragraph{\textbf{Rényi negativity.}}
For Rényi negativities $R_n$, we can use the diagrammatic expression Eq.~\eqref{eq:Rn_diagram} in Sec.~\ref{subsec:SU2}, which also holds for other commutants. 
Explicitly, with $n=3$ as an example and for fixed $\lambda, a, b$, it is
\begin{widetext}
\begin{equation}\label{eq:TrR3_diagram}
\begin{aligned}
    \mathrm{Tr}[(\rho^{T_B}_{\lambda, a, b})^3] =& \frac{1}{(D_0^{(L)} d_\lambda)^3}\sum_{m_1, m_1^\prime, m_2, m_2^\prime, m_3, m_3^\prime} \mathrm{Tr}[\ket{m_1, \bar{m}_1^\prime}\bra{m_1^\prime, \bar{m}_1} \ket{m_2, \bar{m}_2^\prime}\bra{m_2^\prime, \bar{m}_2} \ket{m_3, \bar{m}_3^\prime}\bra{m_3^\prime, \bar{m}_3}] \\
    =& \frac{1}{(D_0^{(L)} d_\lambda)^3} \sum_{m_1, m_1^\prime, m_2, m_2^\prime, m_3, m_3^\prime} \delta_{m_1^\prime m_2}\delta_{\bar{m}_1 \bar{m}_2^\prime}  \delta_{m_2^\prime m_3}\delta_{\bar{m}_2 \bar{m}_3^\prime}  \delta_{m_3^\prime m_1}\delta_{\bar{m}_3 \bar{m}_1^\prime} \\
    \equiv& \,\,\,\TrRthree = \frac{1}{(D_0^{(L)} d_\lambda)^3} \sum_{m_1} 1 =  \frac{1}{(D_0^{(L)})^3 (d_\lambda)^2}.
\end{aligned}
\end{equation}
\end{widetext}
Here we denote $\ket{\lambda, m; a} \ket{\bar{\lambda}, \bar{m}^\prime; b}$ as $\ket{m, \bar{m}^\prime}$, omitting $\lambda, a, b$ for simplicity. 
In the diagrams, each grey block $\TrRblock \equiv \frac{1}{D_0^{(L)} d_\lambda} \ket{m, \bar{m}^\prime}\bra{m^\prime, \bar{m}}$, with the two dots denoting $m$ and $m^\prime$ respectively.
And $\bra{m_1^\prime, \bar{m}_1} \ket{m_2, \bar{m}_2^\prime} = \delta_{m_1^\prime m_2} \delta_{\bar{m}_1 \bar{m}_2^\prime}$ is represented as $\TrRtwoblock$.
With the diagrammatic expressions, for odd $n$, it is easy to check that $\mathrm{Tr}[(\rho^{T_B}_{\lambda, a, b})^n]$ is represented by a single loop (e.g., Eq.~\eqref{eq:TrR3_diagram} for $n=3$). 
Therefore, $\mathrm{Tr}[(\rho^{T_B}_{\lambda, a, b})^n] = \sum_{m} 1/(D_0^{(L)} d_{\lambda})^n = 1/(D_0^{(L)})^n /d_{\lambda}^{n-1}$ for odd $n$. With $\mathrm{Tr}[\rho^n] = \mathrm{Tr} [(\Pi^{\lambda=0}/D_0^{(L)})^n] = 1/(D_0^{(L)})^{n-1}$, the Rényi negativity reads
\begin{equation}
\begin{aligned}
    R_n &= -\log (\frac{1}{D_0^{(L)}}\sum_{\lambda,a,b} \frac{1}{d_{\lambda}^{n-1}})\\
    &=  -\log  (\frac{1}{D_0^{(L)}}\sum_{\lambda} \frac{D^{(L_A)}_{\lambda} D_{\bar{\lambda}}^{(L_B)}}{d_{\lambda}^{n-1}}),
\end{aligned}
\end{equation}
for odd $n$.
For even $n$, $\mathrm{Tr}[(\rho^{T_B}_{\lambda, a, b})^n]$ is represented as two loops, and thus $\mathrm{Tr}[(\rho^{T_B}_{\lambda, a, b})^n] = \sum_{m,m^\prime} 1/(D_0^{(L)} d_\lambda)^n = 1/(D_0^{(L)})^{n} / d_\lambda^{n-2}$. Therefore, $R_{n} = R_{n-1}$ for $n$ even.

In the main text, we introduced generalized Rényi negativity to relate the logarithmic negativity and the Rényi negativity, which is defined as 
\begin{equation}
    \tilde{R}_n = \frac{1}{2-n}\log \left(\frac{\mathrm{Tr}[|\rho^{T_B}|^n]}{\mathrm{Tr}\rho^n}\right),
\end{equation}
for $n\neq 2$. With $\mathrm{Tr}[|\rho^{T_B}|^n] = \sum_i |\lambda_i|^n$, where $\lambda_i$ are the eigenvalues of $\rho^{T_B}$, it is easy to see that $\tilde{R}_n = E_{\mathcal{N}}$ for $n=1$ and $\tilde{R}_n = \frac{1}{n-2}R_n$ for even $n$.
Moreover, for general pure states $|\psi\rangle = \sum_a s_a |\psi_a\rangle |\phi_a\rangle$, which is written as the Schmidt decomposition with $s_a$ as the Schmidt values, the eigenvalues of the partial transposed $(|\psi\rangle\langle \psi|)^{T_B}$ are $s_a s_{a^\prime}$. The trace $\mathrm{Tr}[|(\ket{\psi}\bra{\psi})^{T_B}|^n] = \sum_{a, a^\prime} |s_a s_a^\prime|^n = (\sum_a |s_a|^n)^2$. Therefore, for pure states,
\begin{equation}
    \tilde{R}_n(|\psi\rangle) = \frac{2}{2-n} \log (\sum_{a} s_a^n) =  S_{n/2}(|\psi\rangle),
\end{equation}
where $S_{n}$ is the $n$-th Rényi entropy given by $S_{n} = \frac{1}{1-n} \log \sum_a |s_a|^{2n}$.

For the stationary state $\rho = \Pi^{\lambda=0}/D_0^{(L)}$, we can calculate $\tilde{R}_n$ using the eigenvalues of $\rho^{T_B}$ 
 (calculated for the logarithmic negativity), 
$\mathrm{Tr}|\rho^{T_B}|^n = \sum_{\lambda, a, b} d_{\lambda}^2/ (D_0^{(L)}d_{\lambda})^n = \sum_{\lambda, a, b} 1/(D_0^{(L)})^n / d_{\lambda}^{n-2} $.
With $\mathrm{Tr}(\rho^n) = (D_0^{(L)})^{n-1}$, we obtain
\begin{equation}\label{eq:gen_tilde_Rn_app}
    \tilde{R}_n 
    =  \frac{1}{2-n} \log  (\frac{1}{D_0}\sum_{\lambda} \frac{D^{(L_A)}_{\lambda} D_{\bar{\lambda}}^{(L_B)}}{d_{\lambda}^{n-2}}).
\end{equation}

The $\tilde{R}_n$ in Eq.~\eqref{eq:gen_tilde_Rn_app} can be bounded by the dimension of commutants.
With $\sum_\lambda D_\lambda^{(L_A)} D^{(L_B)}_{\bar{\lambda}}= D_0^{(L)}$, we have $D_\lambda^{(L_A)} D^{(L_B)}_{\bar{\lambda}}\leq D^{(L)}_0$ for all $\lambda$.
For $n<2$ and $d_\lambda\geq 1$, 
\begin{equation}
\begin{aligned}
    \tilde{R}_{n<2} &\leq \frac{1}{2-n}\log\left(\sum_\lambda d_\lambda^{2-n}\right)\leq \frac{1}{2-n}\log\left(\sum_\lambda d_\lambda^2\right)\\&=  \frac{1}{2-n}\log [\mathrm{dim}(\mathcal{C}_{\mathrm{min}})],
\end{aligned}
\end{equation}
where in the last line we used that for $n<2$, $d_\lambda^{2-n}<d_\lambda^{2}$; and the fact that the dimension of the commutant algebra is given by $\mathrm{dim}(\mathcal{C}_{\mathrm{min}})=\sum_\lambda d_\lambda^2$ with $\mathcal{C}_{\mathrm{min}}$ is the commutants defined on the shorter partition of the chain with size $L_{\min} = \min (L_A, L_B)$.
For $\tilde{R}_n$ with $n>2$, 
\begin{equation}
    \tilde{R}_{n>2} \leq  \sum_\lambda \frac{D_{\lambda}^{(L_A)} D_{\bar{\lambda}}^{(L_B)}}{D_0^{(L)}}\log\left(d_\lambda \right) \leq \log (\max(d_\lambda)).
\end{equation}
Moreover, since $\max(d_\lambda)\leq \sqrt{\sum_\lambda d_\lambda^2}=\sqrt{\mathrm{dim}(\mathcal{C}_{\mathrm{min}})}$, we find that $\tilde{R}_n \leq\log [\mathrm{dim}(\mathcal{C}_{\mathrm{min}})]/2$.
This indicates that for general $n$,
\begin{equation}
    \tilde{R}_n \leq c_n \log [\dim (\mathcal{C}_{\mathrm{min}})],
\end{equation}
for $c_n = \frac{1}{2-n}$ with $0<n<2$ and $c_n = \frac{1}{2}$ with $n>2$.  
Specifically, we obtain
\begin{equation}
\begin{aligned}
    & E_{\mathcal{N}} \leq \log [\dim (\mathcal{C}_{\mathrm{min}})], \\
    & R_n = R_{n-1} \leq \frac{n-2}{2} \log [\dim (\mathcal{C}_{\mathrm{min}})], \,\, n \,\, \mathrm{even}.
\end{aligned}
\end{equation}
Since for SU$(N)$ (or more generally, conventional symmetries), the dimension of the largest irrep scales at most $\max (d_\lambda) \sim O(L_{\min})$, as well as $\dim(\mathcal{C}_{\mathrm{min}}) \sim  O(L_{\min})$~\cite{moudgalya_fragment_commutant_2022}, one finds $\tilde{R}_n \sim O(\log L_{\min})$ for general $n$, which scales at most logarithmically with the subsystem size. 
In contrast, for TL$(N)$, $\dim \mathcal{C} \sim e^{L_\mathrm{min}}$, which allow linear scaling of $\tilde{R}_n$ in system size.
As discussed in the main text, we found indeed that for TL$(N)$, $\tilde{R}_n$ scales linearly with system size for $n<2$, which we will derive in App.~\ref{app:fragment}.

\paragraph{\textbf{Operator space entanglement.}}
The OSE is the von Neumann entropy of the vectorized stationary state,
\begin{equation}
\begin{aligned}
    |\rho\rangle\rangle &= \sum_{\lambda, a, b} \sum_{m,m^\prime}  \frac{\eta_{\lambda,m}\eta^{*}_{\lambda,m^\prime}}{D_0 d_\lambda} \\
    &\times |\lambda,m;a\rangle_A |\bar{\lambda},\bar{m};b\rangle_B |\lambda,m^\prime;a\rangle_A |\bar{\lambda},\bar{m^\prime}; b\rangle_B.
\end{aligned}
\end{equation}
The vectorized stationary state can be written as $|\rho\rangle\rangle = \sum_{\lambda} \sum_{A,B} \Psi_{A,B}^{\lambda}|\psi_A\rangle |\psi_B\rangle$. Here $|\psi_{A}\rangle \in \{|\lambda, m; a\rangle |\lambda, m^\prime; a\rangle\}$ for $m,m^\prime = 1,\ldots, d_{\lambda}$, $a = 1,\ldots, D_{\lambda}^{(L_A)}$ which is an orthonormal basis on the left partition. Similarly for the right partition.
The matrix $\Psi^\lambda_{A,B}$ is given by
\begin{equation}
\begin{aligned}
    \Psi_{AB}^{\lambda} &= \frac{1}{\sqrt{D_0^{(L)}}d_{\lambda}}\begin{pmatrix}1 & \ldots & 1\\
    1  & \ddots & \vdots\\
    1  & \ldots & 1
\end{pmatrix}_{D_{\lambda}^{(L_A)}\times D^{(L_B)}_{\bar{\lambda}}} \otimes \begin{pmatrix} &   & 1\\
     & \iddots\\
    1
\end{pmatrix}_{d_{\lambda}^2}\\
&\equiv \frac{1}{\sqrt{D_0^{(L)}}d_{\lambda}} M_{D_{\lambda}^{(L_A)} \times D^{(L_B)}_{\bar{\lambda}}}\otimes N_{d_{\lambda}^2},
\end{aligned}
\end{equation}
where $M_{m \times n}$ is a $m\times n$ matrix with all matrix elements as $1$, and the $N_{n}$ matrix is a $n\times n$ matrix with all off-diagonal values as $1$.
The Schmidt values of $\Psi^{\lambda}_{A,B}$ are given by the square root of the eigenvalues of $(\Psi^{\lambda}_{A,B})^\dagger \Psi^{\lambda}_{A,B}$. 
For the $M$ matrix, $M_{m\times n} ^\dagger M_{m\times n} = m M_{n\times n}$. As $M_{n\times n}^2 = n M_{n\times n}$ with $\mathrm{Tr}(M_{n\times n}) = n$, $M_{n\times n}$ has one eigenvalue $n$. Therefore $M_{D_{\lambda}^{(L_A)} \times D_{\bar{\lambda}}^{(L_B)}}$ has one Schmidt value that squares to $D_{\lambda}^{(L_A)} D_{\bar{\lambda}}^{(L_B)}$.
In addition, the $N_n$ matrix gives $n$ degeneracy of the eigenvalues of $M$.
Therefore, for $\Psi^{\lambda}_{A,B}$, there are $d_\lambda^2$ number of Schmidt values which square to $D_{\lambda}^{(L_A)}D_{\bar{\lambda}}^{(L_B)}/(D_0^{(L)} d_\lambda^2)$.
The OSE is thus given by
\begin{equation}
\begin{aligned}
    S_{\mathrm{OP}} =& -\sum_l s_l^2 \log s_l^2 \\
    =& -\sum_\lambda \frac{D_{\lambda}^{(L_A)}D_{\bar{\lambda}}^{(L_B)}}{D_0^{(L)}} \log \frac{D_{\lambda}^{(L_A)}D_{\bar{\lambda}}^{(L_B)}}{D_0^{(L)} d_\lambda^2 }.
\end{aligned}
\end{equation}

Using the concavity of the logarithm, one finds a general upper bound of the OSE as
\begin{equation}
    S_{\mathrm{OP}}\leq \log [\mathrm{dim}(\mathcal{C}_\mathrm{min})].
\end{equation}

\section{Asymptotic scaling of OSE with U($1$) symmetry}\label{app:U1}
In this section, we derive the asymptotic scaling of OSE with U($1$) symmetry at half-chain bipartition, $L_A=  L_B = L/2$ with $L=4n$, $n\in\mathbb{N}$. 
With $D_M = (\begin{smallmatrix} L\\L/2-M
\end{smallmatrix})$, the square of Schmidt values scale as
\begin{equation}
    \frac{(D_M^{(L/2)})^2}{D_0^{(L)}} \sim \frac{4}{\sqrt{2\pi L}} e^{-\frac{8M^2}{L}}.
\end{equation}
Here we used the asymptotic scaling of binomial coefficients 
\begin{equation}\label{eq:binomial_asy}
    \begin{pmatrix}2n\\
    n+k\end{pmatrix} \sim \frac{2^{2n}}{\sqrt{\pi n}}e^{-k^2/n}, n\rightarrow\infty.
\end{equation}
Using $\sum_M (D_{M}^{(L/2)})^2 = D_0^{(L)}$, OSE simplifies to
\begin{equation}
\begin{aligned}
    S_{\text{OP}} &= -\sum_{M=-L/4}^{L/4} \frac{(D_M^{(L/2)})^2}{D_0^{(L)}} \log \frac{(D_M^{(L/2)})^2}{D_0^{L}} \\
    &\sim \log \frac{\sqrt{2\pi L}}{4} + \sum_M \frac{(D_M^{(L/2)})^2}{D_0^{(L)}} \frac{8 M^2}{L}.\\
\end{aligned}
\end{equation}
Note that the second term is the variance of $M$ because $p_M = (D_{M}^{(L/2)})^2 / D_0^{(L)}$ is the probability of the left-partition in the $M$ sector~\cite{2018_Caha_pairflip}. 
This term can be shown to scales as $O(1)$ as follows:
\begin{equation}
\begin{aligned}
    &\frac{1}{L}\sum_M \frac{(D_M^{(L/2)})^2}{D_0^{(L)}} \frac{8 M^2}{L} \\
    & \,\,\, = \frac{8}{L\left(\begin{smallmatrix}
        L \\ L/2\end{smallmatrix}\right)} \sum_{l=0}^{L/2} \begin{pmatrix}L/2\\
    l\end{pmatrix} (l-\frac{L}{4})^2 \sim \frac{1}{2},
\end{aligned}
\end{equation}
where we used a change of variable $M = l-L/4$.
Therefore, with U($1$) symmetry,
\begin{equation}
    S_{\mathrm{OP}} \sim \frac{1}{2}\log L + \frac{1}{2} + \log \frac{\sqrt{2\pi}}{4}.
\end{equation}

\section{SU($2$)}\label{app:SU2}
\subsection{Asymptotic scaling for the SU($2$) $\lambda=0$ singlet sector}
In this subsection, we derive the SU($2$) asymptotic scaling for different entanglement properties.
For the logarithmic negativity at half-chain entanglement
\begin{equation}
\begin{aligned}
    E_{\mathcal{N}} &= \log \left(\frac{1}{D_0^{(L)}}\sum_{\lambda=0}^{L/2} d_{\lambda} (D_{\lambda}^{(L/2)})^2\right)\\
    &= \log \left(\frac{1}{D_0^{(L)}} \sum_{\lambda=0}^{L/2} \frac{(2\lambda+1)^3}{L+1}\begin{pmatrix}L/2+1\\
    L/4+\lambda+1\end{pmatrix}^2\right)\\
    &= \log\left((\frac{L}{2}+1) \frac{\left(\begin{smallmatrix}L/2\\
    L/4\end{smallmatrix}\right)^2}{\left(\begin{smallmatrix}L\\
    L/2\end{smallmatrix} \right)} \right)\\
    &\sim \frac{1}{2}\log L +\log \sqrt{\frac{2}{\pi}},
\end{aligned}
\end{equation}
where we use Mathematica for the summation in the second line and the asymptotic scaling of binomials Eq.~\eqref{eq:binomial_asy} for the last line.
For the third Rényi negativity, a similar derivation gives 
\begin{equation}
    R_3 \sim \log L - 2\log 2.
\end{equation}

For OSE, using Eq.~\eqref{eq:binomial_asy},
we obtain
\begin{equation}
\begin{aligned}
    S_{\mathrm{OP}} &\sim 16\sqrt{\frac{2}{\pi}} L^{-3/2} \frac{e^{-4\lambda^2/L}}{[1+(\lambda+1)/(L/4)]^2}\\
    &\sim \frac{3}{2}\log L - \log (16\sqrt{\frac{2}{\pi}}) \\
    &\,\,\,+ \sum_{\lambda} \frac{(D_{\lambda}^{(L/2)})^2}{D_0} (2\log(1+\frac{\lambda+1}{L/4}) + 4\frac{\lambda^2}{L})+ O(1).
\end{aligned}
\end{equation}
Now we prove that the last term is vanishing with $L$ or at most scales as $O(1)$. In the remainder of this section, we denote $D_0^{(L)}$ as $D_0$ and $D_{\lambda}^{(L/2)}$ as $D_{\lambda}$ respectively. With the expansion of logarithmic term, the last term is given by
\begin{equation}\label{eq:SU2_SOP_O1_term}
    \sum_\lambda \frac{D_\lambda^2}{D_0} (4\frac{\lambda^2}{L} + \sum_{n=0}^{\infty} \frac{1}{n+1} (\frac{\lambda+1}{L/4})^n).
\end{equation}
We use Mathematica to calculate the following summations,
\begin{equation}\label{eq:app_D0_summations}
\begin{aligned}
    & D_0 = \sum_\lambda D_\lambda^2 \sim O(2^L L^{3/2}),\\
    &\sum_{\lambda} D_{\lambda}^2 (2\lambda + 1) \sim \frac{2^L}{\pi L/4} \sim O(2^L L^{-1}),\\
    &\sum_{\lambda} D_{\lambda}^2 (2\lambda + 1)^2 \sim \frac{3}{\sqrt{2\pi}} \frac{2^L}{\sqrt{L}} \sim O(2^L L^{-1/2}).
\end{aligned}
\end{equation}
Therefore, the $\sum_{\lambda} \frac{D_{\lambda}^2}{D_0} \frac{\lambda^2}{L}$ term scales at most $O(\frac{1}{L}\frac{2^L L^{-1}}{2^L L^{-3/2}}) = O(1)$. The $n$-th order terms in Eq.~\eqref{eq:SU2_SOP_O1_term} is  
\begin{equation}
\begin{aligned}
    &\sum_{\lambda=0}^{L/2} \frac{D_\lambda^2}{D_0} \frac{1}{n+1} (\frac{\lambda+1}{L/4})^n 
    \\
    \leq& \sum_{\lambda=0}^{L/2} \frac{D_\lambda^2}{D_0}\frac{1}{n+1} (2\lambda+1) \frac{(\frac{L}{2}+1)^{n-1}}{(L/4)^{n}} \\
    \sim& O(L^{-1/2}),
\end{aligned}
\end{equation}
which shows that arbitrary $n$-th order terms are vanishing in the thermodynamic limit. 
Therefore, Eq.~\eqref{eq:SU2_SOP_O1_term} scales at most $O(1)$, we prove that $S_{\mathrm{OP}} \sim \frac{3}{2}\log L + O(1)$.

\subsection{Logarithmic negativity for $\lambda_{\mathrm{tot}}>0$}
In Sec.~\ref{subsec:SU2_Haar}, we consider the stationary states corresponding to (i) one $\lambda_{\mathrm{tot}}\neq 0$ subspace $\mathcal{H}^{\mathcal{A}(L)}_{\lambda_{\mathrm{tot}}, m_{\mathrm{tot}}=0}$ or (ii) Haar random initial states sampled uniformly from $\bigoplus_{\lambda_{\mathrm{tot}}=0}^{\lambda_{\mathrm{max}}} \mathcal{H}^{\mathcal{A}(L)}_{\lambda_{\mathrm{tot}}, m_{\mathrm{tot}}=0}$ subspaces. Here we provide the analytic and numerical details.

We consider the stationary states discussed in the main text, which are given by 
\begin{equation}
    \rho = \sum_{\lambda_{\mathrm{tot}}}p_{\lambda_{\mathrm{tot}}} \frac{\Pi^{\lambda_{\mathrm{{tot}}}}_{ m_{\mathrm{tot}}=0}}{D_{\lambda_{\mathrm{tot}}}^{(L)}}
\end{equation}
with $p_{\lambda_{\mathrm{tot}}} = \mathrm{Tr}(\Pi_{\lambda_{\mathrm{tot}}, m_{\mathrm{tot}}=0} \rho_0)$. 
For one total spin sector, the stationary state reduces to $\rho = \frac{\Pi^{\lambda_{\mathrm{tot}}}}{D^{(L)}_{\lambda_{\mathrm{tot}}}}$.
To calculate the bipartite logarithmic negativity, we write the basis states of $\mathcal{H}^{\mathcal{A}(L)}_{\lambda_{\mathrm{tot}}, m_{\mathrm{tot}}=0}$ in the bipartite form,
\begin{equation}
\begin{aligned}
    &|\lambda_{\mathrm{tot}}, m_{\mathrm{tot}=0};\lambda_A, \lambda_B; a, b\rangle\\ 
    &= \sum_{m=-\min{(\lambda_A, \lambda_B)}}^{\min{(\lambda_A,\lambda_B)}} c_{m}(\lambda_{\mathrm{tot}}; \lambda_A, \lambda_B) |\lambda_A, m; a\rangle |\lambda_B, -m; b\rangle,
\end{aligned}
\end{equation}
for $|\lambda_A - \lambda_B|\leq \lambda_{\mathrm{tot}}\leq \lambda_A + \lambda_B$.
The CG coefficients $c_{m}(\lambda_{\mathrm{tot}}; \lambda_A, \lambda_B)$ can be evaluated by analytic expressions.
We observe that the partial transposed matrix $\rho^{T_B}$ decomposes into direct sum of subspaces labeled by $\lambda_A, \lambda_B$ and $a, b$, which are
\begin{equation}
\begin{aligned}
    &\rho^{T_B}_{\lambda_A,\lambda_B, a, b} = \sum_{m,m^\prime} 
    \sum_{\lambda_{\mathrm{tot}}}
    c_m(\lambda_{\mathrm{tot}};\lambda_A, \lambda_B) c^*_{m^\prime}(\lambda_{\mathrm{tot}};\lambda_A, \lambda_B) \\
    &\times \frac{p_{\lambda_{\mathrm{tot}}}}{D_{\lambda_{\mathrm{tot}}}} \ket{\lambda_A, m; a} \ket{\lambda_B, -m^\prime;b}\bra{\lambda_A, m^\prime;a} \bra{\lambda_B, -m ;b}.
\end{aligned}
\end{equation}
The $\rho^{T_B}$ squares to a matrix with only diagonal elements. For fixed $\lambda_{A}, \lambda_B, a, b, m, m^\prime$, we obtain one eigenvalue of $\rho^{T_B}$ with absolute value
$\sum_{\lambda_{\mathrm{tot}}} \frac{p_{{\lambda_{\mathrm{tot}}}}}{D_{\lambda_{\mathrm{tot}}}}
    |c_m(\lambda_{\mathrm{tot}}) c^*_{m^\prime}(\lambda_{\mathrm{tot}})|$, omitting $\lambda_A, \lambda_B$.
The logarithmic negativity is thus given by 
\begin{equation}
\begin{aligned}
    E_{\mathcal{N}} &= \log \sum_{\lambda_{\mathrm{tot}}} \frac{p_{{\lambda_{\mathrm{tot}}}}}{D_{\lambda_{\mathrm{tot}}}} \sum_{\lambda_A,\lambda_B, a, b, m, m^\prime} 
    |c_m(\lambda_{\mathrm{tot}}) c^*_{m^\prime}(\lambda_{\mathrm{tot}})|\\
    &=  \log \sum_{\lambda_{\mathrm{tot}}}  \frac{p_{{\lambda_{\mathrm{tot}}}}}{D^{(L)}_{\lambda_{\mathrm{tot}}}} \sum_{\lambda_A,\lambda_B, m, m^\prime} D_{\lambda_A}^{(L_A)}D_{\lambda_B}^{(L_B)}
    |c_m(\lambda_{\mathrm{tot}}) c^*_{m^\prime}(\lambda_{\mathrm{tot}})|.
\end{aligned}
\end{equation}
We numerically evaluate this $E_{\mathcal{N}}$ for (i) the initial state restricted to one sector $\lambda_{\mathrm{tot}}$, i.e., $p_{\lambda_{\mathrm{tot}}} = 1$ and (ii) the initial states as Haar random states
\begin{equation}
    |\psi (t=0) \rangle = \sum_j (a_j + i b_j)|\psi_j \rangle,
\end{equation}
with $a_j, b_j$ as real numbers sampled from Gaussian distribution (with zero mean and unit variance) and $\{\ket{\psi_j}\}$ as the orthonormal basis of $ \bigoplus_{\lambda_{\mathrm{tot}}=0}^{\lambda_{\mathrm{max}}} \mathcal{H}^{\mathcal{A}(L)}_{\lambda_{\mathrm{tot}}, m_{\mathrm{tot}}=0}$.

We want to comment on special cases where we can analytically evaluate.

For (i) and $\lambda_{\mathrm{tot}}=L/2$, the $\mathcal{H}^{\mathcal{A}(L)}_{\lambda_{\mathrm{tot}}=L/2, m_{\mathrm{tot}}=0}$ is one-dimensional, with the basis state given by 
\begin{equation}
    |\psi\rangle \propto (S^{-}_{\mathrm{tot}})^{L/2} \ket{\uparrow}^{\otimes L} = \frac{1}{\sqrt{D^{\mathrm{U}(1)}_{m_\mathrm{tot}=0}}}\sum \ket{\phi_{m=0}}. 
\end{equation}
It is an equal superposition of $m_{\mathrm{tot}}=0$ states for U($1$) symmetry, which can be given by Eq.~\eqref{eq:U1_basis}.
The logarithmic negativity $E_{\mathcal{N}}(\rho_{\lambda_{\mathrm{tot}}=L/2}) = S_{1/2} (\ket{\psi})$ for pure state $\ket{\psi}$. 
From App.~\ref{app:U1}, we obtain the Schmidt values for $m_{\mathrm{tot}} = 0$ states, which square to $(D_{m}^{\mathrm{U}(1)})^{2}/D_0^{\mathrm{U}(1)}$.
Therefore, 
\begin{equation}
\begin{aligned}
    &E_{\mathcal{N}}(\rho_{\lambda_{\mathrm{tot}}=L/2, m_{\mathrm{tot}}=0}) =  S_{1/2} (\ket{\psi}) \\
    =& \log \sum_m \frac{(D_{m}^{\mathrm{U}(1)})^{2}}{D_0^{\mathrm{U}(1)}}
    \sim \frac{1}{2}\log L +\frac{1}{2} \log \frac{\pi}{2}.
\end{aligned}
\end{equation}
Therefore, $E_{\mathcal{N}}$ scales logarithmically for both $\lambda_{\mathrm{tot}}=0$ and $\lambda_{\mathrm{tot}}=L/2$.

For (ii) with $\lambda_{\mathrm{max}} = L/2$, i.e., the Haar random initial states are sampled from the full $m_{\mathrm{tot}}=0$ subspace.
Starting from Haar random initial states, the weight of the Haar random states in the subspace $\lambda_{\mathrm{tot}}$ and $m_{\mathrm{tot}}=0$ is $p_{\lambda_{\mathrm{tot}}} \sim D_{\lambda_{\mathrm{tot}}}^{(L)}/D_0^{(L)} + \delta$, where $D_{\lambda_{\mathrm{tot}}}^{(L)}/D_{m_{\mathrm{tot}}=0}^{(L)}$ is the probability of being in the subspace $\lambda_{\mathrm{tot}}$, $D_{m_{\mathrm{tot}}=0}^{(L)} = (\begin{smallmatrix}
    L \\ L/2
\end{smallmatrix})$ and $\delta$ as the deviation which is exponentially small with respect to Hilbert space dimension.
For small system sizes when $\delta$ is large, the $E_{\mathcal{N}}$ is non-zero for $\lambda_{\mathrm{max}}$ as observed in Fig.~\ref{fig:SU2_neg_random_Haar}b. 
However, for large system sizes when $\delta$ is negligible, we assume $p_{\lambda_{\mathrm{tot}}} = \frac{D_{\lambda_{\mathrm{tot}}}^{(L)}}{D_{m_{\mathrm{tot}}=0}^{(L)}}$, as $\sum_{\lambda_{\mathrm{tot}}} D_{\lambda_{\mathrm{tot}}}^{(L)}=D_{m_\mathrm{tot}=0}^{(L)}$. The stationary state is given by
\begin{widetext}
\begin{equation}
\begin{aligned}
    \rho
    &= \frac{1}{D_{m_{\mathrm{tot}}=0}^{(L)}} \sum_{\lambda_A} \sum_{\lambda_B}  \sum_{a, b} \sum_{m,m^\prime} \sum_{\lambda_{\mathrm{tot}} = |\lambda_A -\lambda_B|}^{\lambda_A+\lambda_B}  c_m(\lambda_{\mathrm{tot}};\lambda_A, \lambda_B) c_{m^\prime}^*(\lambda_{\mathrm{tot}};\lambda_A, \lambda_B)
    \ket{\lambda_A, m;a} \ket{\lambda_B, -m;b} \bra{\lambda_A, m^\prime;a} \bra{\lambda_B, -m^\prime;b}\\
    &= \frac{1}{D_{m_{\mathrm{tot}}=0}^{(L)}} \sum_{\lambda_A, \lambda_B} \sum_{a, b} \sum_m  \ket{\lambda_A, m;a} \ket{\lambda_B, -m;b} \bra{\lambda_A, m;a} \bra{\lambda_B, -m;b} = \frac{\Pi^{m_{\mathrm{tot}}=0}}{D_{m_{\mathrm{tot}}=0}^{(L)}},
\end{aligned}
\end{equation}
\end{widetext}
where we used the orthonormality relation of CG coefficients $\sum_{\lambda_{\mathrm{tot}} = |\lambda_A -\lambda_B|}^{\lambda_A+\lambda_B}  c_m(\lambda_{\mathrm{tot}};\lambda_A, \lambda_B) c_{m^\prime}^*(\lambda_{\mathrm{tot}};\lambda_A, \lambda_B) = \delta_{m,m^\prime}$. 
Therefore, the stationary state recovers to the case with U($1$) symmetry, which is a separable state.
This explains that in Fig.~\ref{fig:SU2_neg_random_Haar}b, when $\lambda_{\mathrm{max}} = L/2$, the average logarithmic negativity corresponds to Haar random initial states is vanishingly small with the increase of $L$.

\section{SU$(N)$}\label{app:SUN}
\subsection{CG coefficients of SU$(N)$ singlet subspace}
In App.~\ref{app:Hopf}, we prove that for commutants with Hopf algebra structure, the basis states can be written as a bipartite form in Eq.~\eqref{eq:gen_triv0}, which is an equal superposition of states labeled by $\lambda, m$ and its dual $\bar{\lambda}, \bar{m}$ (up to a minus sign).
Here, we introduce another approach to prove rigorously for general SU$(N)$ symmetry in finite-size systems, the basis states can be written as in Eq.~\eqref{eq:gen_triv0}. This approach generalizes the common proof of SU($2$) CG coefficients.

First, we review the proof of SU($2$) symmetry, where the basis states can be labeled by the total spin $J$ and the total magnetization $m$,
\begin{equation}
\begin{aligned}
    &S^2 |J, m\rangle = J(J+1)|J, m\rangle, \\
    &S^z |J, m\rangle = m|J,m\rangle.
\end{aligned}
\end{equation}
Note that we omit the $a, b$ labels corresponding to the bond algebras in the following, as the CG coefficients obtained from the representation theory of SU$(N)$ are independent of them. 
The ladder operators $S^\pm$ acting on the basis states give
\begin{equation}\label{eq:SU2_ladder}
    S^{\pm} |J,m\rangle = \sqrt{J(J+1) - m(m\pm 1)}|J, m\pm1\rangle.
\end{equation}

Consider the singlet sector with $J_{\mathrm{tot}}=0$ (and thus $m_{\mathrm{tot}}=0$). The basis states
which satisfy $S^\pm \ket{J_{\mathrm{tot}}=0} = S^z \ket{J_{\mathrm{tot}}=0} = 0$.
The basis states with $J_{\mathrm{tot}}$ can be represented as the composition of $J$ and $J^\prime$ states, and a general expression is
\begin{equation}
\begin{aligned}
    |J_{\mathrm{tot}}=0&, J, J^\prime\rangle  \\ &= \sum_{m=-J/2}^{J/2} \sum_{m^\prime=-J^\prime/2}^{J^\prime/2} c_{m,m^\prime}(J,J^\prime) \ket{J, m} \ket{J^\prime, m^\prime}.
\end{aligned}
\end{equation}
Using the ladder operators, $S^{\pm} = S^{\pm}_A \otimes I + I \otimes S^{\pm}_B$, 
\begin{equation}
\begin{aligned}
    0 =& S^{\pm}\ket{J_{\mathrm{tot}}, J,J^\prime} \\
    =& \sum_{m=-J/2}^{J/2} \sum_{m^\prime=-J^\prime/2}^{J^\prime/2} c_{m,m^\prime}(J,J^\prime) \\ 
    &\,\,\,\, \times \left(S^{\pm}_A|J,m\rangle \ket{J^\prime,m^\prime} + \ket{J,m} S^{\pm}_B|J^\prime, m^\prime\rangle \right).
\end{aligned}
\end{equation}
With Eq.~\eqref{eq:SU2_ladder} and note the range of sum of $m$ and $m^\prime$, 
we can obtain that
(1) $J=J^\prime$, otherwise $c_{m,m^\prime}(J,J^\prime) \equiv 0$;
(2) $m + m^\prime =0$; and (3) $|c_{m,m^\prime}| = c$, for $c\in \mathbb{R}$.
With normalization $\bra{J_{\mathrm{tot}}=0}\ket{J_{\mathrm{tot}}=0}=1$,
\begin{equation}
    c_{m,m^\prime}(J,J^\prime) = \frac{1}{\sqrt{d_{\lambda}}} \delta_{m,-m^\prime} \delta_{J,J^\prime} \eta_{J,m},
\end{equation}
where $\eta_{J,m} = \pm 1$. This is compatible with the known result $c_{m}(J) = (-1)^{J-m}/\sqrt{2J+1}$ as $d_{J} = 2J+1$ for SU($2$).

For general SU$(N)$, the irrep labels are given by a set of ordered non-negative integers $\lambda = (\lambda_1, \ldots, \lambda_N)$, with $\lambda_1 \geq \lambda_2 \geq \lambda_N \geq 0$. Note that the irreps are the equivalent for $\lambda$ up to a constant $\lambda + c = (\lambda_1 + c, \ldots, \lambda_N+c)$, for $c\in\mathbb{Z}$. The singlet subspace is given by $\lambda = (0, \ldots, 0)$. Therefore, the irreps can also be labeled by $N-1$ numbers, $(p_1, \ldots, p_{N-1}) \equiv (\lambda_1 -\lambda_2, \ldots, \lambda_{N-1}-\lambda_N)$~\cite{sternberg_1995}. For SU($2$), $J=(\lambda_1-\lambda_2)/2$.
A set of operators that we need to construct the basis are
$S^{\pm}_{(l)}$, $S^z_{(l)}$ for $1\leq l \leq N-1$, with commutation relation~\cite{1984_cornwell_group_theory, 2011_alex_numerical_SUN}
\begin{equation}\label{eq:SUN_commutation}
    [S^{+}_{(l)}, S^{-}_{(l)}] = 2 S^z_{(l)}, [ S^z_{(l)}, S^{\pm}_{(l)}] = \pm  S^{\pm}_{(l)}.
\end{equation}
The basis states of the singlet subspace satisfy
\begin{equation}\label{eq:SUN_loweringraising_op}
    S^{\alpha}_{(l)} |\lambda_{\mathrm{tot}}=0\rangle = 0,  \forall 1\leq l \leq N-1, \alpha \in \{\pm,z\}.
\end{equation}
For fixed $\lambda$, the basis states can be uniquely labeled by the Gelfand-Tsetlin (GT) pattern $M=(m_{k,l})$ with $1\leq k\leq l$ and $1\leq l \leq N-1$~\cite{1950_Gelfand_Tsetlin, 1968_Biedenharn_GT, 2011_alex_numerical_SUN}. 
The GT pattern is a down triangle set of numbers, 
\begin{equation}
    M=\begin{pmatrix}m_{1,N} & m_{2,N}  & \ldots & & m_{N,N}\\
 & m_{1,N-1}   & \ldots & m_{N-1,N-1}\\
& \ddots &  \ddots &  \iddots\\
&  & m_{12} \,\,\,\,\,m_{2,2}\\
&  &  m_{1,1}
\end{pmatrix}.
\end{equation}
The irrep labels give the first line, $m_{k, N} = \lambda_k$. And the numbers satisfy the so-called betweenness condition $m_{k,l+1}\geq m_{k,l} \geq m_{k+1, l+1}$ for $1\leq k\leq l$. 
The GT patterns exploit the decomposition of SU$(N)$ irreps to SU$(N-1)$ irreps, and the $l$-th row $(m_{1, l}, m_{2, l}, \ldots, m_{l, l})$ can be viewed as the irrep labels of SU($l$).
We denote the corresponding basis states with GT pattern $M$ as $\ket{M}$. These basis states are the common eigenstates of all $S^z_{(l)}$, and the exact expressions with $S^{\pm}_{(l)}$ similar to Eq.~\eqref{eq:SU2_ladder} are known~\cite{1950_Gelfand_Tsetlin, 1986_raczka_GT_theory, 2011_alex_numerical_SUN}. 
This allows us to employ the same techniques as SU$(2)$.
The $z$-weight of GT pattern $M$ is an analog of magnetization of SU($2$), 
\begin{equation}
    W_z(M) = (w_1^M, w_2^M, \ldots, w_{N-1}^M),
\end{equation}
given by $S_{(l)}^z|M\rangle = w_l^M|M\rangle$.
The $z$-weight is given by the row sums of $M$, with $w_l^M \equiv \sum_{k=1}^l m_{k,l} -\frac{1}{2} (\sum_{k=1}^{l-1} m_{k,l-1} + \sum_{k=1}^{l+1} m_{k,l+1})$. 
Note that the $z$-weight does not uniquely label the basis states for $N>2$, which is different from SU($2$). There is an inner multiplicity of states with the same $z$-weight. For example, these two patterns
\begin{equation}
    \begin{pmatrix}2 &  & 1 &  & 0\\
 & 2 &  & 0\\
 &  & 1
\end{pmatrix},\begin{pmatrix}2 &  & 1 &  & 0\\
 & 1 &  & 1\\
 &  & 1
\end{pmatrix}
\end{equation}
of SU($3$) irrep with $(\lambda_1, \lambda_2, \lambda_3) = (2, 1, 0)$ have the same $z$-weight~\cite{2011_alex_numerical_SUN}.
The raising and lowering operator $S^{(l)}_{\pm}$ acting on the $|M\rangle $ gives a linear combination of states $|M\pm M^{k,l}\rangle$ for $1\leq k\leq l$. The pattern $M - M^{k,l}$ is given by $(m_{1, N}, \ldots, m_{k,l}-1, \ldots, m_{1,1})$, i.e., the element $m_{k,l}$ decrease by one.
To derive the CG coefficients, we will use the coefficients for $S_{(l)}^{-}$~\cite{2011_alex_numerical_SUN}
\begin{widetext}
\begin{equation}\label{eq:SUn_raising_lowering_coeficients}
\begin{aligned}
    a_{k,l}&\equiv\langle M-M^{k,l}|S_{-}^{(l)}|M\rangle,\\
    &=\left(-\frac{\prod_{k^{\prime}=1}^{l+1}(m_{k^{\prime},l+1}-m_{k,l}+k-k^{\prime})\prod_{k^{\prime}=1}^{l-1}(m_{k^{\prime},l-1}-m_{k,l}+k-k^{\prime}-1)}{\prod_{k^{\prime}=1,k^{\prime}\neq k}^{l}(m_{k^{\prime},l}-m_{k,l}+k-k^{\prime})(m_{k^{\prime},l}-m_{k,l}+k-k^{\prime}-1)}\right)^{\frac{1}{2}}.
\end{aligned}
\end{equation}
\end{widetext}

Now we focus on the singlet subspace $\lambda_{\mathrm{tot}} = (0, \ldots, 0)$. The corresponding GT pattern $M_{\lambda_{\mathrm{tot}}=0}$ is given by $m_{k,l} \equiv \mathrm{const.}$ due to the betweenness condition. 
A general expression for the basis states of the singlet subspace can be written as
\begin{equation}\label{eq:SUn_basis_GT}
    \ket{M_{\lambda_{\mathrm{tot}}=0}} = \sum_{M,M^\prime} c_{M,M^\prime} \ket{M}\ket{M^\prime}.
\end{equation}

In the following, we derive the $c_{M,M^\prime}$ and prove that Eq.~\eqref{eq:SUn_basis_GT} is compatible with the general expression for the singlet subspace Eq.~\eqref{eq:gen_triv0}.
More specifically, our goal is to prove that
\begin{itemize}
    \item[(1)] The coefficients $c_{M, M^\prime} = 0$ unless $M^\prime = \bar{M}$ (dual of $M$).
    \item[(2)] All non-zero $c_{M, M^\prime}$ are equal up to a minus sign.
\end{itemize}
Using $S_{(l)}^z \ket{M_{\lambda_{\mathrm{tot}}=0}} = 0$, i.e., the $z$-weight of $\ket{M_{\lambda_{\mathrm{tot}}=0}}$ as $W_z = (0, \ldots, 0)$, we have 
\begin{equation}\label{eq:app_general_z_weight}
    W_z(M) + W_z(M^\prime) = (0, \ldots, 0).
\end{equation}
And the ladder operators 
\begin{equation}
    S^{\pm}_{(l)}  \ket{M_{\lambda_{\mathrm{tot}}=0}} = 0.
\end{equation}

We use these two conditions and start from the lowest row $l=1$ to $l=N-1$.
For $l=1$, the relevant elements of $M$ is $m_{11} \in [m_{22}, m_{12}]$. Therefore, we denote $c_{M,M^\prime}$ with $c_{m_{11}^\prime}^{m_{11}}$ for $l=1$.
Equation~\eqref{eq:app_general_z_weight} gives 
\begin{equation}\label{eq:SUn_CG_A}
    m_{11}+m^{\prime}_{11} = \frac{1}{2}(m_{12} + m_{22} + m^{\prime}_{12} + m^{\prime}_{22}) \equiv A.
\end{equation}
Therefore, $m^{\prime}_{11} = A-{m_{11}}$ determined by $m_{11}$.
The relevant lowering operator $S^-_{(1)}$ gives
\begin{widetext}
\begin{equation}
\begin{aligned}
    0 &= S_{(1)}^{-} \ket{M_{\lambda_{\mathrm{tot}}=0}},\\
      &= \sum_{\substack{m_{11}\in [m_{22}, m_{12}] \\ 
      m^{\prime}_{11}\in [m^{\prime}_{22}, m^{\prime}_{12}]}} c^{m_{11}}_{m^{\prime}_{11}} \left( S_{(1)}^{A, -}|m_{11}\rangle |m^{\prime}_{11}\rangle +  |m_{11}\rangle S_{(1)}^{B,-}|m^{\prime}_{11}\rangle \right), \\
      &= \sum_{\substack{m_{11}\in [m_{22}+1, m_{12}] \\ 
      m^{\prime}_{11}\in [m^{\prime}_{22}, m^{\prime}_{12}]}} c^{m_{11}}_{m^{\prime}_{11}} a_{11}^{m_{11}}|m_{11}-1\rangle |m^{\prime}_{11}\rangle + \sum_{\substack{m_{11}\in [m_{22}, m_{12}] \\ 
      m^{\prime}_{11}\in [m^{\prime}_{22}+1, m^{\prime}_{12}]}} c^{m_{11}}_{m^{\prime}_{11}} (a^{\prime}_{11})^{m^{\prime}_{11}}|m_{11}\rangle |m^{\prime}_{11}-1\rangle, \\
      &= \sum_{\substack{m_{11}\in [m_{22}, m_{12}-1] \\ 
      m^{\prime}_{11}\in [m^{\prime}_{22}, m^{\prime}_{12}]}} c^{m_{11}+1}_{m^{\prime}_{11}} a_{11}^{m_{11}+1} |m_{11}\rangle |m^{\prime}_{11}\rangle + \sum_{\substack{m_{11}\in [m_{22}, m_{12}] \\ 
      m^{\prime}_{11}\in [m^{\prime}_{22}, m^{\prime}_{12}-1]}} c^{m_{11}}_{m^{\prime}_{11}+1} (a^{\prime}_{11})^{m^{\prime}_{11}+1} |m_{11}\rangle |m^{\prime}_{11}\rangle,
\end{aligned}
\end{equation}
\end{widetext}
where $a_{11}$ and $a_{11}^{\prime}$ are calculated from Eq.~\eqref{eq:SUn_raising_lowering_coeficients}.
For all $m_{11}\in [m_{22}, m_{12}-1]$ and $m^{\prime}_{11} \in [m^{\prime}_{22}, m^{\prime}_{12}-1]$,
\begin{equation}\label{eq:CG_l1}
    0 = c^{m_{11}+1}_{m^{\prime}_{11}} a_{11}^{m_{11}+1} + c^{m_{11}}_{m^{\prime}_{11}+1} (a^{\prime}_{11})^{{m}^\prime_{11}+1}.
\end{equation}
From Eq.~(\ref{eq:CG_l1}), by iteration starting from $m_{11} = m_{12}-1$, and $m^{\prime}_{11} = A-m_{12}$, 
\begin{equation}
    c^{m_{12}-x}_{A-m_{12}+x} a_{11}^{m_{12}-x} + c^{m_{12}-x-1}_{A-m_{12}+x+1} (a^{\prime}_{11})^{A-m_{12}+x+1} = 0.
\end{equation}
for all $x \in [0, m_{12}-m_{22}-1]$.
This requires that $m^{\prime}_{11} = A-m_{12}+x+1 \in [m^{\prime}_{22}, m^{\prime}_{12}]$. If $m^{\prime}_{11}$ is out of range for some $x$, this gives $c_{A-m_{12}-x}^{m_{12}+x} = 0$ for all $x$. Therefore, the non-zero coefficients are given when $m_{12} - m_{22} = m_{12}^\prime - m^\prime_{22}$, or equivalently, $A = m_{12} + m^\prime_{22} = m_{12} + m_{12}^\prime = m_{11} + m_{11}^\prime$.
Using this condition, we obtain $a_{11}^{m_{12}-x} = (a^{\prime}_{11})^{A-m_{12}+x+1}$ with Eq.~\eqref{eq:SUn_raising_lowering_coeficients}.
Therefore, we conclude from the $l=1$, 
\begin{equation}
\begin{aligned}
    &|c_{m_{11}^\prime}^{m_{11}}| = c, \mathrm{with}\, \, m_{11} + m_{11}^\prime = A, \\
    & A = m_{12} + m^\prime_{22} = m_{12}^\prime + m_{22}.
\end{aligned}
\end{equation}
This is exactly the selection rule for the singlet subspace of SU($2$), as $J = m_{12} - m_{22} = m^\prime_{12} - m^{\prime}_{22} = J^\prime$.

Following the same strategy and by iteration from lower rows to higher, we can prove that for general $k$ and $l$,
\begin{equation}
    m_{k, l} + m^\prime_{l-k+1, l} = A.
\end{equation}
Up to the first row $l=N$, i.e., $(m_{1,N}, m_{2,N} ,\ldots, m_{N,N}) = (\lambda_1, \lambda_2, \ldots, \lambda_{N})$, 
\begin{equation}
    A = \lambda_1 + \lambda^\prime_N = \lambda_2 + \lambda^\prime_{N-1} = \ldots = \lambda_{N} + \lambda^\prime_{1}.
\end{equation}
Therefore, for fixed $\lambda = (\lambda_1, \ldots, \lambda_N)$, we obtain $\lambda^\prime = (A-\lambda_N, \ldots, A - \lambda_1) = \bar{\lambda}$ (for certain constant $A$), which is the dual of $\lambda$.
Note that the Schur-Weyl duality requires that $\lambda_1 + \lambda_2 + \ldots \lambda_N = L_A$ and $\lambda_1^\prime + \lambda_2^\prime + \ldots \lambda_N^\prime = L_B$, thus determining the value of $A=L/N$ and possible range of $\lambda$.

To sum up, for fixed GT pattern $M = (m_{k,l})$, there is a unique $M^\prime = \bar{M}$ with $m^\prime_{k,l} = L/N-m_{l-k+1, l}$, such that $c_{M,M^\prime}$ is non-zero, 
Moreover, the CG coefficients  $c_{M, \bar{M}}$ are constant up to a minus sign for fixed $\lambda$. 
With the normalization of the basis states, 
\begin{equation}
    |c_{M, M^\prime}| = \frac{1}{\sqrt{d_M}} \delta_{M^\prime, \bar{M}}.
\end{equation}
Therefore, denote $\ket{M} = \ket{\lambda, m}$, where the irrep labels $\lambda$ and $m$ are given by the GT patterns $M$, we have proven that
\begin{equation}
    |\lambda_{\mathrm{tot}}=0; \lambda, m\rangle = \frac{1}{\sqrt{d_\lambda}} \sum_m \eta_{\lambda,m}|\lambda, m\rangle |\bar{\lambda}, \bar{m}\rangle,
\end{equation}
with $|\eta_{\lambda,m}|=1$.

\subsection{Asymptotic scaling for SU$(N)$}
For SU$(N)$ symmetry on a chain with local Hilbert space dimension $N$, the dimension of irreps is given by
\begin{equation}\label{eq:SUN_irrep_dim_app}
\begin{aligned}
    d_\lambda &= \frac{1}{(N-1)!(N-2)!\ldots 1!}\prod_{1\leq i < j\leq N} (\tilde{\lambda}_i - \tilde{\lambda}_j),\\
    D_\lambda^{(L)} &= \frac{L!}{\tilde{\lambda}_1!\tilde{\lambda}_2! \ldots \tilde{\lambda}_N!} \prod_{1\leq i < j\leq N} (\tilde{\lambda}_i - \tilde{\lambda}_j),
\end{aligned}
\end{equation}
where $\tilde{\lambda}_i = \lambda_i + N-i$.

First, we prove the lower and upper bound of scaling coefficients of $R_3$ shown in Fig.~\ref{fig:R3_SUd}. 
With Eq.~\eqref{eq:general_Rn} and Eq.~\eqref{eq:SUN_irrep_dim_app}, the $R_3$ of SU$(N)$ on half-chain bipartition of a chain with length $L=2nN$, $n\in \mathbb{N}$ is
\begin{equation}\label{eq:R3_SUN_exact}
    R_3 = - \log \frac{1}{D_0} \sum_{\lambda} \frac{[(L/2)! (N-1)!(N-2)!\ldots1!]^2}{\Tilde{\lambda}_1!\ldots\Tilde{\lambda}_N!\Tilde{\bar{\lambda}}_1!\ldots\Tilde{\bar{\lambda}}_N!},
\end{equation}
where $\bar{\lambda}_i = \frac{L}{N} - \lambda_i$ and $\tilde{\bar{\lambda}}_i = \bar{\lambda}_i + N-i$. 
The dimension $D_0^{(L)}$ scales as 
\begin{equation}\label{eq:SUn_D0_scaling}
\begin{aligned}
    D_0^{(L)} 
    &= \frac{L!(N-1)! (N-2)! \ldots 1!}{(\frac{L}{N}+N-1)!(\frac{L}{N}+N-2)!\ldots\frac{L}{N}!}, \\
    &= \frac{(N-1)! (N-2)! \ldots 1!}{(\frac{L}{N}+N-1)\ldots(\frac{L}{N}+1)^{N-1}} \begin{pmatrix}L\\
    \frac{L}{N}, \frac{L}{N}, \ldots, \frac{L}{N}
    \end{pmatrix},\\
    &\sim O( N^L L^{-N(N-1)/2} L^{1/2-N/2}),
\end{aligned}
\end{equation}
where we use the asymptotic of multinomial coefficients,
\begin{equation}\label{eq:asy_multinomial}
\begin{aligned}
    \begin{pmatrix}n\\
    x_1, \ldots , x_N
    \end{pmatrix} &\sim (2\pi n)^{1/2-N/2} N^{n+N/2} \\
    &\,\,\,\times \exp \left(-\frac{N}{2n} \sum_{i=1}^{N} (x_i-n/N)^2\right).
\end{aligned}
\end{equation}
The terms inside the summation of Eq.~\eqref{eq:R3_SUN_exact} can be written as multinomials. Define
\begin{equation}\label{eq:multinomial_na}
    M_{n} = \sum_{x_1>\ldots>x_N} \begin{pmatrix}n\\
    x_1, \ldots , x_N
    \end{pmatrix}\begin{pmatrix}n\\
    a-x_1, \ldots, a-x_N
    \end{pmatrix},
\end{equation}
with $x_i = \tilde{\lambda}_i$, $a=L/N + N-1$, and $n = \frac{L+(N-1)N}{2}$. 
The value $M_n$ is upper bounded by summation without restriction, 
\begin{equation}\label{eq:upperbound_Mn}
\begin{aligned}
     M_{n} &\leq \sum_{x_1,\ldots,x_N} \begin{pmatrix}n\\
    x_1, \ldots , x_N
    \end{pmatrix}\begin{pmatrix}n\\
    a-x_1, \ldots, a-x_N
    \end{pmatrix},\\
    &= \begin{pmatrix}2n\\
    a, a, \ldots, a
    \end{pmatrix},\\
    &\sim (4\pi n)^{1/2-N/2} N^{2n+N/2},\\
    &= O(N^L L^{1/2-N/2}),
\end{aligned}
\end{equation}
where we use the Vandemondes's identity.
An asymptotic lower bound for $M_n$ is given by
\begin{equation}
\begin{aligned}
    M_{n} 
    &\geq \begin{pmatrix}n\\
    \frac{L}{2N}+N-1, \frac{L}{2N}+N-2, \ldots, \frac{L}{2N}
    \end{pmatrix}^2, \\
    &\sim (2\pi n)^{1-} N^{2n+N},\\
    &= O(N^L L^{1-N}),
\end{aligned}
\end{equation}
since all the terms are positive, and $x_i = \frac{L}{2N}+N-i$.
We omit the exponential term in Eq.~\eqref{eq:asy_multinomial} for $N \ll L$.

With the upper and lower bound of $M_n$, we obtain
\begin{equation}
\begin{aligned}
    R_3 &= -\log \frac{1}{D_0^{(L)}} [\frac{1}{(\frac{L}{N}+\frac{N(N-1)}{2})...(\frac{L}{2}+1)}]^2 M_{n}, \\
    &\sim -\log \frac{1}{D_0^{(L)}} (\frac{L}{N})^{-N(N-1)} M_n,\\
    &= c_N^{R_3} \log L + O(1),
\end{aligned}
\end{equation}
with $\frac{N(N-1)}{2}\leq c_N^{R_3} \leq \frac{N^2-1}{2}$.

Next, we prove an upper bound of the logarithmic negativity on a half-chain bipartition,
\begin{equation}
\begin{aligned}
    E_{\mathcal{N}} &= \log \frac{1}{D_0^{(L)}} \sum_\lambda \frac{1}{(N-1)!(N-2)!\ldots 1!}\\
     & \times\left[\prod_{1\leq i \leq j \leq N} (\tilde{\lambda}_i -\tilde{\lambda}_j)\right]^3  \frac{(L/2)!}{\tilde{\lambda}_1! \ldots \tilde{\lambda}_N!}  \frac{(L/2)!}{\tilde{\bar{\lambda}}_1!\ldots \tilde{\bar{\lambda}}_N!}.
\end{aligned}
\end{equation}
We adopt the notation as before, $x_i = \tilde{\lambda}_i$, $n=\frac{L+N(N-1)}{2}$, and $a=L/N+N-1$.
Note that for $x_1 > x_2 > \ldots > x_N$ and $x_1 + x_2 + \ldots + x_N = n$,
\begin{equation}
\begin{aligned}
    \left[\prod_{1\leq i <j\leq N} (x_i - x_j)\right]^3 
    &\leq \prod_{1\leq i <j\leq N} |x_i - x_j|^3 \\
    &\leq \prod_{1\leq i <j\leq N} n^3\\
    &\leq n^{\frac{3N(N-1)}{2}},
\end{aligned}
\end{equation}
as the product only has $N(N-1)/2$ terms, and $|x_i-x_j|\leq n$.
Therefore,
\begin{equation}
\begin{aligned}
    &\sum_{x_1 > x_2 > \ldots > x_N} \left[\prod_{1\leq i <j\leq N} (x_i - x_j)\right]^3 \\ 
    &\times \,\,\, \begin{pmatrix}n\\
    x_1, ... , x_N
    \end{pmatrix}\begin{pmatrix}n\\
    a-x_1, ..., a-x_N
    \end{pmatrix} \\
    \leq & n^{\frac{3N(N-1)}{2}} \begin{pmatrix}2n\\
    a,a,\ldots,a
    \end{pmatrix}\\
    \sim& L^{\frac{3N(N-1)}{2}} N^L L^{1/2-N/2} + O(1),
\end{aligned}
\end{equation}
where we used the unrestricted summation as in Eq.~\eqref{eq:upperbound_Mn} as an upper bound.
Therefore, with the scaling of $D_0^{(L)}$ (Eq.~\eqref{eq:SUn_D0_scaling}), the logarithmic negativity is upper bounded by 
\begin{equation}
    E_{\mathcal{N}} \leq N(N-1) \log L + O(1).
\end{equation}
This bound is compatible with the general bound $E_{\mathcal{N}}\leq \log [\dim \mathcal{C}_{\mathrm{SU}(N)} (L/2)] \sim (N^2-1)\log L$.

For OSE, we can use the same upper bound of $M_n$, which gives a lower bound of OSE as
\begin{equation}
    S_{\mathrm{OP}} \geq \frac{N^2-1}{2} \log L + O(1).
\end{equation}
Note that the OSE is also upper bounded by the dimension of the commutant, thus
\begin{equation}
    S_{\mathrm{OP}} \leq (N^2-1) \log L + O(1). 
\end{equation}
These analytic bounds for $E_{\mathcal{N}}$ and $S_{\mathrm{OP}}$ are tested numerically.

\section{Numerical results of dynamics}\label{app:numerics_dynamics}
\begin{figure*}[bt]
\includegraphics[width=18cm, scale=1]{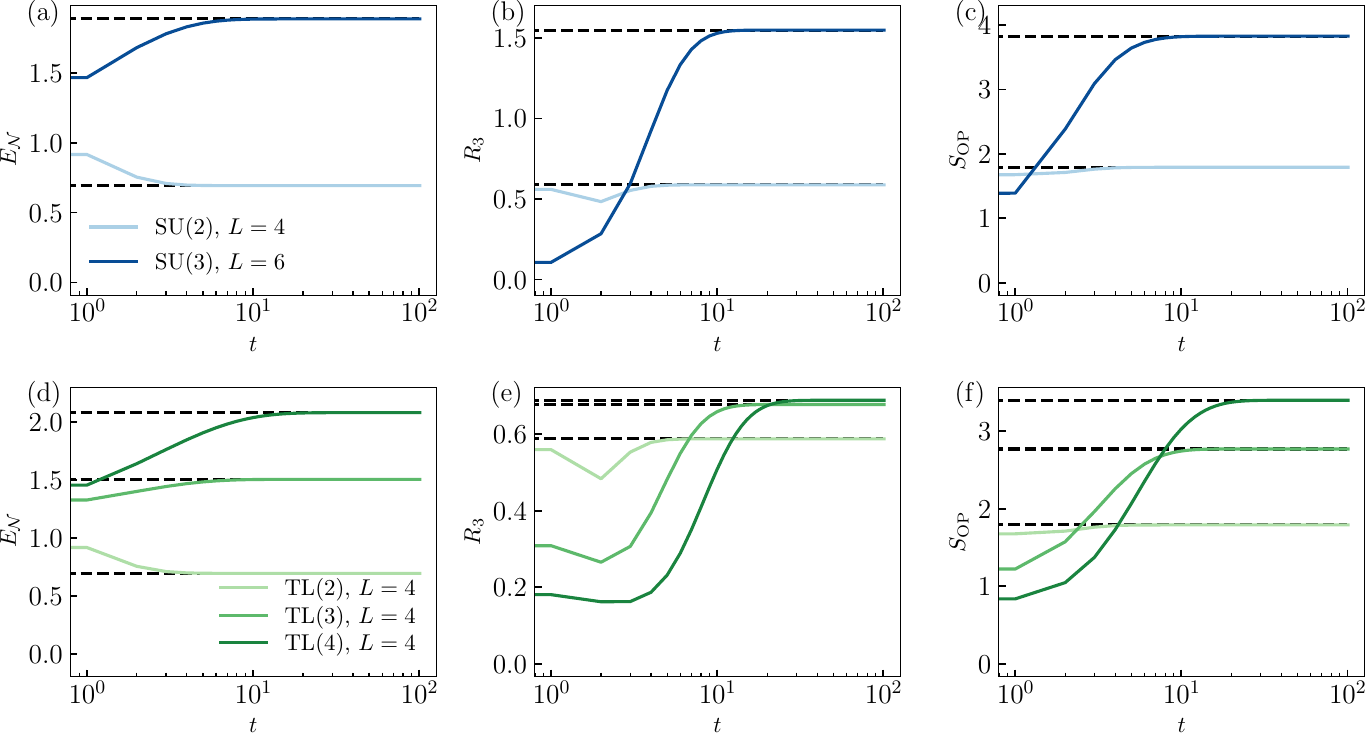}
\caption{\label{fig:Entanglement_dynamics}\textbf{Entanglement dynamics under symmetric quantum channels.} (a-c) Dynamics of $E_{\mathcal{N}}$, $R_3$, and $S_{\mathrm{OP}}$ under quantum channels with SU$(N)$ symmetry, on a spin chain with system size $L=2N$ and local Hilbert space dimension $N$.
(d-f) Dynamics of the entanglement under quantum channels that preserves the $\mathcal{C}_{\mathrm{TL}}(N)$ commutant, on a spin chain with system size $L=4$ with local Hilbert space dimension $N$.
The dashed lines show the values obtained from exact expressions in Sec.~\ref{sec:exact_entanglement}.}
\end{figure*}

In this section, we provide numerical results of quantum channels with SU$(N)$ symmetries and the RS commutant (i.e., dynamics generated by TL($N$) algebra) as strong symmetries, and show that the entanglement indeed saturates to the value given by exact expressions in Sec.~\ref{sec:exact_entanglement}.
For SU$(N)$ symmetry and  he RS commutant, the bond algebras can be generated by 
\begin{equation}
\begin{aligned}
    &\mathcal{A}_{\mathrm{SU}(N)}(L) = \langle \{ P_{j,j+1} \}, \mathbb{1} \rangle,\\ 
    &\mathcal{A}_{\mathrm{TL}(N)}(L) = \langle \{ e_{j,j+1} \},\mathbb{1} \rangle,
\end{aligned}
\end{equation}
respectively, where $P_{j,j+1}$ is the permutation operators of the two neighboring spins, and $e_{j,j+1}$ is the singlet projector.
Therefore, we can choose the Kraus operators for the quantum channels as 
\begin{equation}
\begin{aligned}
   & K_{j,\alpha}^{\mathrm{SU}(N)} = \{ (\mathbb{1} + P_{j,j+1})/2, (\mathbb{1} - P_{j,j+1})/2 \},\\
    &K_{j,\alpha}^{\mathrm{TL}(N)} = \{ e_{j,j+1}/3, \mathbb{1} - e_{j,j+1}/3\}.
\end{aligned}
\end{equation}
Note that they satisfy $\sum_{\alpha} K_{j,\alpha}^\dagger K_{j,\alpha} = \mathbb{1}$.
Figure~\ref{fig:Entanglement_dynamics} shows the dynamics of logarithmic negativity $E_\mathcal{N}$, third Rényi negativity $R_3$ and operator space entanglement $S_{\mathrm{OP}}$ at half-chain bipartition. For SU$(N)$ with $N=2, 3$, the system size is $L=2N$ with local Hilbert space dimension $N$. For TL($N$) with $N=2, 3, 4$, the system size is $L=4$ with local Hilbert space dimension $N$.
The data shows that these entanglement quantities saturate exactly to the analytic values obtained from the general expressions.
In addition, note that the TL($2$) model maps to the SU($2$) model via an onsite unitary transform. Therefore, the saturation values of all entanglement quantities are equal.

\section{Quantum fragmentation with Read-Saleur commutant}\label{app:fragment}

\subsection{CG for small system sizes}
In App.~\ref{app:Hopf}, we prove analytically that for the trivial one-dimensional irrep associated with the Hopf algebras, the basis states can be written as Eq.~\eqref{eq:gen_triv0}. 
Here we provide some examples of bipartition of basis states for systems with RS commutants and smaller system sizes.

Consider $\mathrm{TL}(3)$ bond algebra generated by $e_{j,j+1}$. The basis states are given by singlet and dot patterns, i.e., each basis state can be written as e.g.,  $\ket{\twodimer\,\,\,\,\twodot\,\,\,\dimer \,\,\ldots}$, which can be separated into direct products of the singlet part and the dot pattern part~\cite{2007_Read_Commutant,2023_sanjay_commutant_symmetries}. The singlet patterns consist of $\ket{\dimer}_{j,k} = \frac{1}{\sqrt{3}} (\ket{00} + \ket{11} + \ket{22})_{j,k}$. And the dot patterns are annihilated by all $e_{j,j+1}$. Therefore, all states with $L$ dots ($\lambda = L/2$) on a chain with length $L$ are the ground states of the frustration-free Hamiltonian $H_{\mathrm{TL}} = \sum_j e_{j,j+1}$, which can be numerically obtained by exact diagonalization.

For $L=2$, there is one state in the singlet subspace ($\lambda = 0$ and $d_{\lambda=0} = 1$), $\ket{\dimer}$, and eight dot states $\ket{\twodot}$ of two dots ($\lambda = 1$ and $d_{\lambda=1} = 8$). The dot states can be chosen as six product states, $\ket{\sigma \sigma^\prime}$ with $\sigma, \sigma^\prime = 1, 2, 3$ and $\sigma \neq \sigma^\prime$, as well as two entangled dot states $\frac{1}{\sqrt{2}}(\ket{00} + \ket{11})$ and $\frac{1}{\sqrt{6}} (\ket{00} - \ket{11} + 2\ket{22})$.
For $L=4$, the singlet subspace is two dimensional $D_{\lambda=0}^{(L=4)}=2$, spanned by two linearly-independent states $\ket{\dimer\,\,\,\dimer}$ and $\ket{\twodimer}$. 
An orthogonal basis of the singlet subspace is
\begin{equation}
\begin{aligned}
    &\ket{L=4, \lambda_{\mathrm{tot}} = 0; \lambda = 0} = \ket{\dimer\,\,\,\dimer}, \\
    &\ket{L=4, \lambda_{\mathrm{tot}} = 0; \lambda =1} = \frac{1}{\sqrt{6}} (3\ket{\twodimer} -  \ket{\dimer\,\,\,\dimer}).
\end{aligned}
\end{equation}
We can see that the first state is indeed given by the direct product of $\lambda=0$ singlets, $\ket{\dimer}$. Also, we can verify by hand that the second state equals to sum of dot patterns, $\sum \frac{1}{\sqrt{8}} \ket{\twodot}\otimes \ket{\overline{\twodot}}$, with $\ket{\overline{\twodot}} = \ket{\twodot}$ for the two entangled dot states and $\ket{\overline{\twodot}} = \ket{\sigma^\prime \sigma}$ for $\ket{{\twodot}} = \ket{\sigma \sigma^\prime}$.
Below we sketch a numerical method to verify these coefficients, which we perform for system size $L=8$.
First, we obtain the dot patterns with $\lambda = L/2$ on $L$ sites as the ground states of $H_{\mathrm{TL}}$. The basis states can be constructed by iteratively acting $e_{j,j+1}$ on the root state $\ket{\dimer\,\,\,\ldots \twodot \ldots}$.
Denote the basis states as $\ket{L, \lambda, n, a}$, where $2\lambda$ is the number of dots, $n=1,\ldots,d_\lambda$ as the degeneracy, and $a=1,\ldots, D_\lambda$ as the different states in the same Krylov subspace.
Note that we use $n$ here, as the numerically obtained orthonormal basis for $\lambda$ subspace might not have a definite $m$ value.
Second, calculate the coefficient matrix $\tilde{C}^{(\lambda, a, b)}$ for fixed $\lambda, a, b$, with matrix elements
\begin{equation}
    \tilde{C}_{n n^\prime}^{(\lambda, a, b)} = \bra{L, \lambda_{\mathrm{tot}}=0, m_{\mathrm{tot}=0}; c}\ket{L_A, \lambda, n; a}\ket{L_B, \lambda, n^\prime; b},
\end{equation}
where $c=1,\ldots,D^{(L)}$ goes over all basis states in the singlet subspace. 
We can verify that the matrix $\tilde{C}^{\lambda,a,b}$ is either a zero matrix or a matrix with $d_\lambda$ eigenvalues that squares to $1/d_{\lambda}$.
In the former case, it means that the left and right bipartition of the singlet state are not labeled by $\lambda$.
For the latter case, it means that by diagonalization of $C^{\lambda,a,b} = U\tilde{C}^{\lambda,a,b}U^\dagger$, we obtain a new basis $\ket{L_A(B), \lambda,m;a(b)}$ in which the $c_{m,m^\prime}(\lambda,\lambda) = (C^{\lambda,a,b})_{m,m^\prime} = \pm\delta_{m,\bar{m}}$ for certain $m,\bar{m}$. The singlet state recovers the general form of bipartition of basis state, and it corresponds to $\ket{L, \lambda_{\mathrm{tot}}=0, m_{\mathrm{tot}}=0, c} = \ket{\lambda_{\mathrm{tot}}=0;\lambda; a, b}$. 
For example, for the state $\ket{\dimer\,\,\,\dimer}$,  $C_{nn^\prime}^{(\lambda=0, a, b)} = 1$, while $C_{nn^\prime}^{(\lambda=1, a, b)}$ is a zero matrix.

\subsection{Asymptotic scaling for RS commutant $\mathcal{C}_{\mathrm{TL}(N)}$}\label{app:fragment_asym}
For the TL($N$) model, the dimension of Krylov subspaces $D_{\lambda}$ is identical to SU($2$) symmetry, with a much larger degeneracy
\begin{equation}
    d_{\lambda} = [2\lambda + 1]_q, 
\end{equation}
where $[n]_q = \frac{q^n - q^{-n}}{q - q^{-1}}$, and $q$ is defined as $q + q^{-1} = N$. 

The logarithmic negativity is given by $E_{\mathcal{N}} = \log \frac{1}{D_0^{(L)}} \sum_\lambda A_\lambda$, with $A_{\lambda} \equiv d_{\lambda} (D_{\lambda}^{(L/2)})^2$ at half-chain bipartition for $L=4n$. 
Therefore, a lower bound is given by $E_{\mathcal{N}} \geq \log A_{\lambda_{\mathrm{max}}}$ for certain $\lambda_{\mathrm{max}}$ as all $A_{\lambda}$ are positive. We choose the largest term $A_{\lambda_{\mathrm{max}}}$ such that an optimal lower bound is obtained.
Consider a term $\lambda = a L$ with $0<a<1$ and $a=O(1)$. Using the exact expressions of $E_{\mathcal{N}}$ and the asymptotic scaling of binomial coefficients when $k=O(L)$,
\begin{equation}
    \begin{pmatrix}n\\
    k
    \end{pmatrix} \sim \sqrt{\frac{n}{2\pi k(n-k)}} \frac{n^n}{k^k (n-k)^{n-k}},
\end{equation}
we obtain $\log(A_{a L}/D_0^{(L)}) \sim c_a L + \frac{1}{2}\log L + O(1)$, with 
\begin{equation}\label{eq:TLN_max_linear_c}
\begin{aligned}
    c_a =& -(\frac{1}{2}+2a)\log(\frac{1}{4}+a)-(\frac{1}{2}-2a)\log(\frac{1}{4}-a)\\
    &\,\, +2a\log q -2\log2.
\end{aligned}
\end{equation}
Note that we use the approximation $d_{aL} \propto (q^{aL} - q^{-aL}) \sim q^{aL}$ as $L\rightarrow\infty$.  
To obtain the largest lower bound, we calculate $0 \stackrel{!}{=} \frac{d c_a}{dc}$, which gives a maximal $c_{a}$ for Eq.~\eqref{eq:TLN_max_linear_c}
obtained when
\begin{equation}\label{eq:TL_amax}
    a_{\mathrm{max},q} = \frac{1}{4}\frac{q-1}{q+1}.
\end{equation}
Therefore, the logarithmic negativity is lower bounded by 
\begin{equation}
    E_{\mathcal{N}} \geq c_{\mathrm{TL}(N)}^{E_{\mathcal{N}}} L + O(\log L)
\end{equation}
where $c_{\mathrm{TL}(N)}^{E_{\mathcal{N}}} = c_{a_{\mathrm{max}, q}}$ given by Eq.~\eqref{eq:TLN_max_linear_c} and Eq.~\eqref{eq:TL_amax}, which is dependent on $N = q + q^{-1}$.
As shown in Fig.~\ref{fig:Entanglement_frag}, the logarithmic negativity is indeed lower bounded by a linear scaling with $c_{\mathrm{TL}(3)}^{E_{\mathcal{N}}} \approx 0.1116$.

For the Rényi negativity $R_n$, Eq.~\eqref{eq:general_Rn} shows that for $n > 2$, 
\begin{equation}\label{eq:general_Rn_app}
    R_n = -\log \left(\frac{1}{D_0^{(L)}} \sum_{\lambda} \frac{(D_{\lambda}^{(L/2)})^2}{d_{\lambda}^{n-1}} \right) \leq R_{n+1},
\end{equation}
as $d_{\lambda} \geq 1$.
This is also valid for non-integer $n>2$.
For $n\rightarrow\infty$, $R_n$ is only contributed by $d_{\lambda = 0} = 1$, the non-degenerate singlet subspace, which means 
\begin{equation}
    R_{\infty} = - \log \left(\frac{(D_{0}^{(L/2)})^2}{D_0^{(L)}} \right) \sim \frac{3}{2}\log L + O(1).
\end{equation}
Therefore, we have $R_n \leq R_{\infty} \sim \frac{3}{2} \log L + O(1)$ for integer $n>2$. It shows that Rényi negativities scales at most logarithmically.

As discussed in the main text, we introduce $\tilde{R}_n$ to understand the transition from linear law in logarithmic negativity to logarithmic law in Rényi negativities.
First, for real numbers $n>2$, note that Eq.~\eqref{eq:general_Rn_app} is also valid, i.e., $(n-2) \tilde{R}_n \leq (m-2) \tilde{R}_m \leq R_{\infty}$ for real $n<m$. Therefore, 
\begin{equation}
    \tilde{R}_n \leq \frac{3}{2(n-2)}\log L, \, n >2.
\end{equation}
Now consider $n<2$, $\tilde{R}_n = \frac{1}{2-n} \log (\frac{1}{D_0^{(L)}} \sum_{\lambda} A_{\lambda_{\mathrm{max}}, n}) \geq \frac{1}{2-n} \log (\frac{1}{D_0^{(L)}} A_{\lambda, n})$, with $A_{\lambda,n} \equiv d_{\lambda}^{2-n} (D_{\lambda}^{(L/2)})^2$.
Similar to the case of logarithmic negativity, take $\lambda = aL$. As $L \rightarrow \infty$, $d_{2 a L} \sim q^{a L}$, thus $A_{cL, n}(q) \approx A_{cL} (q^{2-n})$, where $q$ is modified by a power of $2-n$.
Using the conclusion from $E_{\mathcal{N}}$, take $\lambda = \tilde{a} L$, we have $\log (A_{\tilde{a} L}/D_0^{(L)}) \sim \tilde{c}_{\tilde{a},n} L + O(\log L)$, with
\begin{equation}\label{eq:TLN_max_linear_c_Rn}
\begin{aligned}
    \tilde{c}_{\tilde{a},n} =& -(\frac{1}{2}+2\tilde{a})\log(\frac{1}{4}+\tilde{a})-(\frac{1}{2}-2\tilde{a})\log(\frac{1}{4}-\tilde{a})\\
    &\,\, +2(2-n)\tilde{a} \log q -2\log2,
\end{aligned}
\end{equation}
The maximum of $\tilde{c}_{\tilde{a},n}$ is obtained when
\begin{equation}\label{eq:TLN_amax_Rn}
    \tilde{a}_{\mathrm{max}, q, n} = \frac{1}{4} \frac{q^{2-n}-1}{q^{2-n}+1}.
\end{equation}
We can obtain that
\begin{equation}\label{eq:tilde_Rn_scaling_app}
    \tilde{R}_n \geq \tilde{c}_{N,n}^{\mathrm{lin}} L + O(\log L), \text{ with } n <2,
\end{equation}
with $\tilde{c}_{N,n}^{\mathrm{lin}} = \frac{1}{2-n} \tilde{c}_{\tilde{a}_{\mathrm{max}, q, n}}$ given by Eq.~\eqref{eq:TLN_max_linear_c_Rn} and Eq.~\eqref{eq:TLN_amax_Rn}.
Note that $\tilde{c}_{N,n}^{\mathrm{lin}} \rightarrow 0$ as $n \rightarrow 2$.
Therefore, we prove that $\tilde{R}_n$ has a transition across $n=2$ from linear scaling $(n<2)$ to logarithmic scaling $n>2$, for general $N$. 

Lastly, consider the operator space entanglement. From Eq.~\eqref{eq:general_SOP}, 
\begin{equation}
\begin{aligned}
    S_{\mathrm{OP}}^{\mathrm{TL}(N)} &= S_{\mathrm{OP}}^{\mathrm{SU}(2)} - \sum_{\lambda} \frac{(D_{\lambda}^{(L/2)})^2 }{D_0^{(L)}} \log (2\lambda+1)^2 \\
    &\,\,+ \sum_{\lambda} \frac{(D_{\lambda}^{(L/2)})^2 }{D_0^{(L)}} \log d_\lambda^2.
\end{aligned}
\end{equation}
The first term $S_{\mathrm{OP}}^{\mathrm{SU}(2)} \sim \frac{3}{2}\log L$, and second term is upper bounded by $0$ and lower bounded by $2\log(L/2+1)$, which are of the order $O(\log L)$. For the third term, notice that $d_{\lambda} \sim q^{2\lambda+1}$ because it grows exponentially fast with $\lambda$. Therefore, the third term 
\begin{equation}
\begin{aligned}
    \sum_{\lambda} \frac{(D_{\lambda}^{(L/2)})^2 }{D_0^{(L)}} 2\log d_\lambda &\sim \sum_\lambda \frac{(D_{\lambda}^{(L/2)})^2 }{D_0^{(L)}} 2(2\lambda+1) \log q, \\
    &\sim 2\log q \frac{2^L}{\pi L/4} \left(\frac{1}{L/2+1}\frac{2^L}{\sqrt{\pi L/2}}\right)^{-1},\\
    &\sim (\sqrt{\frac{8}{\pi}} \log q) \sqrt{L},
\end{aligned}
\end{equation}
where we used Eq.~\eqref{eq:app_D0_summations}.
Thus for TL($N$),
\begin{equation}
    S_{\mathrm{OP}} \sim (\sqrt{\frac{8}{\pi}} \log q) \sqrt{L} + O(\log L),
\end{equation}
with $q + q^{-1} = N$, 
which scales as $O(\sqrt{L})$ for general $N$.
%

%
==

\end{document}